\pdfoutput=1

\documentclass[11pt,twoside,a4paper,cmspaper,final,collab]{cms-tdr}
\def\svnVersion{260985}\def\cmsCernNo{2013-037}\def\cmsCernDate{\today}\def\cmsMessage{Published in the Journal of High Energy Physics as \href{http://dx.doi.org/10.1007/JHEP09(2014)087}{\doi{10.1007/JHEP09(2014)087.}}}

\begin{document}\cmsNoteHeader{HIG-13-029}

\hyphenation{had-ron-i-za-tion}
\hyphenation{cal-or-i-me-ter}
\hyphenation{de-vices}
\RCS$Revision: 255100 $
\RCS$HeadURL: svn+ssh://svn.cern.ch/reps/tdr2/papers/HIG-13-029/trunk/HIG-13-029.tex $
\RCS$Id: HIG-13-029.tex 255100 2014-08-07 18:49:40Z alverson $
\newlength\cmsFigWidth
\ifthenelse{\boolean{cms@external}}{\setlength\cmsFigWidth{0.85\columnwidth}}{\setlength\cmsFigWidth{0.4\textwidth}}
\ifthenelse{\boolean{cms@external}}{\providecommand{\cmsLeft}{top}}{\providecommand{\cmsLeft}{left}}
\ifthenelse{\boolean{cms@external}}{\providecommand{\cmsRight}{bottom}}{\providecommand{\cmsRight}{right}}

\newcolumntype{M}{>{\centering\arraybackslash}m{\dimexpr.28\linewidth-2\tabcolsep}}
\newcolumntype{N}{>{\centering\arraybackslash}m{\dimexpr.23\linewidth-2\tabcolsep}}

\newcommand{\ttH}{\ensuremath{\ttbar\PH}\xspace}
\newcommand{\muTTH}{\ensuremath{\mu_{\ttH}}\xspace}
\newcommand{\sigmaTTH}{\ensuremath{\sigma_{\ttH}}\xspace}
\newcommand{\sigmaTTHSM}{\ensuremath{\sigma_{\ttbar\PH_{\mathrm{SM}}}}\xspace}
\newcommand{\kV}{\ensuremath{\kappa_{\mathrm{V}}}\xspace}
\newcommand{\kf}{\ensuremath{\kappa_{\mathrm{f}}}\xspace}
\providecommand{\tauh}{\ensuremath{\tau_\mathrm{h}}\xspace}

\cmsNoteHeader{HIG-13-029}
\title{Search for the associated production of the Higgs boson with a top-quark pair}

\date{\today}

\abstract{A search for the standard model Higgs boson produced in
association with a top-quark pair ($\ttbar \PH$) is presented, using
data samples corresponding to integrated luminosities of up to 5.1\fbinv and 19.7\fbinv collected in
pp collisions at center-of-mass energies of
7\TeV and 8\TeV respectively.  The search is
based on the following signatures of the Higgs boson decay: $\PH
\to$ hadrons, $\PH \to$ photons, and $\PH \to$ leptons.  The results
are characterized by an observed $\ttbar\PH$ signal
strength relative to the standard model cross section,
$\mu=\sigma/\sigma_{\mathrm{SM}}$, under the assumption that the
Higgs boson decays as expected in the standard model.  The best fit
value is $\mu=2.8 \pm 1.0$ for a Higgs boson mass of 125.6\GeV.}

\hypersetup{%
pdfauthor={CMS Collaboration},%
pdftitle={Search for the associated production of the Higgs boson with a top-quark pair},%
pdfsubject={CMS},%
pdfkeywords={CMS, physics, Higgs, top}}

\maketitle

\section{Introduction}
\label{sec:intro}

Since the discovery of a new boson by the CMS and ATLAS
Collaborations~\cite{cms_higgs,atlas_higgs} in 2012, experimental
studies have focused on determining the consistency of this particle's
properties with the expectations for the standard model (SM) Higgs
boson~\cite{Englert:1964et,Higgs:1964ia,Higgs:1964pj,Guralnik:1964eu,Higgs:1966ev,Kibble:1967sv}.
To date, all measured properties, including couplings, spin, and
parity are consistent with the SM expectations within experimental
uncertainties~\cite{Chatrchyan:2013lba, Chatrchyan:2012jja,
  Aad:2013wqa, Aad:2013xqa, Aaltonen:2013kxa}.

\begin{figure}[!hbtp]
  \begin{center}
  \includegraphics[height=4cm]{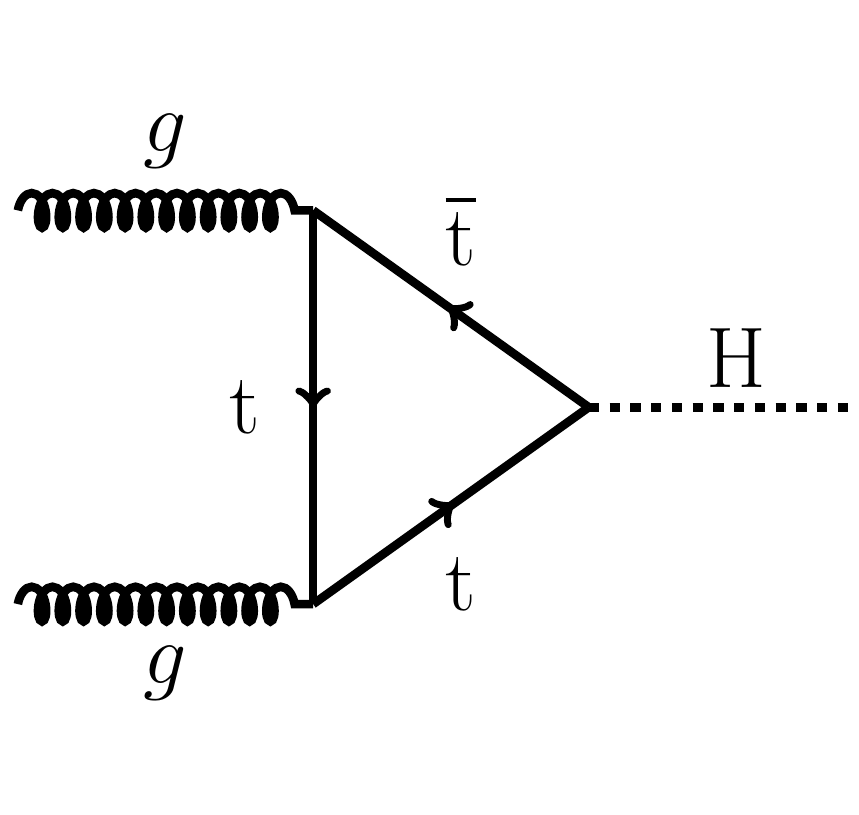}\hspace{1.2cm}
  \includegraphics[height=4cm]{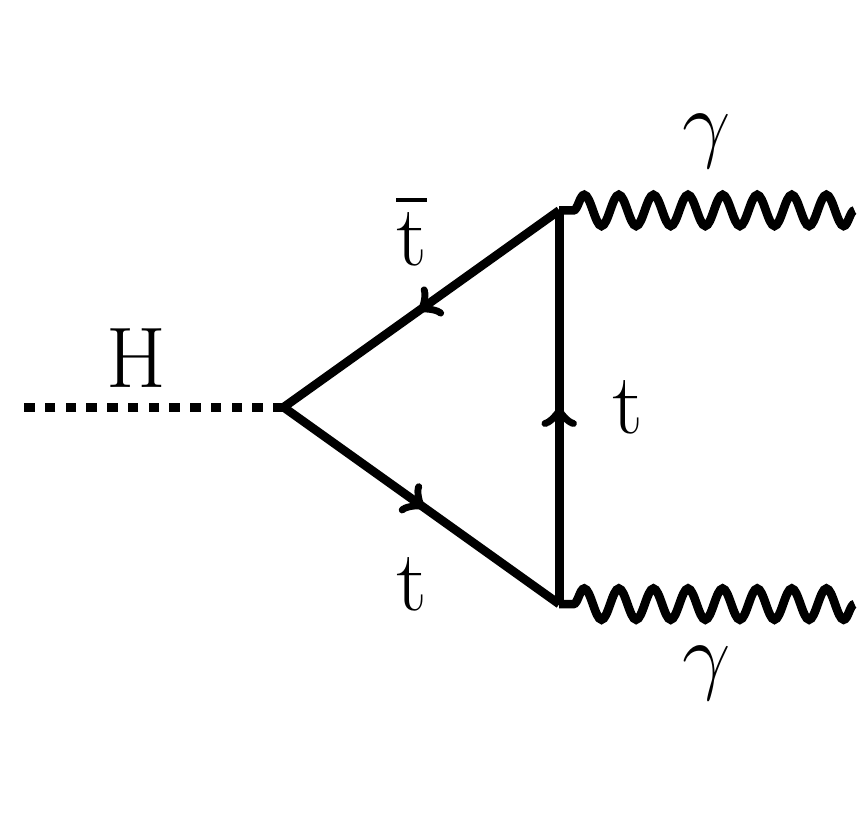}\hspace{1.2cm}
  \includegraphics[height=4.1cm]{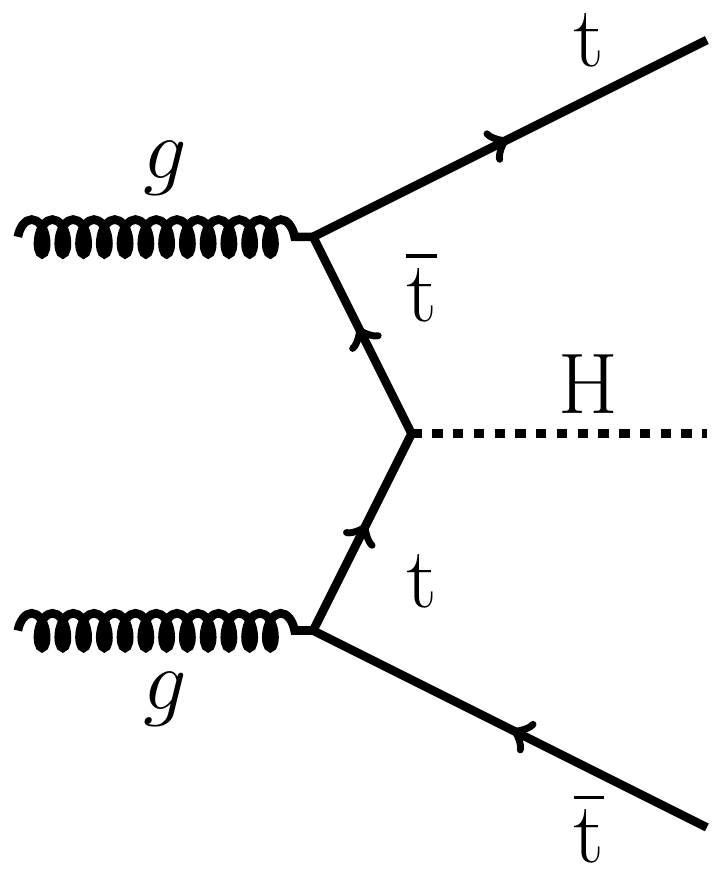}
 \caption{Feynman diagrams showing the gluon fusion production
   of a Higgs boson through a top-quark loop (left), the
   decay of a Higgs boson to a pair of photons through a top-quark
   loop (center), and the production of a Higgs boson in association with
   a top-quark pair (right).  These diagrams are representative of SM processes with sensitivity to the coupling between the top quark and the Higgs boson.}
\label{fig:feynman} \end{center}
\end{figure}

One striking feature of the SM Higgs boson is its strong coupling to
the top quark relative to the other SM fermions.  Based on its large
mass~\cite{tevatron_top_mass} the top-quark Yukawa coupling is
expected to be of order one.  Because the top quark is heavier than
the Higgs boson, its coupling cannot be assessed by measuring Higgs
boson decays to top quarks.  However, the Higgs boson's coupling to
top quarks can be experimentally constrained through measurements
involving the gluon fusion production mechanism that proceeds via a
fermion loop in which the top quark provides the dominant contribution
(left panel of figure~\ref{fig:feynman}), assuming there is no physics
beyond the standard model (BSM) contributing to the loop.  Likewise
the decay of the Higgs boson to photons involves both a fermion loop
diagram dominated by the top-quark contribution (center panel of
figure~\ref{fig:feynman}), as well as a $\PW$ boson loop contribution.
Current measurements of Higgs boson production via gluon fusion are
consistent with the SM expectation for the top-quark Yukawa coupling
within experimental uncertainties~\cite{Chatrchyan:2013lba,
  Chatrchyan:2012jja, Aad:2013wqa, Aad:2013xqa}.

Probing the top-quark Yukawa coupling directly requires a process
that results in both a Higgs boson and top quarks explicitly
reconstructed via their final-state decay products.  The production of
a Higgs boson in association with a top-quark pair ($\ttbar \PH$)
satisfies this requirement (right panel of figure~\ref{fig:feynman}).  A
measurement of the rate of $\ttbar \PH$ production provides a direct
test of the coupling between the top quark and the Higgs boson.
Furthermore, several new physics
scenarios~\cite{ArkaniHamed2001232, 1126-6708-2002-07-034,
  Contino:2006qr} predict the existence of heavy top-quark partners,
that would decay into a top quark and a Higgs boson. Observation of a
significant deviation in the $\ttbar \PH$ production rate with respect
to the SM prediction would be an indirect indication of unknown
phenomena.

The results of a search for $\ttbar \PH$ production using the CMS
detector~\cite{Chatrchyan:2008zzk} at the LHC are described in this
paper.  The small $\ttbar \PH$ production cross section---roughly 130\unit{fb} at $\sqrt{s} = 8\TeV$~\cite{Raitio:1978pt, Ng:1983jm,
  Kunszt:1984ri, Beenakker:2001rj, Beenakker:2002nc, Dawson:2002tg,
  Dawson:2003zu,Garzelli:2011vp,YellowReport,YellowReport3}---makes
measuring its rate experimentally challenging.  Therefore, it is
essential to exploit every accessible experimental signature.  As the
top quark decays with nearly 100\% probability to a $\PW$ boson and a
b quark, the experimental signatures for top-quark pair production are
determined by the decay of the $\PW$ boson.  When both $\PW$ bosons
decay hadronically, the resulting final state with six jets (two of
which are b-quark jets) is referred to as the all-hadronic final
state.  If one of the $\PW$ bosons decays leptonically, the final
state with a charged lepton, a neutrino, and four jets (two of which
are b-quark jets) is called lepton + jets.  Finally, when both $\PW$
bosons decay leptonically, the resulting dilepton final state has two
charged leptons, two neutrinos, and two b-quark jets.  All three of
these top-quark pair signatures are used in the search for $\ttbar
\PH$ production in this paper.  Although in principle, electrons,
muons, and taus should be included as ``charged leptons,''
experimentally, the signatures of a tau lepton are less distinctive
than those of the electron or muon.  For the rest of this paper, the
term ``charged lepton'' will refer only to electrons or muons,
including those coming from tau lepton decays.

Within the SM, the observed mass of the Higgs boson near 125\GeV~\cite{Chatrchyan:2013lba, Chatrchyan:2013mxa,hgg_inclusive}
implies that a variety of Higgs boson decay modes are experimentally
accessible.  At this mass, the dominant decay mode, $\PH \to
\bbbar$, contributes almost 60\% of the total Higgs boson decay width.
The next largest contribution comes from $\PH \to \PW \PW$ with a
branching fraction around 20\%.  Several Higgs boson decay channels
with significantly smaller branching fractions still produce
experimentally accessible signatures, especially $\PH \to \Pgg \Pgg$,
$\PH \to \Pgt \Pgt$, and $\PH \to \cPZ \cPZ$.

The experimental searches for $\ttbar \PH$ production presented here
can be divided into three broad categories based on the Higgs boson
signatures: $\PH \to$ hadrons, $\PH \to$ photons, and $\PH \to$
leptons.  There are two main Higgs boson decay modes that contribute
to the $\PH \to$ hadrons searches: $\PH \to \bbbar$ and $\PH \to \Pgt
\Pgt$, where both $\Pgt$ leptons decay hadronically.  Note that events with $\Pgt$ pairs include both direct $\PH \to \Pgt \Pgt$
decays and those where the $\Pgt$ leptons are produced by the decays
of $\PW$ or $\cPZ$ bosons from $\PH \to \PW \PW$ and $\PH \to \cPZ
\cPZ$ decays.  Events used in the $\PH \to$ hadrons searches have one
or more isolated charged lepton from the $\PW$ boson decays from the
top quarks, which means these searches focus on the lepton + jets and
dilepton $\ttbar$ final states, using single-lepton or dilepton
triggers, respectively.  Multivariate analysis (MVA) techniques are
employed to tag the jets coming from b-quark or $\Pgt$-lepton decays
and to separate $\ttbar \PH$ events from the large $\ttbar+$jets
backgrounds.

In contrast, the $\PH \to$ photons search focuses exclusively on the
$\PH \to \Pgg \Pgg$ decay mode.  In this case, the photons provide the
trigger, and all three $\ttbar$ decay topologies are included in the
analysis.  The CMS detector's excellent $\Pgg \Pgg$ invariant mass
resolution~\cite{Chatrchyan:2013dga} is used to separate the $\ttbar
\PH$ signal from the background, and the background model is entirely
based on data.

Finally, in the $\PH \to$ leptons search, the leptons arise as
secondary decay products from $\PH \to \PW \PW$, $\PH \to \cPZ \cPZ$,
and $\PH \to \Pgt \Pgt$ decays, as well as from the $\PW$ bosons
produced in the top quark decays.  To optimize the
signal-to-background ratio, events are required to have either a pair
of same-sign charged leptons, or three or more charged leptons.  The
events are required to pass the dilepton or trilepton triggers.
Multivariate analysis techniques are used to separate leptons arising
from $\PW$-boson, $\cPZ$-boson and $\Pgt$-lepton decays, referred to
as signal leptons, from background leptons, which come from b-quark or
c-quark decays, or misidentified jets.  MVA techniques are also used
to distinguish $\ttbar \PH$ signal events from background events that
are modeled using a mixture of control samples in data and Monte Carlo
(MC) simulation.  Table~\ref{tab:channelSummary} summarizes the main
features of each search channel described above.

\begin{table}[!htbp]
\begin{center}
  \topcaption{Summary of the search channels used in the $\ttbar \PH$
    analysis.  In the description of the signatures, an $\ell$ refers
    to any electron or muon in the final state (including those coming
    from leptonic $\Pgt$ decays).  A hadronic $\Pgt$ decay is
    indicated by $\tauh$.  Finally, $\mathrm{j}$ represents
    a jet coming from any quark or gluon, or an unidentified hadronic
    $\Pgt$ decay, while $\cPqb$ represents a b-quark jet.  Any element
    in the signature enclosed in square brackets indicates that the
    element may not be present, depending on the specific decay mode
    of the top quark or Higgs boson.  The minimum transverse momentum
    $\pt$ of various objects is given to convey some sense of the
    acceptance of each search channel; however, additional
    requirements are also applied.  Jets labeled as b-tagged jets have
    been selected using the algorithm described in
    section~\ref{sec:objects}.  More details on the triggers used to
    collect data for each search channel are given in
    section~\ref{sec:samples}.  Selection of final-state objects
    (leptons, photons, jets, etc.) is described in general in
    section~\ref{sec:objects}, with further channel-specific details
    included in sections~\ref{sec:tth-hadrons}--\ref{sec:tth-leptons}.
    In this table and the rest of the paper, the number of b-tagged
    jets is always included in the jet count. For example, the notation
    4~jets + 2~b-tags means four jets of which two jets are b-tagged.
  }
    \begin{tabular}{|l|l|l|l|} \hline
Category                     & Signature          & Trigger       & Signature \\ \hline \hline
                             & Lepton + Jets      & Single Lepton & 1~$\Pe$/$\Pgm$, $\pt > 30\GeV$ \\
{\bf \boldmath$\PH \to$ Hadrons} & ($\ttbar \PH \to \ell \Pgn \mathrm{jj} \cPqb\cPqb\cPqb \cPqb$)& & ${\geq}4$~jets + ${\geq}2$~b-tags, $\pt > 30\GeV$ \\
\cline{2-4}
~~~$\PH \to \bbbar$ & Dilepton  & Dilepton      & 1~$\Pe$/$\Pgm$, $\pt > 20\GeV$  \\
~~~$\PH \to \tauh\tauh$ & ($\ttbar \PH \to \ell \Pgn \ell \Pgn \cPqb\cPqb\cPqb \cPqb$) & & 1~$\Pe$/$\Pgm$, $\pt > 10\GeV$  \\
~~~$\PH \to \PW \PW$            &                    &               & ${\geq}3$~jets + ${\geq}2$~b-tags, $\pt > 30\GeV$ \\
\cline{2-4}
                             & Hadronic $\Pgt$    & Single Lepton & 1 $\Pe$/$\Pgm$, $\pt > 30\GeV$ \\
                             & ($\ttbar \PH \to \ell \Pgn \tauh [\Pgn] \tauh [\Pgn] \mathrm{jj} \cPqb\cPqb$) & & 2~$\tauh$, $\pt > 20\GeV$ \\
                             &                    &               & ${\geq}2$~jets + 1-2~b-tags, $\pt > 30\GeV$ \\
\hline
                             & Leptonic           & Diphoton      & 2~$\Pgg$, $\pt > m_{\Pgg\Pgg}/2\,(25)\GeV$ for 1$^{\mathrm{st}}$ (2$^{\mathrm{nd}}$) \\
{\bf \boldmath$\PH \to $ Photons} & ($\ttbar \PH \to \ell \Pgn \mathrm{jj} \cPqb\cPqb\Pgg \Pgg$, & & ${\geq}1$~$\Pe$/$\Pgm$, $\pt > 20\GeV$ \\
~~~$\PH \to \Pgg \Pgg$       &  $\ttbar \PH \to \ell \Pgn \ell \Pgn \cPqb\cPqb \Pgg \Pgg$)& & ${\geq}2$~jets + ${\geq}1$~b-tags, $\pt > 25\GeV$ \\
\cline{2-4}
                             & Hadronic           & Diphoton      & 2~$\Pgg$, $\pt > m_{\Pgg\Pgg}/2\,(25)\GeV$ for 1$^{\mathrm{st}}$ (2$^{\mathrm{nd}}$) \\
                             & ($\ttbar \PH \to \mathrm{jj} \mathrm{jj} \cPqb\cPqb\Pgg \Pgg$)& & 0~$\Pe$/$\Pgm$, $\pt > 20\GeV$ \\
                             &                    &               & ${\geq}4$~jets + ${\geq}1$~b-tags, $\pt > 25\GeV$ \\
\hline
                             & Same-Sign Dilepton & Dilepton      & 2~$\Pe$/$\Pgm$, $\pt > 20\GeV$  \\
{\bf \boldmath $\PH \to $ Leptons} & ($\ttbar \PH \to \ell^\pm \Pgn \ell^\pm [\Pgn] \mathrm{jjj[j]} \cPqb\cPqb$) & & ${\geq}4$~jets + ${\geq}1$~b-tags, $\pt > 25\GeV$ \\
\cline{2-4}
~~~$\PH \to \PW\PW$          & 3 Lepton           & Dilepton,     & 1~$\Pe$/$\Pgm$, $\pt > 20\GeV$  \\
~~~$\PH \to \Pgt\Pgt$        & ($\ttbar \PH \to \ell \Pgn \ell [\Pgn] \ell [\Pgn] \mathrm{j[j]} \cPqb\cPqb$)&  Trielectron & 1~$\Pe$/$\Pgm$, $\pt > 10\GeV$  \\
~~~$\PH \to \cPZ\cPZ$        &                    &               & 1~$\Pe$($\Pgm$), $\pt > 7(5)\GeV$  \\
                             &                    &               & ${\geq}2$~jets + ${\geq}1$~b-tags, $\pt > 25\GeV$ \\
\cline{2-4}
                             & 4 Lepton           & Dilepton,     & 1~$\Pe$/$\Pgm$, $\pt > 20\GeV$  \\
                             & ($\ttbar \PH \to \ell \Pgn \ell \Pgn \ell [\Pgn] \ell [\Pgn] \cPqb\cPqb$) &  Trielectron & 1~$\Pe$/$\Pgm$, $\pt > 10\GeV$  \\
                             &                    &               & 2~$\Pe$($\Pgm$), $\pt > 7(5)\GeV$  \\
                             &                    &               & ${\geq}2$~jets + ${\geq}1$~b-tags, $\pt > 25\GeV$ \\
\hline
\end{tabular}
    \label{tab:channelSummary}
\end{center}
\end{table}

To characterize the strength of the $\ttbar \PH$ signal relative to
the SM cross section ($\mu=\sigma/\sigma_{\mathrm{SM}}$) a fit is
performed simultaneously in all channels.  The fit uses specific
discriminating distributions in each channel, either a kinematic
variable like the diphoton invariant mass in the $\PH \to$ photons
channel or an MVA discriminant as in the $\PH \to$ hadrons and $\PH
\to$ leptons cases.  The uncertainties involved in the background
modeling are introduced in the fit as nuisance parameters, so that the
best-fit parameters provide an improved description of the background.

This paper is structured as follows. Sections~\ref{sec:cms}
and~\ref{sec:samples} describe the CMS detector, and the data and
simulation samples, respectively.  Section~\ref{sec:objects} discusses
the common object reconstruction and identification details shared
among the different search channels.
Sections~\ref{sec:tth-hadrons}, \ref{sec:tth-gamma},
and~\ref{sec:tth-leptons} outline the selection, background modeling,
and signal extraction techniques for the $\PH \to$ hadrons, $\PH \to$
photons, and $\PH \to$ leptons analyses, respectively.
Section~\ref{sec:systematics} details the impact of systematic
uncertainties on the searches.  Finally, the combination procedure and
results are presented in section~\ref{sec:results}, followed by a
summary in section~\ref{sec:summary}.

\section{The CMS detector}
\label{sec:cms}
The central feature of the CMS apparatus is a superconducting solenoid
of 6\unit{m} internal diameter, providing an axial magnetic field of
3.8\unit{T} parallel to the beam direction. Within the superconducting
solenoid volume, there are a silicon pixel and strip tracker, a lead
tungstate crystal electromagnetic calorimeter (ECAL), and a
brass/scintillator hadron calorimeter (HCAL).  The tracking detectors
provide coverage for charged particles within $\abs{\eta} < 2.5$. The ECAL
and HCAL calorimeters provide coverage up to $\abs{\eta} < 3.0$. The ECAL
is divided into two distinct regions: the barrel region, which covers
$\abs{\eta} < 1.48$, and the endcap region, which covers $1.48 < \abs{\eta} <
3.0$.  A quartz-fiber forward calorimeter extends the coverage further
up to $\abs{\eta} < 5.0$.  Muons are measured in gas-ionization detectors
embedded in the steel flux-return yoke outside the solenoid. The first
level (L1) of the CMS trigger system, composed of custom hardware
processors, uses information from the calorimeters and muon detectors
to select the most interesting events in a fixed time interval of less
than 4\mus. The high-level trigger (HLT) processor farm further
decreases the event rate from around $100\unit{kHz}$ to less than
$1\unit{kHz}$, before data storage. A more detailed description of the
CMS detector, together with a definition of the coordinate system used
and the relevant kinematic variables, can be found in
Ref.~\cite{Chatrchyan:2008zzk}.

\section{Data and simulation samples}
\label{sec:samples}

This search is performed with samples of proton-proton collisions at
$\sqrt{s} = 7\TeV$, collected with the CMS detector in 2011 (referred
to as the 7\TeV dataset), and at $\sqrt{s} = 8\TeV$, collected in
2012 (referred to as the 8\TeV dataset).  All of the search
channels make use of the full CMS 8\TeV dataset, corresponding to
an integrated luminosity that ranges from 19.3\fbinv to 19.7\fbinv, with a 2.6\% uncertainty~\cite{CMS-PAS-LUM-13-001}.  The
luminosity used varies slightly because the different search channels
have slightly different data quality requirements, depending on the
reconstructed objects and triggers used.  In addition, the $\PH \to$
photons analysis makes use of data collected at $\sqrt{s} = 7\TeV$,
corresponding to an integrated luminosity of 5.1\fbinv.  Finally,
the $\ttbar \PH$ search in the $\PH \to \bbbar$ final state based on
the 7\TeV dataset with an integrated luminosity of 5.0\fbinv,
described in Ref.~\cite{Chatrchyan:2013yea}, is combined with the
8\TeV analysis to obtain the final $\ttbar \PH$ result.  The
uncertainty on the 7\TeV luminosity is 2.2\%~\cite{CMS:lumi}.  In the $\PH
\to$~hadrons and $\PH \to$~leptons analyses, events are selected by
triggering on the presence of one or more leptons.  For the $\PH
\to$~photons analysis, diphoton triggers are used.

Single-lepton triggers are used for channels with one lepton in the
final state. The single-electron trigger requires the presence of an
isolated, good-quality electron with transverse momentum $\pt > 27\GeV$. The single-muon trigger requires a muon candidate isolated from
other activity in the event with $\pt > 24\GeV$.  Dilepton triggers
are used for channels with two or more leptons in the final state. The
dilepton triggers require any combination of electrons and muons, one
lepton with $\pt > 17\GeV$ and another with $\pt > 8\GeV$.  In the
$\PH \to$~leptons analysis, a trielectron trigger is used, with
minimum $\pt$ thresholds of 15\GeV, 8\GeV, and 5\GeV.  The
$\PH \to$~photons analysis uses diphoton triggers with two different
photon identification schemes. One requires calorimetric
identification based on the electromagnetic shower shape and isolation
of the photon candidate.  The other requires only that the photon has
a high value of the $R_{9}$ shower shape variable, where $R_9$ is
calculated as the ratio of the energy contained in a $3{\times}3$
array of ECAL crystals centered on the most energetic deposit in the
supercluster to the energy of the whole supercluster.  The
superclustering algorithm for photon reconstruction is explained in
more detail in section~\ref{sec:objects}.  The $\ET$ thresholds at
trigger level are 26 (18)\GeV and 36 (22)\GeV on the leading
(trailing) photon depending on the running period.  To maintain high
trigger efficiency, all four combinations of thresholds and selection
criteria are used.

Expected signal events and, depending on the analysis channel, some
background processes are modeled with MC simulation. The $\ttbar\PH$
signal is modeled using the \PYTHIA generator~\cite{PYTHIA} (version
6.4.24 for the 7\TeV dataset and version 6.4.26 for the 8\TeV
dataset).  Separate samples were produced at nine different values of
$m_{\PH}$: 110, 115, 120, 122.5, 125, 127.5, 130, 135, and 140\GeV, and
are used to interpolate for intermediate mass values.  The background
processes $\ttbar\PW$, $\ttbar\cPZ$, $\ttbar+$jets, Drell--Yan$+$jets,
$\PW+$jets, $\cPZ\cPZ+$jets, $\PW\PW+$jets, and $\PW\cPZ+$jets are all
generated with the \MADGRAPH 5.1.3~\cite{MADGRAPH} tree-level matrix
element generator, combined with \PYTHIA for the parton shower and
hadronization.  For the $\PH \to$ leptons analysis, the rare
$\PW\PW\cPZ$, $\PW\PW\PW$, $\ttbar+\Pgg+$jets, and $\ttbar\PW\PW$
processes are generated similarly.  Single top quark production
(t+$\cPq$, t+b, and t+$\PW$) is modeled with the next-to-leading-order
(NLO) generator \POWHEG 1.0
\cite{Nason:2004rx,Frixione:2007vw,Alioli:2010xd,Re:2010bp,Alioli:2009je,Melia:2011tj}
combined with \PYTHIA.  Samples that include top quarks in the final
state are generated with a top quark mass of $172.5\GeV$.  For the
$\PH \to$ photons analysis, the gluon fusion ($\Pg\Pg\to\PH$) and
vector boson fusion ($\cPq\cPaq\to\cPq\cPaq\PH$) production modes are
generated with \POWHEG at NLO, and combined with \PYTHIA for the
parton shower and hadronization.  Higgs boson production in
association with weak bosons ($\cPq\cPaq\to\PW\PH/ Z\PH$) is simulated
with \PYTHIA.  Samples generated with a leading order generator use
the CTEQ6L1 parton distribution function (PDF)~\cite{Pumplin:2002vw}
set, while samples generated with NLO generators use the CTEQ6.6M PDF
set~\cite{PhysRevD.78.013004}.

The CMS detector response is simulated using the \GEANTfour software
package \cite{Agostinelli:2002hh}.  All events from data and simulated
samples are required to pass the same trigger conditions and are
reconstructed with identical algorithms to those used for collision
data.  Effects from additional $\Pp\Pp$ interactions in the same bunch
crossing (pileup) are modeled by adding simulated minimum bias events
(generated with \PYTHIA) to the generated hard interactions.  The
pileup interaction multiplicity distribution in simulation reflects
the luminosity profile observed in pp collision data.  Additional
correction factors are applied to individual object efficiencies and
energy scales to bring the MC simulation into better agreement with
data, as described in section~\ref{sec:objects}.

\section{Object reconstruction and identification}
\label{sec:objects}

A global event description is obtained with the CMS particle-flow (PF)
algorithm~\cite{CMS-PAS-PFT-10-001, CMS-PAS-PFT-10-002}, which
optimally combines the information from all CMS sub-detectors to
reconstruct and identify each individual particle in the $\Pp\Pp$
collision event. The particles are classified into mutually exclusive
categories: charged hadrons, neutral hadrons, photons, muons, and
electrons.  The primary collision vertex is identified as the
reconstructed vertex with the highest value of $\sum \pt^{2}$, where the
summation includes all particles used to reconstruct the vertex.
Although the separate $\ttbar \PH$ search channels share the same
overall object reconstruction and identification approach, there are
differences in some of the selection requirements.  Generally
speaking, the requirements in the $\PH \to$ hadrons channel are more
stringent than in the $\PH \to$ photons or leptons because of the
larger backgrounds in the first channel and the smaller amount of
signal in the other ones.

Photon candidates are reconstructed from the energy deposits in the
ECAL, grouping the individual clusters into a supercluster.  The
superclustering algorithms achieve an almost complete reconstruction
of the energy of photons (and electrons) that convert into
electron-positron pairs (emit bremsstrahlung) in the material in front
of the ECAL.  In the barrel region, superclusters are formed from
five-crystal-wide strips in $\eta$, centered on the locally most
energetic crystal (seed), and have a variable extension in $\phi$.  In
the endcaps, where the crystals are arranged according to an $x$-$y$
rather than an $\eta$-$\phi$ geometry, matrices of $5{\times}5$ crystals
(which may partially overlap) around the most energetic crystals are
merged if they lie within a narrow $\phi$ road.  The photon candidates
are collected within the ECAL fiducial region $\abs{\eta}<2.5$, excluding
the barrel-endcap transition region $1.44 < \abs{\eta} < 1.57$ where
photon reconstruction is sub-optimal.  Isolation requirements are
applied to photon candidates by looking at neighboring particle
candidates reconstructed with the PF event reconstruction
technique~\cite{CMS-PAS-PFT-10-001}.  Additional details on photon
reconstruction and identification can be found in
Ref.~\cite{hgg_inclusive}.

Electrons with $\pt > 7\GeV$ are reconstructed within the geometrical
acceptance of the tracker, $\abs{\eta} < 2.5$.  The reconstruction
combines information from clusters of energy deposits in the ECAL and
the electron trajectory reconstructed in the inner
tracker~\cite{Baffioni:2006cd,CMS-PAS-EGM-10-004,CMS_DPS_2011-003,CMS_DP_2013-003}.
The track-cluster matching is initiated either ``outside-in'' from
ECAL clusters, or ``inside-out'' from track candidates.  Trajectories
in the tracker volume are reconstructed using a dedicated modeling of
the electron energy loss and fitted with a Gaussian sum
filter~\cite{Baffioni:2006cd}. The electron momentum is determined
from the combination of ECAL and tracker measurements.  Electron
identification relies on a multivariate technique that combines
observables sensitive to the amount of bremsstrahlung along the
electron trajectory, the spatial and momentum matching between the
electron trajectory and associated clusters, and shower shape
observables.  In order to increase the lepton efficiency, the $\PH
\to$~leptons analysis uses a looser cut on the multivariate
discriminant than do the other analysis channels.  Although the
minimum $\pt$ requirement on electrons is $\pt > 7\GeV$, the
different $\ttbar \PH$ search channels, particularly the $\PH \to$
hadrons channel, use a higher threshold on some of the selected
electrons depending on the trigger requirements and to help control
backgrounds (see
sections~\ref{sec:tth-hadrons}--\ref{sec:tth-leptons} for more
details).

Muons are reconstructed within $\abs{\eta} < 2.4$ and for $\pt >
5\GeV$~\cite{Chatrchyan:2012xi}.  The reconstruction combines
information from both the silicon tracker and the muon spectrometer.
The matching between the inner and outer tracks is initiated either
``outside-in'', starting from a track in the muon system, or
``inside-out'', starting from a track in the silicon tracker.  The PF
muons are selected among the reconstructed muon track candidates by
applying minimal requirements on the track components in the muon and
tracker systems and taking into account matching with energy deposits
in the calorimeters~\cite{CMS-PAS-PFT-10-003}.  Depending on the level
of backgrounds in a given analysis channel, different requirements can
be placed on the distance of closest approach for the muon to the
collision vertex---referred to as the impact parameter (IP)---in
both the $z-$direction ($d_z$) and the $x-y$ plane ($d_{xy}$) to
reject background muons.  As in the electron case, the $\pt$ threshold
for some or all of the muons is set higher than the $5\GeV$ default,
depending on the trigger requirements used by a particular search
channel and to control backgrounds.

An important quantity for distinguishing signal and background leptons
is isolation.  Although conceptually similar, isolation is defined
slightly differently for muons and electrons depending on the analysis
channel.  Muon isolation is assessed by calculating the sum of the
transverse energy of the other particles in a cone of
$\Delta R = \sqrt{(\Delta \eta)^2 + (\Delta \phi)^2} = 0.4$ around the
muon direction, excluding the muon itself, where $\Delta \eta$ and
$\Delta \phi$ are the angular differences between the muon and the
other particles in the $\eta$ and $\phi$ directions.  To
correct for the effects of pileup, charged contributions not
originating from the primary collision vertex are explicitly removed
from the isolation sum, and the neutral contribution is corrected
assuming a ratio of 0.5 for the contribution of neutral to charged
objects to the pileup activity.  The ratio of the corrected isolation
sum to the muon $\pt$ is the relative isolation of the muon.  For the
$\PH \to$ leptons search, electron isolation is calculated identically
to muon isolation.  For the $\PH \to$ hadrons and $\PH \to$ photons
searches, there are two differences.  The first is that the electron
isolation sum only takes into account charged and neutral particles in
a cone of $\Delta R = 0.3$.  Second, the correction for pileup effects
to the neutral contribution in the isolation sum is made using the
average $\pt$ density calculated from neutral particles
multiplied by the effective area of the isolation cone.  The relative
isolation is the ratio of this corrected isolation sum to the electron
$\pt$.

Jets are reconstructed by clustering the charged and neutral PF
particles using the anti-$\kt$ algorithm with a distance parameter of
0.5~\cite{Cacciari:2005hq,Cacciari:2008gp}. For the $\PH \to$ hadrons
search, particles identified as isolated muons and electrons are
expected to come from $\PW$-boson decays and are excluded from the
clustering.  Non-isolated muons and electrons are expected to come
from b-quark decays and are included in the clustering.  The $\PH \to$
leptons and $\PH \to$ photons searches do not exclude the isolated
leptons from the jet clustering, but require selected jets to be
separated by $\Delta R > 0.5$ from the selected leptons. The
choice not to exclude leptons from the clustered jets in the $\PH \to$
leptons search is an integral part of the non-prompt lepton rejection
strategy.  When a lepton is clustered into a jet, that information is
used to help determine whether the lepton originated from a
semileptonic decay of a heavy (bottom or charm) quark (see
section~\ref{sec:tth-leptons} for more details).

Jets are required to have at least two PF constituents and more than
1\% of their energy in both the electromagnetic and hadronic
components to reject jets arising from instrumental effects.  For the
$\PH \to$ leptons and $\PH \to$ photons searches, additional
requirements are applied to remove jets coming from pileup
vertices~\cite{CMS-PAS-JME-13-005}.  For the $\PH \to$ hadrons and
$\PH \to$ leptons analyses, charged PF particles not associated with
the primary event vertex are ignored when clustering the jets to
reduce the contribution from pileup.  The momentum of the clustered
jet is corrected for a variety of effects~\cite{cmsJEC}.  The
component coming from pileup activity---in the case of $\PH \to$
hadrons or leptons, just the neutral part---is removed by applying a
residual energy correction following the area-based procedure
described in Refs.~\cite{Cacciari:2008gn,Cacciari:2007fd}. Further
corrections based on simulation, $\Pgg/\cPZ+$jets data, and dijet data
are then applied, as well as a correction to account for
residual differences between data and simulation~\cite{cmsJEC}.  Selected jets
are required to have $\abs{\eta}<2.4$, and $\pt>25$\GeV ($\PH \to$ leptons
and $\PH \to$ photons) or $\pt>30$\GeV ($\PH \to$ hadrons).  The
higher $\pt$ requirement in the latter case arises from the larger
amount of background in that sample.

Jets are identified as originating from a b-quark using the combined
secondary vertex (CSV) algorithm~\cite{CMS:2012hd,CMS-PAS-BTV-13-001}
that utilizes information about the impact parameter of tracks and
reconstructed secondary vertices within the jets in a multivariate
algorithm. The CSV algorithm provides a continuous output ranging from
0 to 1; high values of the CSV discriminant indicate that the jet
likely originates from a b quark, while low values indicate the jet is
more consistent with light-flavor quarks or gluons.  The efficiency to
tag b-quark jets and the rate of misidentification of non-b-quark jets
depend on the working point chosen.  For the medium working point of
the CSV algorithm, the b-tagging efficiency is around 70\% (20\%) for
jets originating from a b\,(c) quark and the probability of mistagging
for jets originating from light quarks or gluons is approximately
2\%. For the loose working point, the efficiency to tag jets from b\,(c) quarks is approximately 85\% (40\%) and the probability to tag
jets from light quarks or gluons is about 10\%.  These efficiencies
and mistag probabilities vary with the $\pt$ and $\eta$ of the jets,
and the values quoted are indicative of the predominant jets in this
analysis.

The hadronic decay of a $\Pgt$ lepton ($\tauh$) produces a
narrow jet of charged and neutral hadrons---almost all pions. Each
neutral pion subsequently decays into a pair of photons. The
identification of $\tauh$ jets begins with the formation of
PF jets by clustering charged hadron and photon objects via the
anti-\kt algorithm. Then, the hadron-plus-strips
(HPS)~\cite{PTDR2,Chatrchyan:2012zz} algorithm tests each of the most
common $\tauh$ decay mode hypotheses using the
electromagnetic objects found within rectangular bands along the
azimuthal direction. In the general algorithm, combinations of charged
hadrons and photons (one charged hadron, one charged hadron + photons,
and three charged hadrons) must lead to invariant masses consistent
with the appropriate intermediate resonances~\cite{Chatrchyan:2012zz}.
For this analysis, only the decays involving exactly one charged hadron
are used.

The missing transverse energy vector is calculated
as the negative vector $\pt$ sum of all PF candidates
identified in the event. The magnitude of this vector is denoted as
$\MET$. Since pileup interactions degrade the performance of the \MET
variable, the $\PH \to$ leptons search also uses the
$H_\mathrm{T}^\text{miss}$ variable. This variable is computed in the
same way as the $\MET$, but uses only the selected jets and
leptons. The $H_\mathrm{T}^\text{miss}$ variable has worse resolution
than $\MET$ but it is more robust as it does not rely on soft objects
in the event.  A linear discriminator is computed based on the two variables,
\begin{equation}
 L_\mathrm{D} = 0.60 \MET + 0.40 H_\mathrm{T}^\text{miss},
 \label{eq:reconstruction_isolation}
\end{equation}
exploiting the fact that $\MET$ and $H_\mathrm{T}^\text{miss}$ are
less correlated in events with missing transverse energy from
instrumental mismeasurement than in events with genuine missing
transverse energy.  The linear discriminant is constructed to optimize
separation between $\ttbar \PH$ and $\cPZ+$jets in simulation.

To match the performance of reconstructed objects between data and
simulation, the latter is corrected with the following data-MC scale
factors: Leptons are corrected for the difference in trigger
efficiency, as well as in lepton identification and isolation
efficiency. For the $\PH \to$ leptons channel, corrections accounting
for residual differences between data and simulation are applied to
the muon momentum, as well as to the ECAL energy before combining with
the tracking momentum for electrons. All lepton corrections are
derived using tag-and-probe techniques~\cite{WZxs7TeV} based on
samples with $\cPZ$ boson and $\JPsi$ decays into two leptons.  Jet
energy corrections as described above are applied as a function of the
jet $\pt$ and $\eta$~\cite{cmsJEC}.  Standard efficiency scale factors
for the medium and loose b-tagging working
points~\cite{CMS:2012hd,CMS-PAS-BTV-13-001} are applied for light- and
heavy-flavor jets in the $\PH \to$ leptons and $\PH \to$ photons
searches, while the $\PH \to$ hadrons search uses a more sophisticated
correction to the CSV shape (see section~\ref{sec:tth-hadrons} for
more details).

\section{\texorpdfstring{$\PH \to$~hadrons}{H to hadrons}}
\label{sec:tth-hadrons}

\subsection{Event selection}

Events in the $\PH \to$ hadrons analysis are split into three
different channels based on the decay modes of the top-quark pair and
the Higgs boson: the lepton+jets channel ($\ttbar \to \ell\cPgn \cPq
\cPaq^{\prime}\bbbar$, $\PH \to \bbbar$), the dilepton channel
($\ttbar \to \ell^{+} \cPgn \ell^{-} \cPagn\bbbar$, $\PH \to \bbbar$),
and the $\tauh$ channel ($\ttbar \to \ell\cPgn \cPq
\cPaq^{\prime}\bbbar$, $\PH \to \tauh\tauh$),
where a lepton is an electron or a muon.  For the lepton+jets channel,
events containing an energetic, isolated lepton, and at least four
energetic jets, two or more of these jets must be b-tagged, are
selected.  For the dilepton channel, a pair of oppositely charged
leptons and three or more jets, with at least two of the jets being
b-tagged, are required.  For the $\tauh$ channel, beyond
the two identified hadronically decaying $\Pgt$ leptons, at least two
jets, one or two of which must be b-tagged, are required.  The event
selections are designed to be mutually exclusive.  For all figures
(figures~\ref{fig:bb_lj_input_8TeV}--\ref{fig:bb_tau_finalMVA_8TeV})
and tables (tables
\ref{tab:yield_bb_lj_8TeV}--\ref{tab:yield_bb_tau_8TeV}) of the $\PH
\to$ hadrons analysis, the b-tagged jets are included in the jet
count.

In addition to the baseline selection detailed in
section~\ref{sec:objects}, two additional sets of selection criteria are
applied to leptons in the $\PH \to$ hadrons analysis: tight and loose,
described below.  All events are required to contain at least one
tight electron or muon.  Loose requirements are only applied to the
second lepton in the dilepton channel.

Tight and loose muons differ both in the identification and kinematic
requirements.  For events in the lepton+jets and $\tauh$
channels, tight muons are required to have $\pt>30\GeV$ and
$\abs{\eta}<2.1$ to ensure that the trigger is fully efficient with
respect to the offline selection. Tight muons in the dilepton channel
have a lower $\pt$ threshold at 20\GeV. Loose muons must have $\pt
> 10\GeV$ and $\abs{\eta}<2.4$.  For tight (loose) muons, the relative
isolation is required to be less than 0.12\,(0.2).  Tight muons must
also satisfy additional quality criteria based on the number of hits
associated with the muon candidate in the pixel, strip, and muon
detectors. To ensure the muon is from a $\PW$ decay, it is required to
be consistent with originating from the primary vertex with an impact
parameter in the $x-y$ plane $d_{xy} < 0.2\cm$ and distance from the
primary vertex in the $z$-direction $d_z < 0.5\cm$.  For loose muons,
no additional requirements beyond the baseline selection are applied.

Tight electrons in the lepton+jets and $\tauh$ channels
are required to have $\pt>30\GeV$, while the dilepton channel
requires $\pt>20\GeV$.  Loose electrons are required to have $\pt>10\GeV$.  All electrons must have $\abs{\eta}<2.5$, and those that fall into
the transition region between the barrel and endcap of the ECAL
($1.44<\abs{\eta}<1.57$) are rejected.  Tight electrons must have a
relative isolation less than 0.1, while loose electrons must have a
relative isolation less than 0.2.  In a manner similar to tight muons,
tight electrons are required to have $d_{xy} < 0.02\cm$ and $d_z < 1\cm$, while loose electrons must have $d_{xy} < 0.04 \cm$.

For $\Pgt$ leptons decaying hadronically, only candidates with
well-reconstructed decay modes~\cite{Chatrchyan:2012zz} that contain
exactly one charged pion are accepted.  Candidates must have
$\pt>20\GeV$ and $\abs{\eta}<2.1$, and the $\pt$ of the charged pion must
be greater than $5\GeV$.  Candidates are additionally required to
fulfill criteria that reject electrons and muons mimicking hadronic
$\Pgt$-lepton decays.  These include requirements on the consistency
of information from the tracker, calorimeters, and muon detectors,
including the absence of large energy deposits in the calorimeters for
muons and bremsstrahlung pattern recognition for electrons.  A
multivariate discriminant, which takes into account the effects of
pileup, is used to select loosely isolated $\tauh$
candidates~\cite{CMS:utj}. Finally, the $\tauh$ candidates
must be separated from the single tight muon or electron in the event
by a distance $\Delta R>0.25$. Events are required to contain at least
one pair of oppositely charged $\tauh$ candidates. In the
case that multiple valid pairs exist, the pair with the most isolated
$\tauh$~signatures, based on the aforementioned MVA
discriminant, is chosen.

While the basic jet $\pt$ threshold is 30\GeV, in the lepton+jets
channel, the leading three jets must have $\pt>40\GeV$. Jets originating
from b quarks are identified using the CSV medium working point.

\subsection{Background modeling}

All the backgrounds in the $\PH\to$ hadrons analysis are normalized
using NLO or better inclusive cross section
calculations~\cite{Czakon:2013goa,Campbell:2012dh,Garzelli:2012bn,Kidonakis:2012rm,Campbell:2010ff,Li:2012wna}.
To determine the contribution of individual physics processes to
exclusive final states as well as to model the kinematics, the MC
simulations described in section~\ref{sec:samples} are used.  The main
background, $\ttbar +$ jets, is generated using \MADGRAPH inclusively,
with tree-level diagrams for up to $\ttbar + 3$ extra partons.  These
extra partons include both b and c quarks.  However, as there are
significantly different uncertainties in the production of additional
light-flavor (lf) jets compared to heavy-flavor (hf), the $\ttbar
+$jets sample is separated into subsamples based on the quark flavor
associated with the reconstructed jets in the event.  Events where at
least two reconstructed jets are matched at the generator level to
extra b quarks (that is b quarks not originating from a top-quark
decay) are labeled as $\ttbar + \bbbar$ events.  If only a single jet
is matched to a b quark, the event is classed as $\ttbar + $b.  These
cases typically arise because the second extra b quark in the event is
either too far forward or too soft to be reconstructed as a jet, or
the two extra b quarks have merged into a single jet.  Finally, if at
least one reconstructed jet is matched to a c quark at the generator
level, the event is labeled as $\ttbar + \ccbar$.  Different
systematic uncertainties affecting both rates and shapes are applied
to each of the separate subsets of the $\ttbar +$jets sample, as
described in section~\ref{sec:systematics}.

Besides the common corrections to MC samples described in
section~\ref{sec:objects}, additional correction factors are applied
for samples modeling the backgrounds for this analysis channel.  A
correction factor to $\ttbar +$jets MC samples is applied so that the
top-quark $\pt$ spectrum from \MADGRAPH agrees with the distribution
observed in data and predicted by higher-order calculations.  These
scale factors, which range from roughly 0.75 to 1.2, were derived from
a fully corrected measurement of the $\ttbar$ differential cross
section as function of the top-quark $\pt$ using the $\sqrt{s} = 8
\TeV$ dataset obtained using the same techniques as described in
Ref.~\cite{Chatrchyan:2012saa}.

Furthermore, a dedicated correction
to the CSV b-tagging rates is applied to all the MC samples.  The CSV
discriminant is used to identify b-quark jets, and the CSV
discriminant shape is used in the signal extraction technique to
distinguish between events with additional genuine b-quark jets and
those with mistags. Therefore, a correction for the efficiency
difference between data and simulation over the whole range of
discriminator values is applied.  The scale factors---which are
between 0.7 and 1.3 for the bulk of the jets---are derived separately
for light-flavor (including gluons) and b-quark jets using two
independent samples of 8\TeV data in the dilepton channel. Both
control samples are also orthogonal to the events used in the signal
extraction. The light-flavor scale factor derivation uses a control
sample enriched in events with a $\cPZ$ boson, selected by requiring a
pair of opposite-charge, same-flavor leptons and exactly two jets.
The b-quark scale factor is derived in a sample dominated by
dileptonic $\ttbar$, a signature that includes exactly two b-quark
jets, by selecting events with two leptons that are not consistent
with a $\cPZ$ boson decay and exactly two jets.  Using these control
samples, a tag-and-probe approach is employed where one jet (``tag'')
passes the appropriate b-tagging requirement for a light-flavor or
b-quark jet. The CSV discriminant of the other jet (``probe'') is
compared between the data and simulation, and the ratio gives a scale
factor for each jet as a function of CSV discriminant value, $\pt$ and
$\eta$.  Each light-flavor or b-quark jet is then assigned an
appropriate individual scale factor.  The CSV output shape for c-quark
jets is dissimilar to that of both light-flavor and b-quark jets;
hence, in the absence of a control sample of c-quark jets in data, a
scale factor of 1 is applied, with twice the relative uncertainty
ascertained from b-quark jets (see
section~\ref{sec:systematics}). These CSV scale factors are applied to
simulation on an event-by-event basis where the overall scale factor
is the product of the individual scale factors for each jet in the
event.  This procedure was checked using control samples.

Tables~\ref{tab:yield_bb_lj_8TeV}, ~\ref{tab:yield_bb_dil_8TeV}, and
~\ref{tab:yield_bb_tau_8TeV} show the predicted event yields compared
to data after the selection in the lepton+jets, dilepton, and $\tauh$
channels, respectively.  The tables are sub-divided into the different
jet and b-tag categories used in each channel. The signal yield is the
SM prediction ($\mu$ fixed to 1).  In these tables, background yields and
uncertainties use the best-fit value of all nuisance parameters, with
$\mu$ fixed at 1.  For more details about the statistical treatment
and the definition of $\mu$, see section~\ref{sec:results}.
The expected and observed yields agree well in all final states across
the different jet and b-tag categories.

\begin{figure}[!hbtp]
 \begin{center} \hspace{0.32\textwidth}
   \subfigure{
     \includegraphics[width=0.31\textwidth]{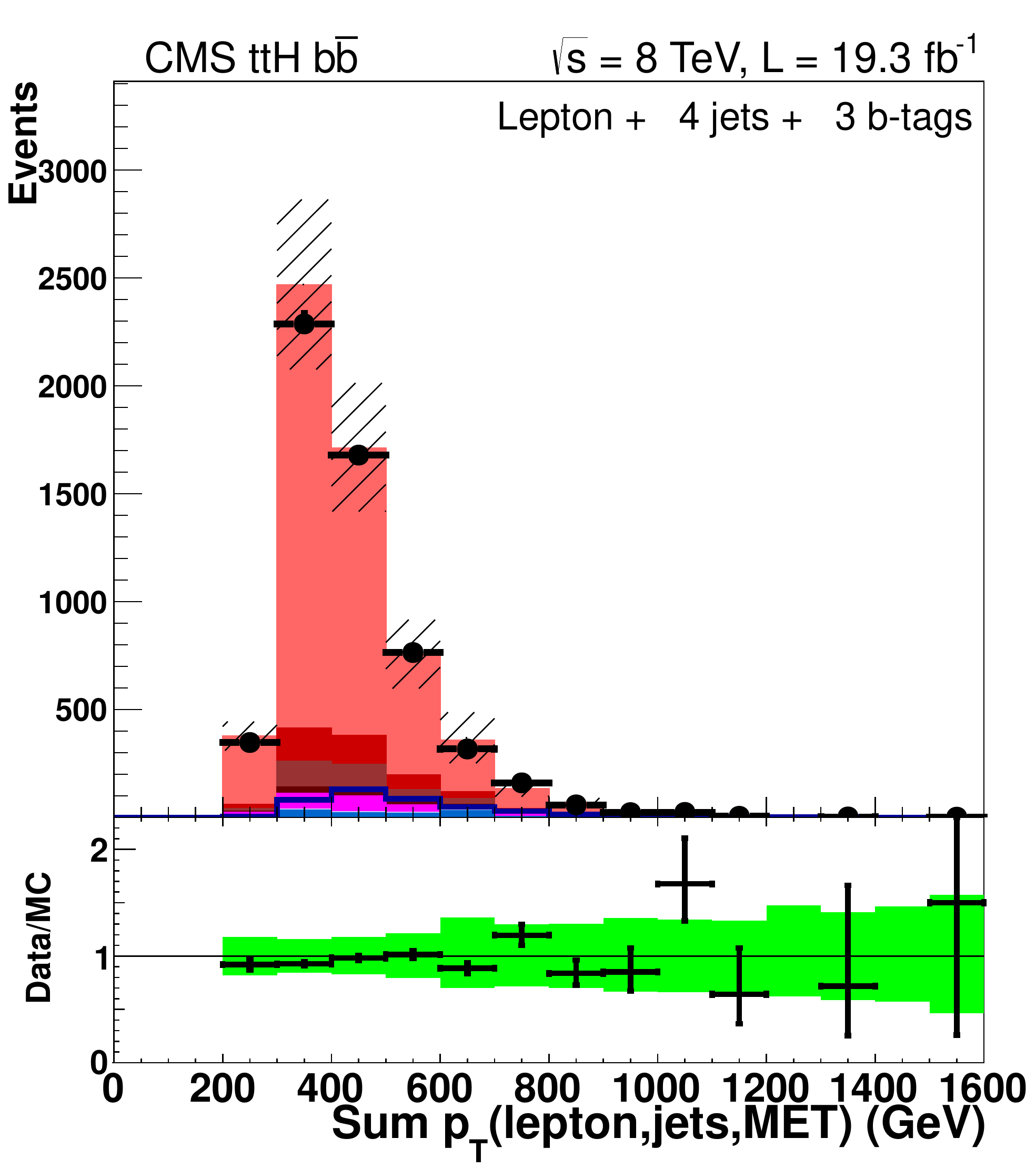}
   }
   \includegraphics[width=0.31\textwidth]{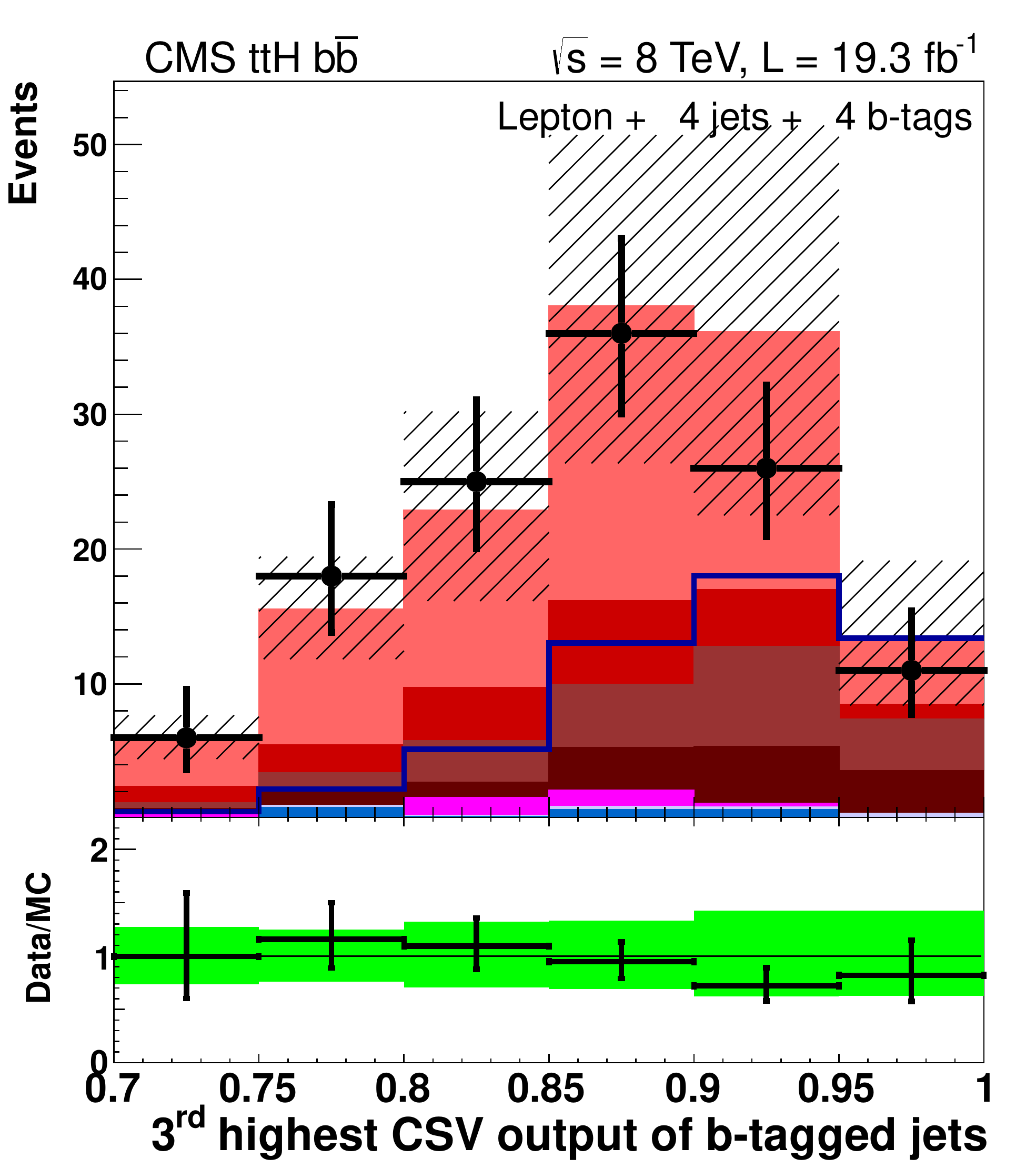}
   \vspace{0.2cm}

   \subfigure{
     \includegraphics[width=0.27\textwidth]{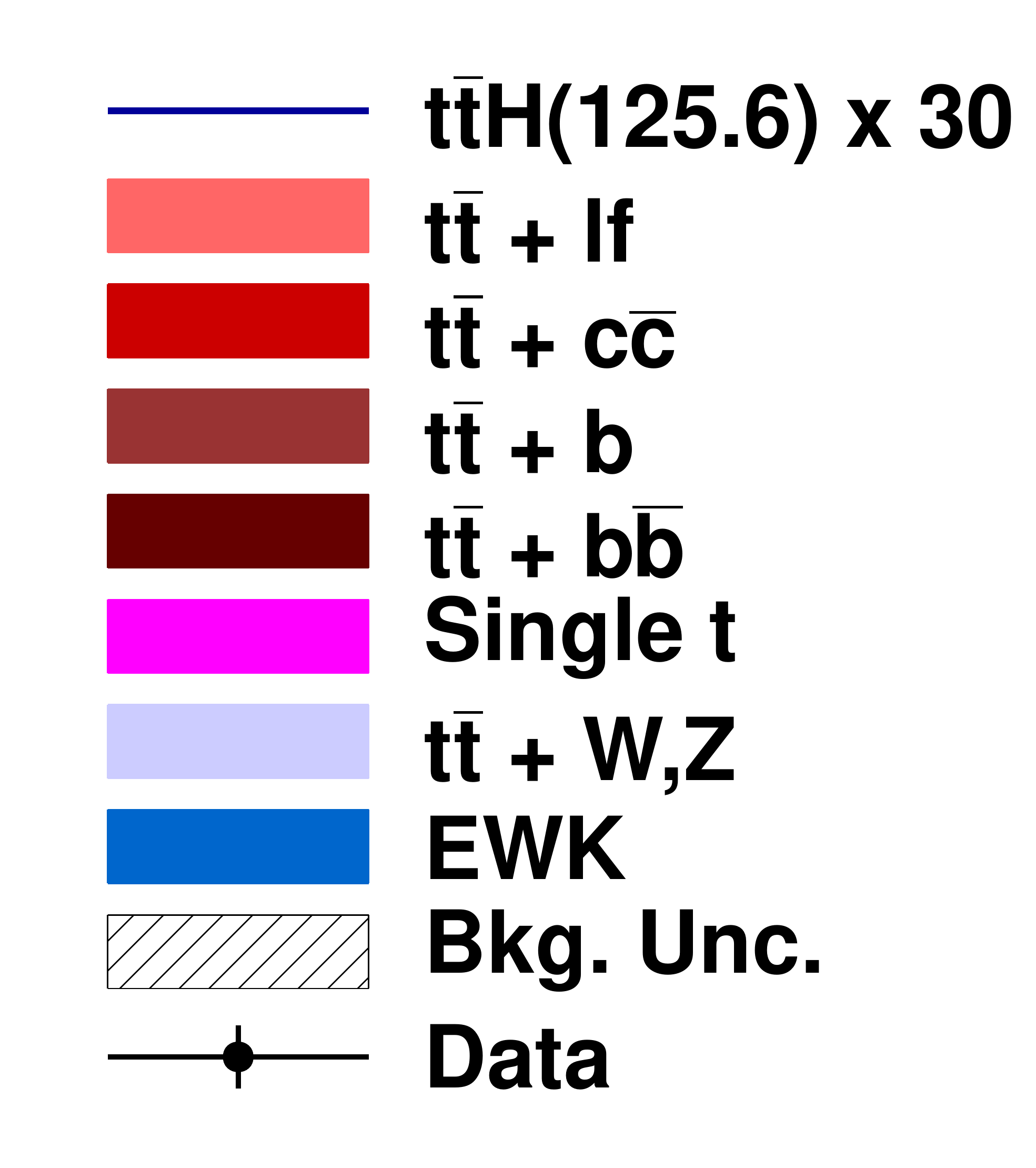}
   }
   \hspace{0.04\textwidth}
   \subfigure{
     \includegraphics[width=0.31\textwidth]{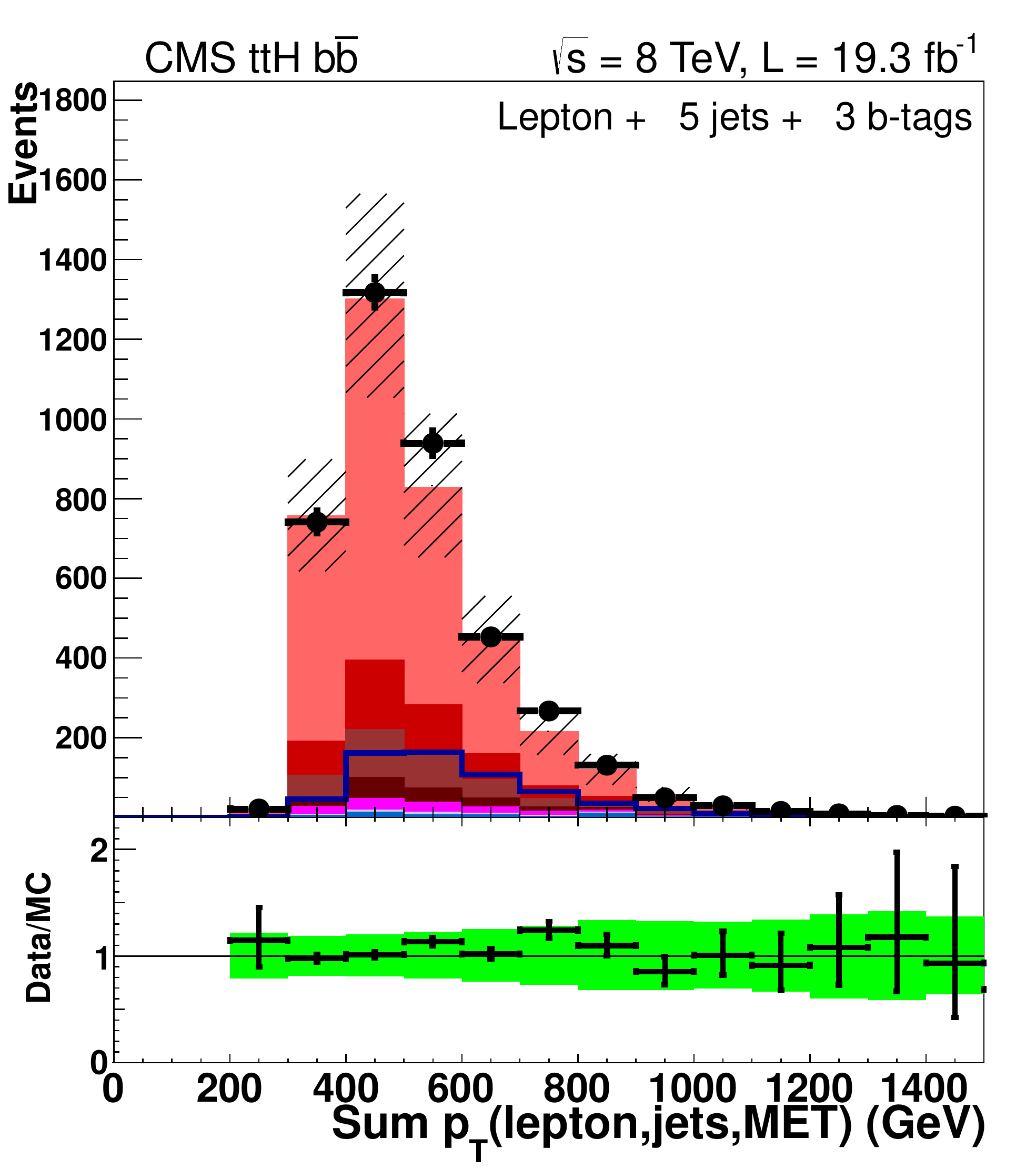}
   }
   \subfigure{
     \includegraphics[width=0.31\textwidth]{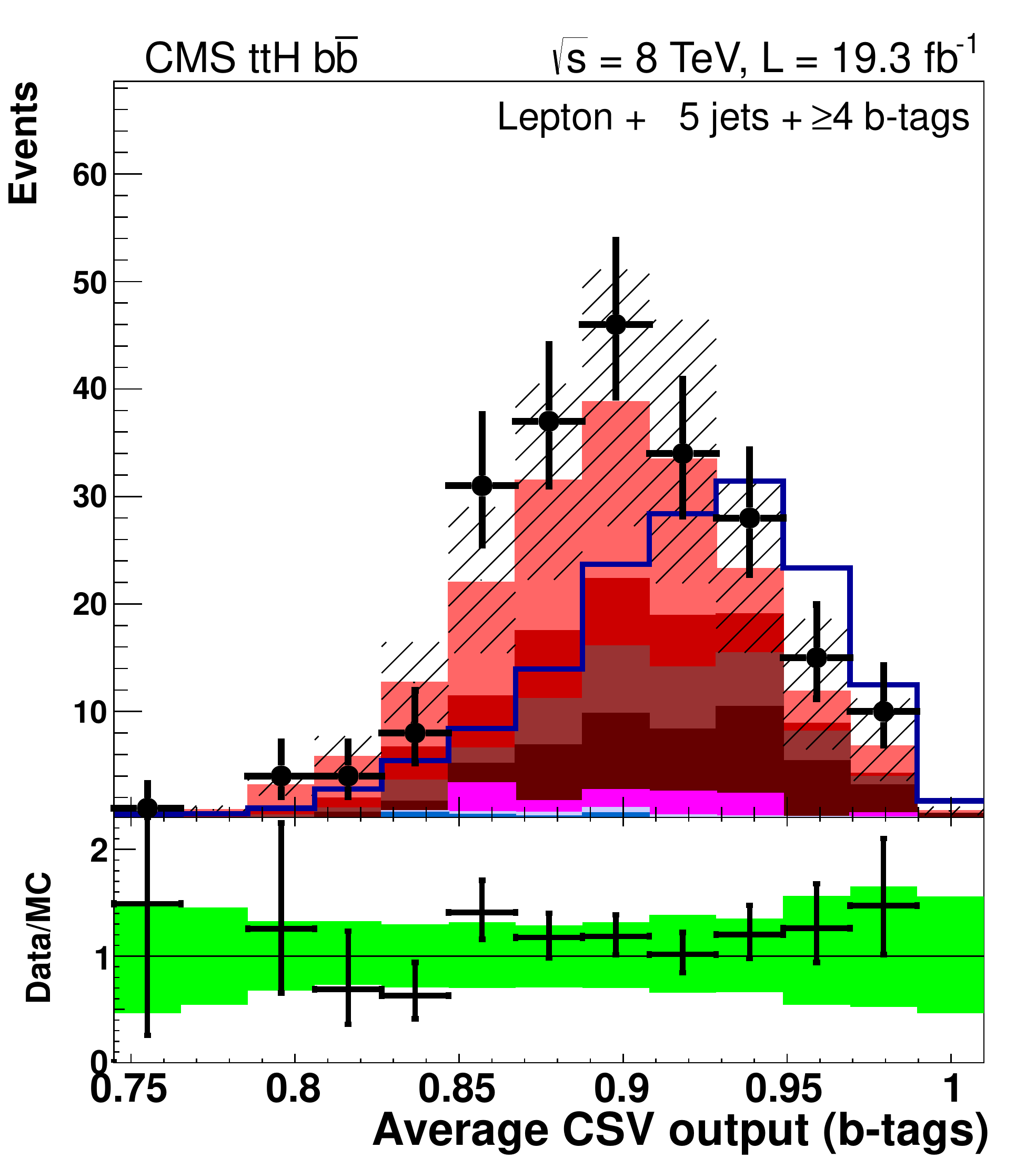}
   }
   \vspace{0.2cm}
   \includegraphics[width=0.31\textwidth]{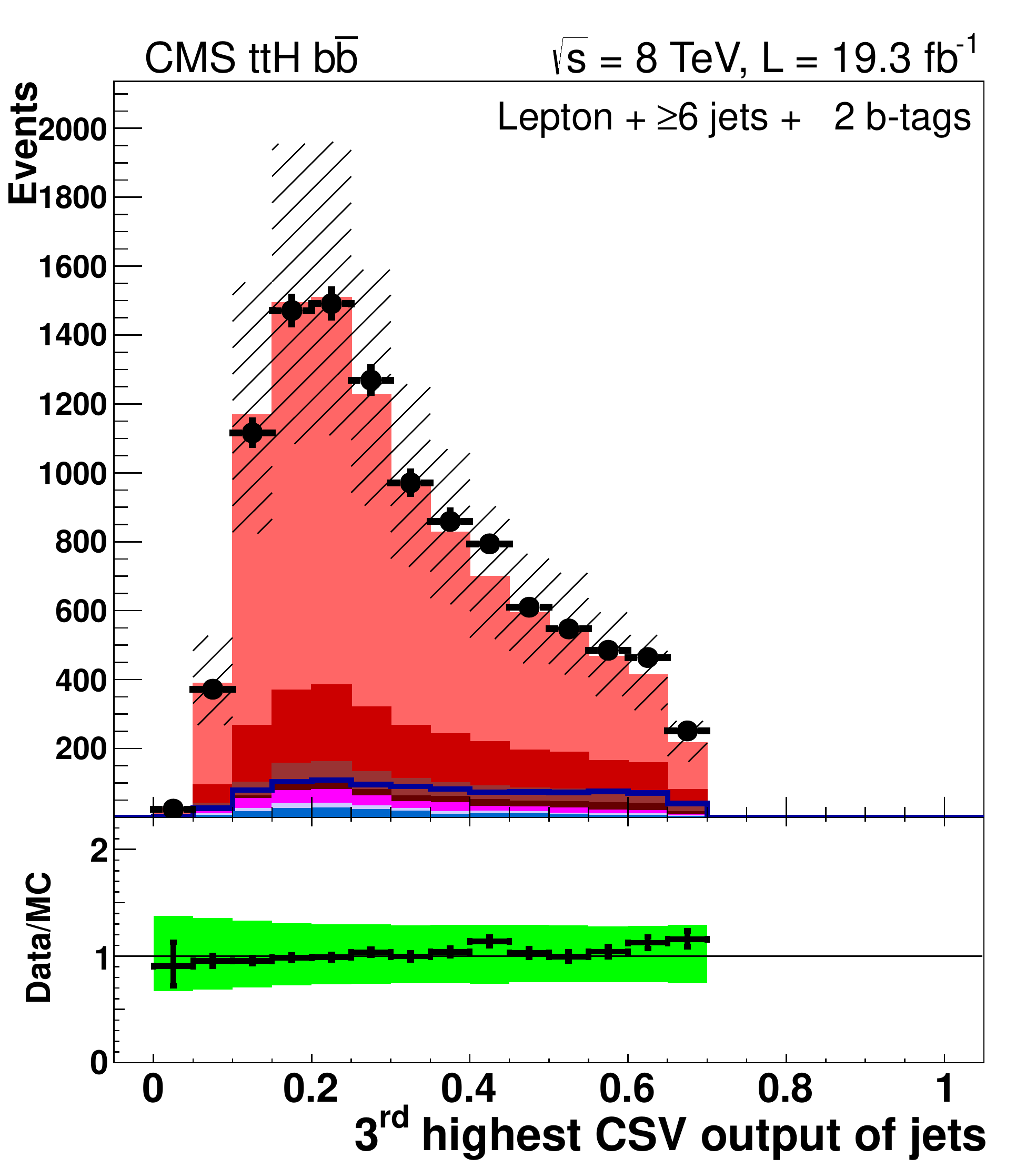}
   \includegraphics[width=0.31\textwidth]{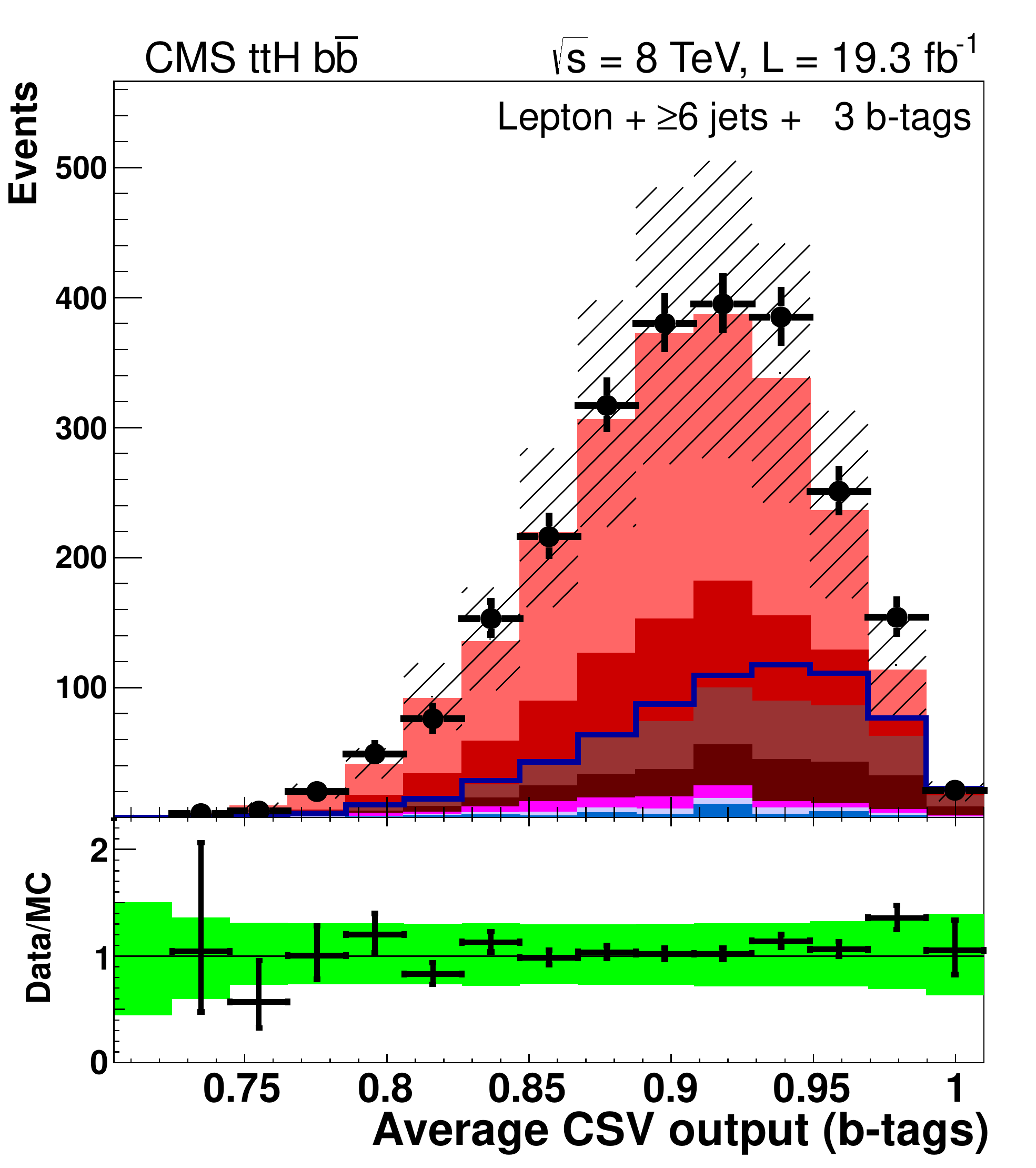}
   \subfigure{
     \includegraphics[width=0.31\textwidth]{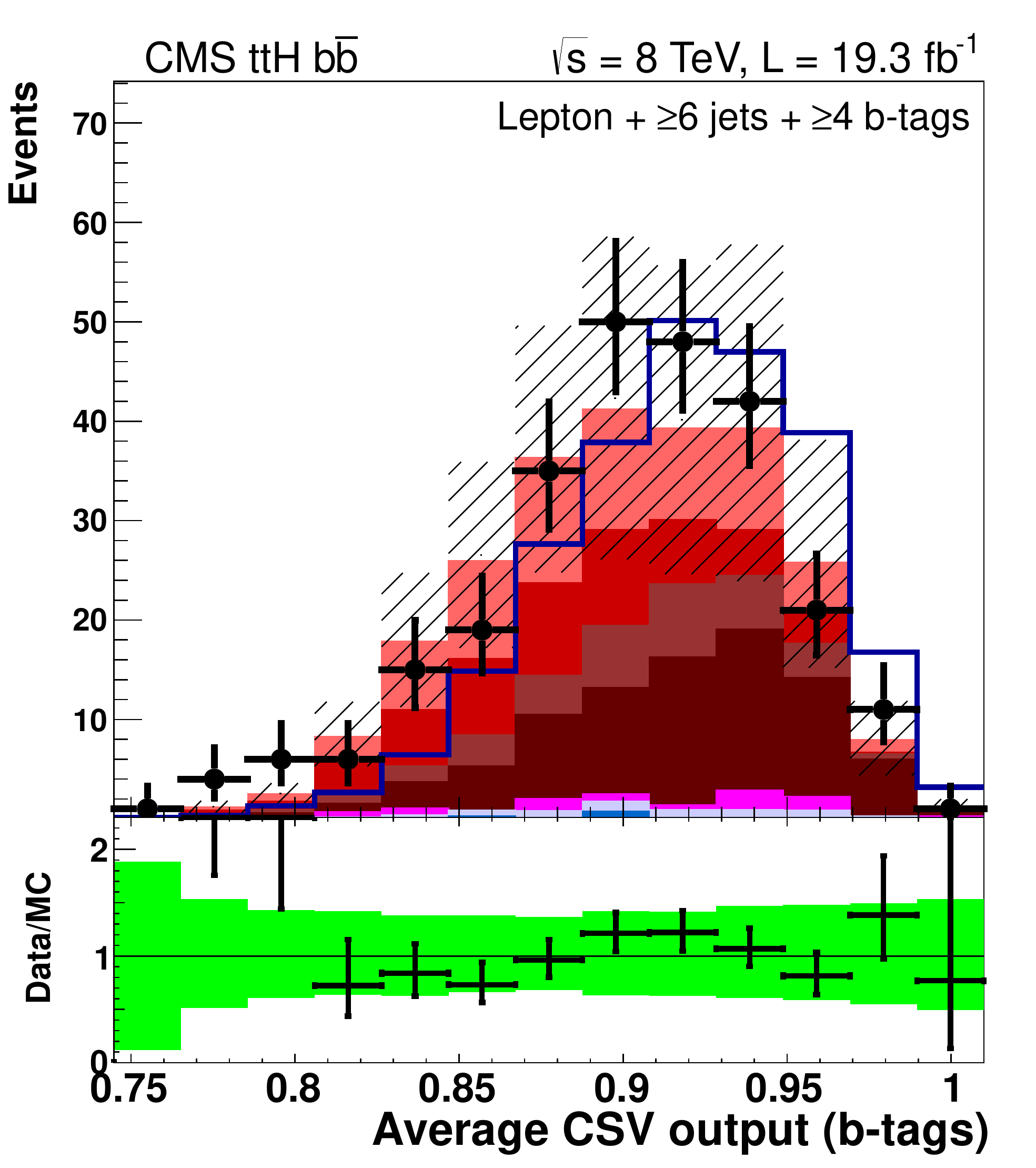}
   }
   \caption{Input variables that give the best signal-background
     separation for each of the lepton+jets categories used in the
     analysis at $\sqrt{s} = 8\TeV$.  The top, middle, and bottom rows
     show the events with 4, 5, and $\ge$6 jets, respectively, while
     the left, middle, and right columns are events with 2, 3, and
     $\ge$4 b-tags, respectively.  More details regarding these plots
     are found in the text.}
     \label{fig:bb_lj_input_8TeV}
   \end{center}
\end{figure}

\begin{figure}[!hbtp]
\begin{center}
  \includegraphics[width=0.70\textwidth]{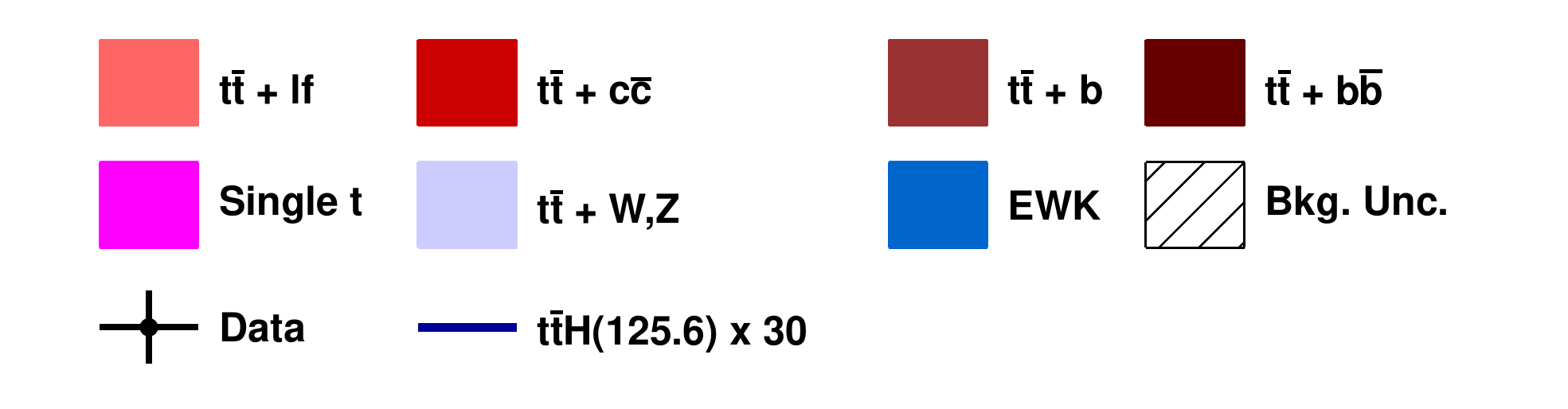}
  \includegraphics[width=0.31\textwidth]{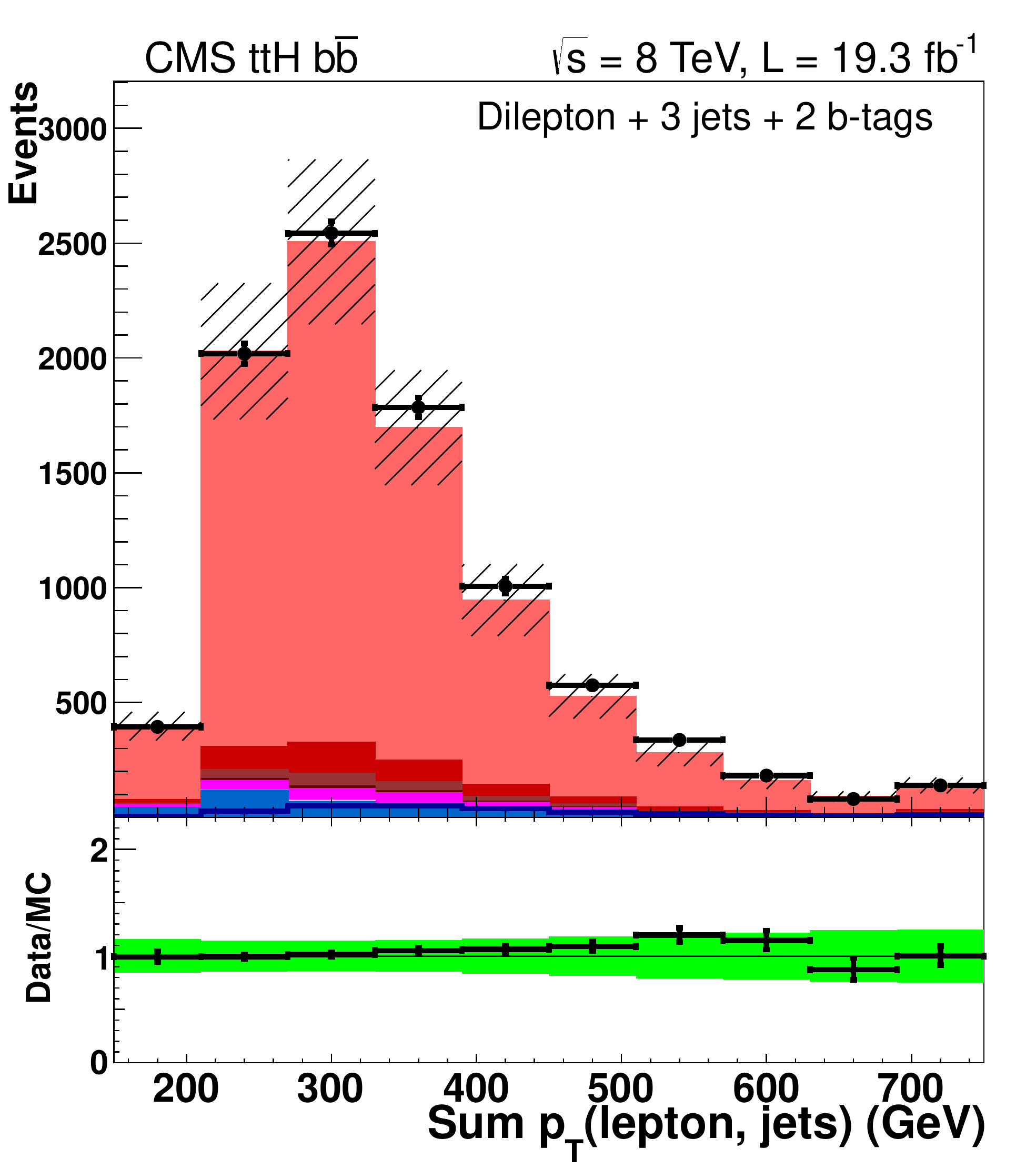}
  \includegraphics[width=0.31\textwidth]{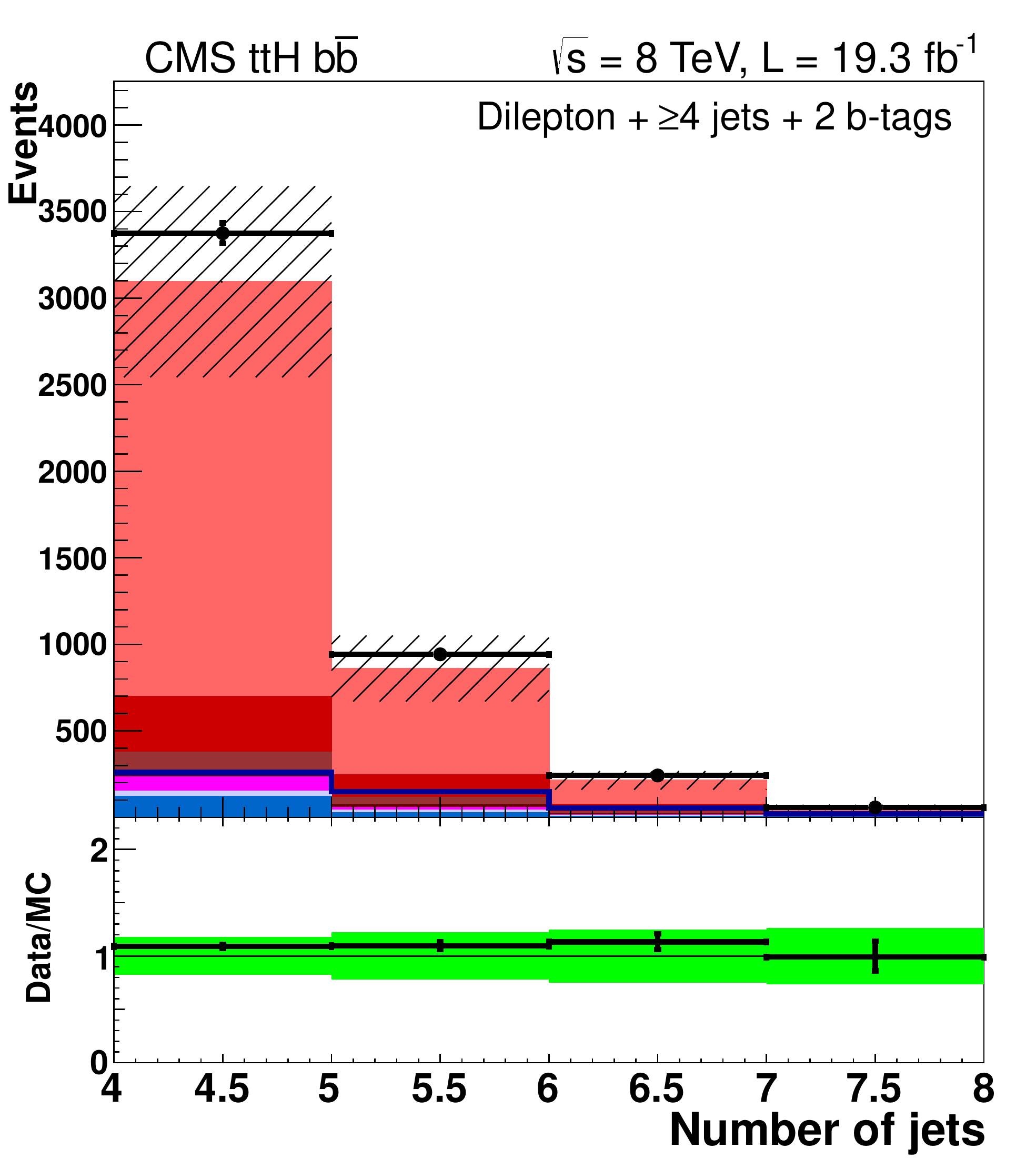}
  \includegraphics[width=0.31\textwidth]{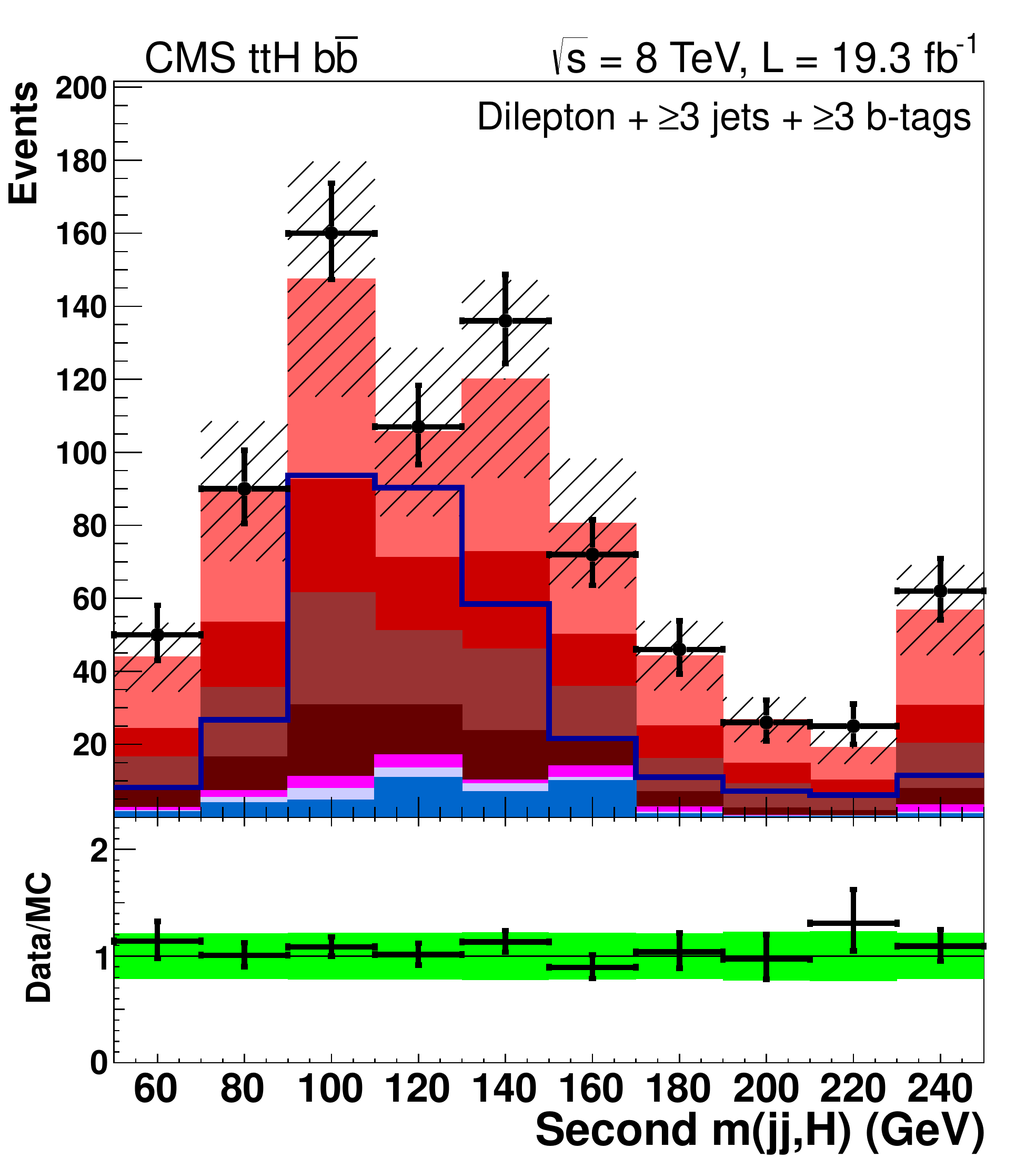}
  \caption{Input variables that give the best signal-background
    separation for each of the dilepton categories used in the
    analysis at $\sqrt{s} = 8\TeV$.  The left, middle, and right
    panels show the events with 3 jets and 2 b-tags, $\ge$4 jets and 2
    b-tags, and $\ge$3 b-tags, respectively.  More details regarding
    these plots are found in the text.}
\label{fig:bb_dil_input_8TeV} \end{center}
\end{figure}

\begin{figure}[!hbtp]
  \begin{center}
    \subfigure{\includegraphics[width=0.21\textwidth]{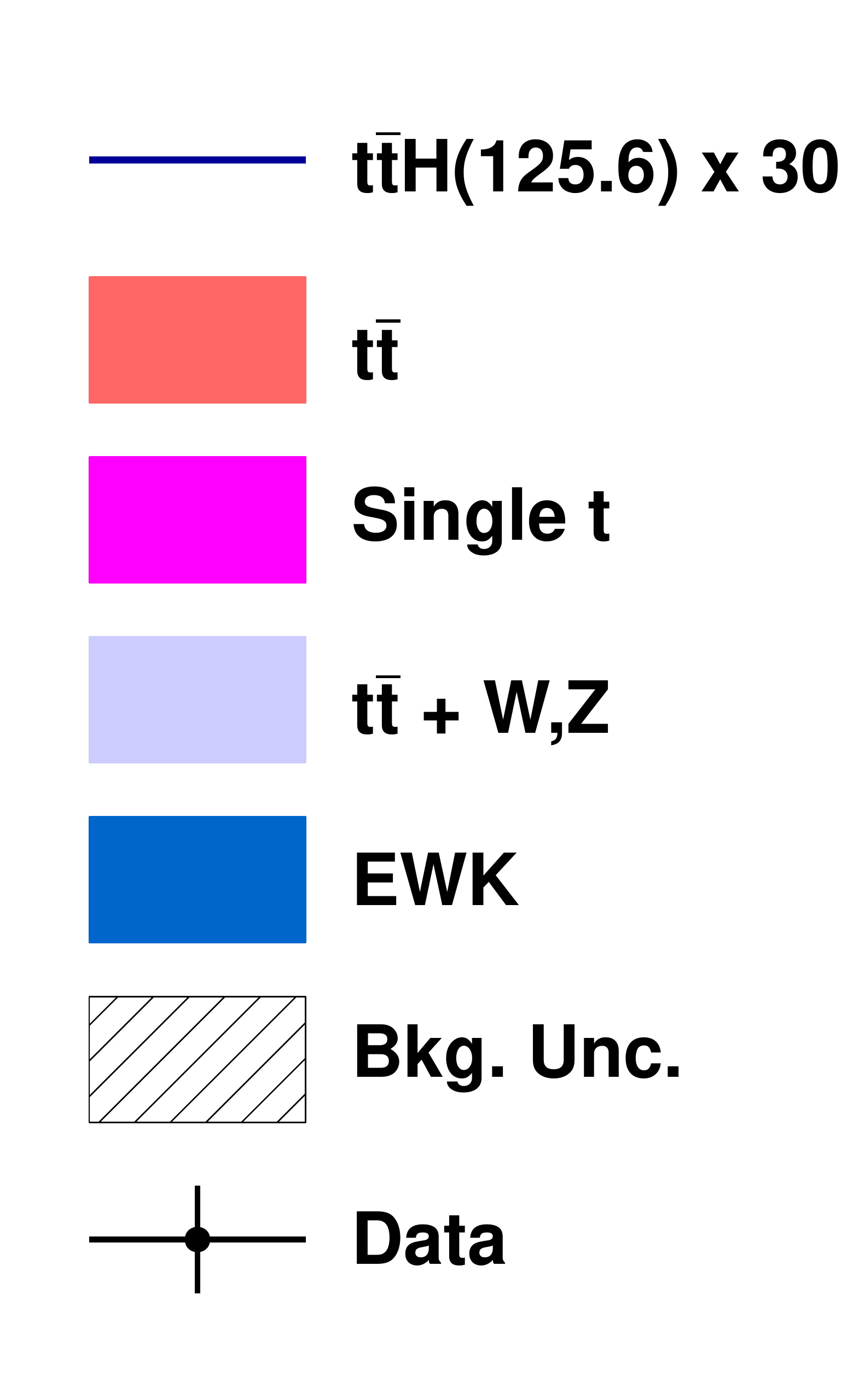}}
    \hspace{0.11\textwidth}
    \subfigure{\includegraphics[width=0.31\textwidth]{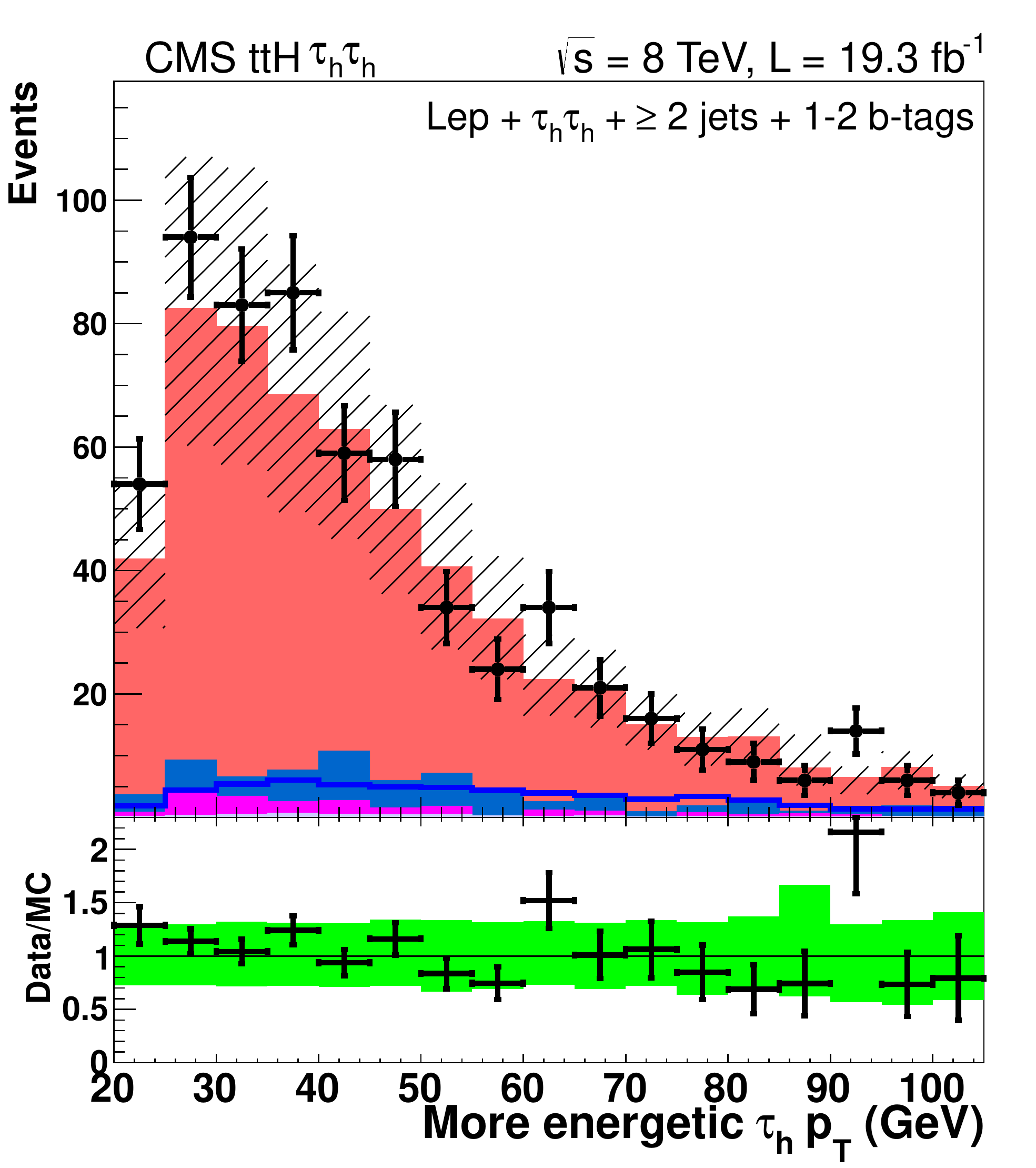}}
    \subfigure{\includegraphics[width=0.31\textwidth]{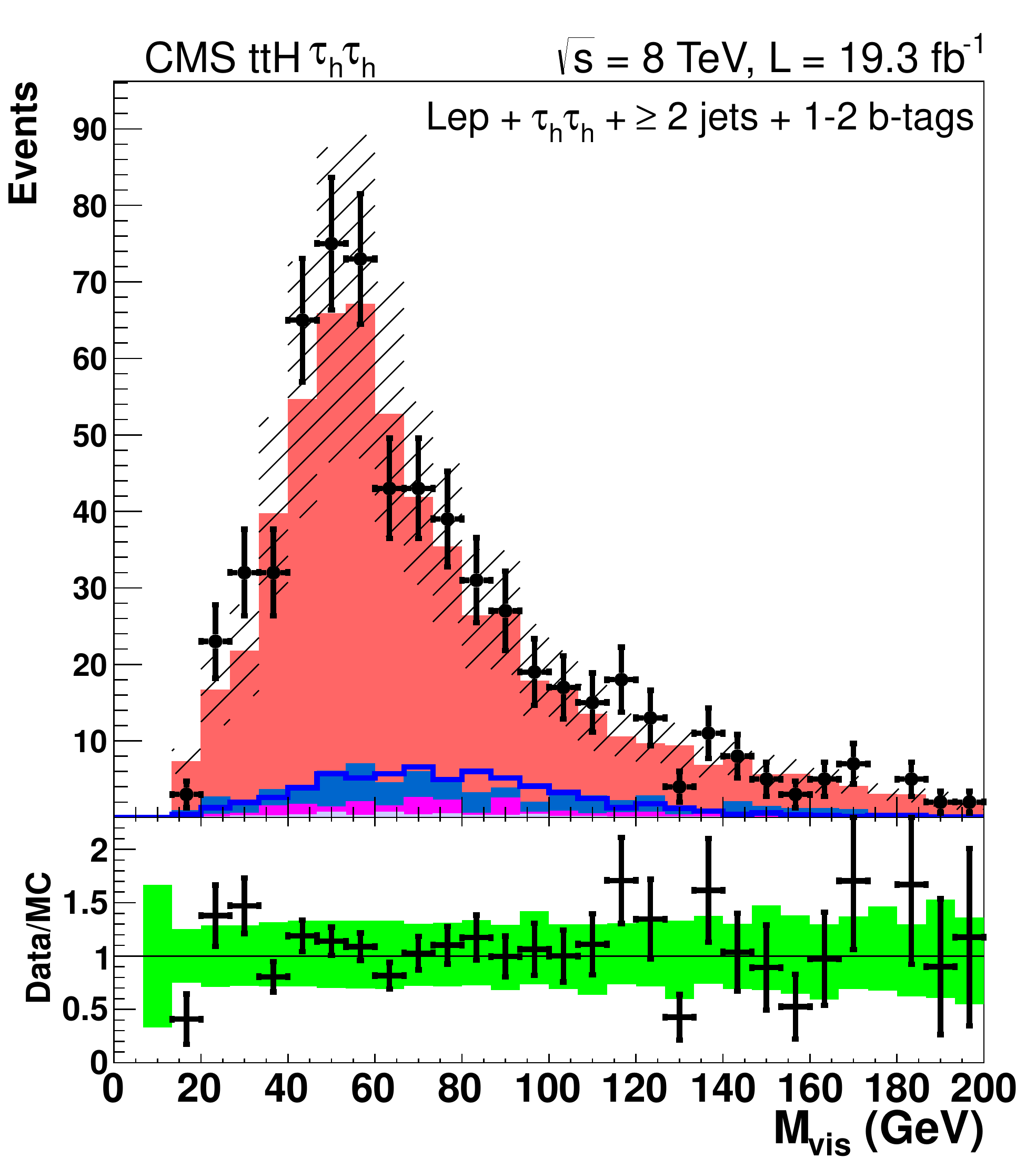}}
    \caption{ Examples of input variables that give the best
      signal-background separation in the analysis of the
      $\tauh$ channels at $\sqrt{s} = 8\TeV$.  The left plot
      shows the $\pt$ of the more energetic $\tauh$, while
      the right plot displays $M_\mathrm{vis}$, the mass of the
      visible $\tauh$ decay products. Events of all
      categories are shown.  More details regarding these plots are
      found in the text.}
\label{fig:bb_tau_input_8TeV} \end{center}
\end{figure}

Figures~\ref{fig:bb_lj_input_8TeV},~\ref{fig:bb_dil_input_8TeV},
and~\ref{fig:bb_tau_input_8TeV} show the data-to-simulation
comparisons of variables that give the best signal-background
separation in each of the lepton+jets, dilepton, and
$\tauh$ channels, respectively.  In these plots, the
background is normalized to the SM expectation; the uncertainty band
(shown as a hatched band in the stack plot and a green band in the
ratio plot) includes statistical and systematic uncertainties that
affect both the rate and shape of the background distributions. For
the ratio plots shown below each distribution, only the background
expectation (and not the signal) is included in the denominator of the
ratio.  The contribution labeled ``EWK'' is the sum of the diboson and
$\PW/\cPZ+$jets backgrounds.  The $\ttbar\PH$ signal ($m_{\PH} = 125.6\GeV$) is not included in the stacked histogram, but is shown as
a separate open histogram normalized to 30 times the SM expectation
($\mu = 30$).  To calculate the variable second $m$(jj,H), the
invariant masses of all jet pairs with at least one b-tagged jet are
calculated and the jet pair whose mass is the second closest to the
Higgs boson mass is chosen.  Within the uncertainties, the simulation
reproduces well the shape and the normalization of the distributions.

\begin{table}[!htbp]
\begin{center}
\topcaption{Expected event yields for signal
($m_{\PH} = 125.6\GeV$) and backgrounds in the lepton+jets channel.
Signal and background normalizations used for this table are
described in the text.}
\noindent
\footnotesize
    \begin{tabular}{|l|c|c|c|c|c|c|c|} \hline
& $\geq$6 jets + & 4 jets + & 5 jets + & $\geq$6 jets + & 4 jets + & 5 jets + & $\geq$6 jets + \\
& 2 b-tags & 3 b-tags & 3 b-tags & 3 b-tags & 4 b-tags & $\geq$4 b-tags & $\geq$4 b-tags \\ \hline \hline
$\ttbar\PH(125.6\GeV)$ & 28.5 $\pm$ 2.5 & 12.4 $\pm$ 1.0 & 18.1 $\pm$ 1.5 & 18.9 $\pm$ 1.5 & 1.5 $\pm$ 0.2 & 4.4 $\pm$ 0.4 & 6.7 $\pm$ 0.6 \\
 \hline
$\ttbar+$lf & 7140 $\pm$ 310 & 4280 $\pm$ 150 & 2450 $\pm$ 130 & 1076 $\pm$ 74 & 48.4 $\pm$ 10.0 & 54 $\pm$ 12 & 44 $\pm$ 11 \\
$\ttbar+$b & 570 $\pm$ 170 & 364 $\pm$ 94 & 367 $\pm$ 98 & 289 $\pm$ 87 & 20.0 $\pm$ 5.5 & 28.6 $\pm$ 8.0 & 33 $\pm$ 10 \\
$\ttbar+\bbbar$ & 264 $\pm$ 59 & 123 $\pm$ 29 & 193 $\pm$ 42 & 232 $\pm$ 49 & 15.8 $\pm$ 3.6 & 45.2 $\pm$ 9.7 & 86 $\pm$ 18 \\
$\ttbar+\ccbar$ & 2420 $\pm$ 300 & 690 $\pm$ 130 & 800 $\pm$ 130 & 720 $\pm$ 110 & 29.7 $\pm$ 5.6 & 55 $\pm$ 11 & 81 $\pm$ 13 \\
$\ttbar$+W/Z & 85 $\pm$ 11 & 15.0 $\pm$ 2.0 & 20.9 $\pm$ 2.8 & 24.7 $\pm$ 3.3 & 1.0 $\pm$ 0.2 & 2.1 $\pm$ 0.4 & 4.7 $\pm$ 0.8 \\
Single t & 236 $\pm$ 18 & 213 $\pm$ 17 & 101.7 $\pm$ 10.0 & 47.7 $\pm$ 6.7 & 2.8 $\pm$ 1.4 & 7.5 $\pm$ 3.8 & 6.7 $\pm$ 2.6 \\
W/Z+jets & 75 $\pm$ 27 & 46 $\pm$ 30 & 13 $\pm$ 12 & 7.7 $\pm$ 8.8 & 1.1 $\pm$ 1.2 & 0.9 $\pm$ 1.0 & 0.3 $\pm$ 0.8 \\
Diboson & 4.5 $\pm$ 1.0 & 5.4 $\pm$ 0.9 & 2.0 $\pm$ 0.5 & 1.0 $\pm$ 0.4 & 0.2 $\pm$ 0.2 & 0.1 $\pm$ 0.1 & 0.2 $\pm$ 0.1 \\
 \hline
Total bkg & 10790 $\pm$ 200 & 5730 $\pm$ 110 & 3935 $\pm$ 74 & 2394 $\pm$ 65 & 119.0 $\pm$ 8.2 & 193.4 $\pm$ 10.0 & 256 $\pm$ 16 \\
 \hline
Data & 10724 & 5667 & 3983 & 2426 & 122 & 219 & 260 \\
\hline
\end{tabular}
    \label{tab:yield_bb_lj_8TeV}
\end{center}
\end{table}

\begin{table}[!htbp]
\begin{center}
\topcaption{ Expected event yields for signal
($m_{\PH} = 125.6\GeV$) and backgrounds in the dilepton channel.
Signal and background normalizations used for this table are
described in the text.}
    \begin{tabular}{|l|c|c|c|} \hline
& 3 jets + 2 b-tags & $\geq$4 jets + 2 b-tags & $\geq$3 b-tags  \\ \hline \hline
$\ttbar\PH(125.6\GeV)$ & 7.4 $\pm$ 0.6 & 14.5 $\pm$ 1.2 & 10.0 $\pm$ 0.8 \\
 \hline
$\ttbar+$lf & 7650 $\pm$ 170 & 3200 $\pm$ 120 & 227 $\pm$ 35 \\
$\ttbar+$b & 210 $\pm$ 55 & 198 $\pm$ 57 & 160 $\pm$ 43 \\
$\ttbar+\bbbar$ & 50 $\pm$ 13 & 76 $\pm$ 17 & 101 $\pm$ 21 \\
$\ttbar+\ccbar$ & 690 $\pm$ 110 & 761 $\pm$ 97 & 258 $\pm$ 46 \\
$\ttbar$+W/Z & 29.5 $\pm$ 3.8 & 50.5 $\pm$ 6.4 & 10.9 $\pm$ 1.5 \\
Single t & 218 $\pm$ 16 & 95.2 $\pm$ 8.8 & 14.6 $\pm$ 3.6 \\
W/Z+jets & 217 $\pm$ 52 & 98 $\pm$ 28 & 21 $\pm$ 15 \\
Diboson & 9.5 $\pm$ 0.9 & 2.9 $\pm$ 0.4 & 0.6 $\pm$ 0.1 \\
 \hline
Total bkg & 9060 $\pm$ 130 & 4475 $\pm$ 82 & 793 $\pm$ 28 \\
 \hline
Data & 9060 & 4616 & 774 \\
\hline
\end{tabular}
    \label{tab:yield_bb_dil_8TeV}
\end{center}
\end{table}

\begin{table}[!htbp]
  \begin{center}
\topcaption{Expected event yields for signal
($m_{\PH} = 125.6\GeV$) and backgrounds in the $\tauh$ channel. Signal
and background normalizations used for this table are described in
the text.}
\small
    \begin{tabular}{|l|c|c|c|c|c|c|} \hline
& 2 jets + & 3 jets + & $\geq$4 jets + & 2 jets + & 3 jets + & $\geq$4 jets + \\
& 1 b-tag & 1 b-tag & 1 b-tag & 2 b-tags & 2 b-tags & 2 b-tags \\ \hline \hline
$\ttbar\PH(125.6\GeV)$ & 0.4 $\pm$ 0.1 & 0.6 $\pm$ 0.1 & 0.5 $\pm$ 0.1 & 0.1 $\pm$ 0.0 & 0.2 $\pm$ 0.0 & 0.3 $\pm$ 0.0 \\
 \hline
$\ttbar+$lf & 266 $\pm$ 12 & 144.7 $\pm$ 7.1 & 72.1 $\pm$ 4.1 & 55.0 $\pm$ 3.4 & 45.2 $\pm$ 2.8 & 28.8 $\pm$ 2.1 \\
$\ttbar$+W/Z & 1.1 $\pm$ 0.2 & 1.3 $\pm$ 0.2 & 1.3 $\pm$ 0.3 & 0.5 $\pm$ 0.1 & 0.6 $\pm$ 0.1 & 0.9 $\pm$ 0.2 \\
Single t & 12.9 $\pm$ 2.1 & 3.5 $\pm$ 1.2 & 0.7 $\pm$ 0.6 & 2.2 $\pm$ 0.9 & 1.2 $\pm$ 0.5 & 0.4 $\pm$ 0.7 \\
W/Z+jets & 22.9 $\pm$ 6.3 & 7.7 $\pm$ 2.8 & 2.1 $\pm$ 1.2 & 1.0 $\pm$ 0.6 & 0.3 $\pm$ 0.2 & 0.2 $\pm$ 0.4 \\
Diboson & 0.9 $\pm$ 0.2 & 0.7 $\pm$ 0.2 & 0.1 $\pm$ 0.0 & 0.0 $\pm$ 0.0 & 0.1 $\pm$ 0.0 & 0.0 $\pm$ 0.1 \\
 \hline
Total bkg & 304 $\pm$ 14 & 158.0 $\pm$ 7.5 & 76.4 $\pm$ 4.2 & 58.7 $\pm$ 3.6 & 47.3 $\pm$ 2.9 & 30.4 $\pm$ 2.3 \\
 \hline
Data & 292 & 171 & 92 & 41 & 48 & 35 \\
\hline
\end{tabular}
    \label{tab:yield_bb_tau_8TeV}
  \end{center}
\end{table}

\subsection{Signal extraction}

Boosted decision trees (BDTs)~\cite{bdt_nim} are used to further
improve signal sensitivity. In the lepton+jets and dilepton channels,
BDTs are trained separately for each category, using the $\ttbar \PH$
sample with $m_{\PH} = 125\GeV$.  The three dilepton categories use a
single BDT. Of the seven lepton+jets categories, four categories use a
single BDT, while three categories each use two BDTs in a tiered
configuration.  The tiered configuration includes one BDT that is
trained specifically to discriminate between $\ttbar \PH$ and
$\ttbar\bbbar$ events, the output of which is then used as an
input variable in the second, more general, $\ttbar \PH$ versus
$\ttbar$+jets BDT. This tiered approach allows better discrimination
between the $\ttbar \PH$ process and the difficult $\ttbar\bbbar$
component of $\ttbar$+jets production, resulting in better control of
$\ttbar$+hf systematics and a lower expected limit on $\mu$.  In the
$\tauh$ channel, due to the low event counts, a single BDT
is used for all categories, using an event selection equivalent to the
union of all categories with more than one untagged jet.

All BDTs utilize variables involving the kinematics of the
reconstructed objects, the event shape, and the CSV b-tag
discriminant. Ten variables are used as inputs to the final BDTs in
all lepton+jets categories, while 10 or 15 variables are used in the
first BDT in categories employing the tiered-BDT system (the $\geq$6
jets + $\geq$4 b-tags and $\geq$6 jets + 3 b-tags categories use 15
variables, and the 5 jets + $\geq$4 b-tags category uses ten variables
due to lower available training statistics in that category). The
dilepton channel uses four variables for the 3 jets + 2 b-tags category
and six in each of the other categories.  In the $\tauh$ channel, almost all
variables used to train the BDT are related to the $\tauh$
system, such as the mass of the visible $\Pgt$ decay products, the
$\pt$, the isolation, and the decay mode of both $\tauh$,
and the $\abs{\eta}$ and distance to the lepton of the more energetic
$\tauh$.  In addition, the $\pt$ of the most energetic jet,
regardless of the b-tagging status, is used in the BDT.

To train the BDTs, the $\tauh$ channel uses simulated
$\ttbar\PH$, $\PH \to \Pgt\Pgt$ ($m_{\PH} = 125\GeV$) events
with generator-level matched $\tauh$ pairs as the signal,
whereas both the lepton+jets and dilepton channels uses $\ttbar\PH$
($m_{\PH}=125\GeV$) events, with inclusive Higgs boson decays.  All
three channels use $\ttbar$+jets events as background when training.
An equal number of signal and background events are used for a given
category and channel. The signal and background events are evenly
divided into two subsamples: one set of events is used to do the
actual training, and the other is used as a test sample to monitor
against overtraining.  The specific BDT method used is a ``gradient
boost'', available as part of the TMVA package~\cite{Hocker:2007ht} in
ROOT~\cite{Brun:1997pa}. The tree architecture consists of five
nodes, a few hundred trees form a forest, and the learning rate is set
to 0.1.

\begin{figure}[!hbtp]
 \begin{center} \hspace{0.32\textwidth} \subfigure{
   \label{bb_lj_finalMVA_8TeV_1}
   \includegraphics[width=0.31\textwidth]{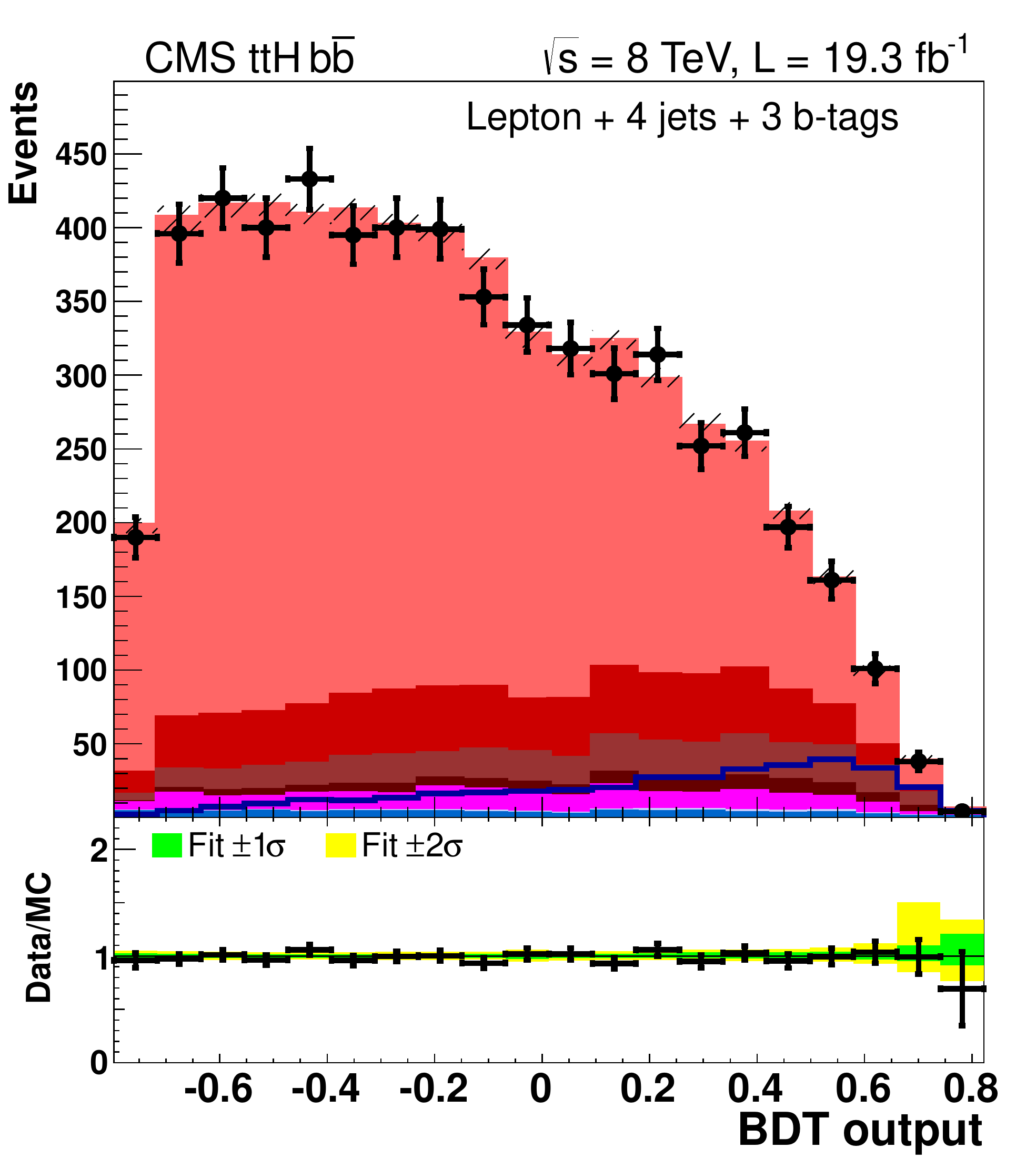}
   } \subfigure{ \label{bb_lj_finalMVA_8TeV_2}
   \includegraphics[width=0.31\textwidth]{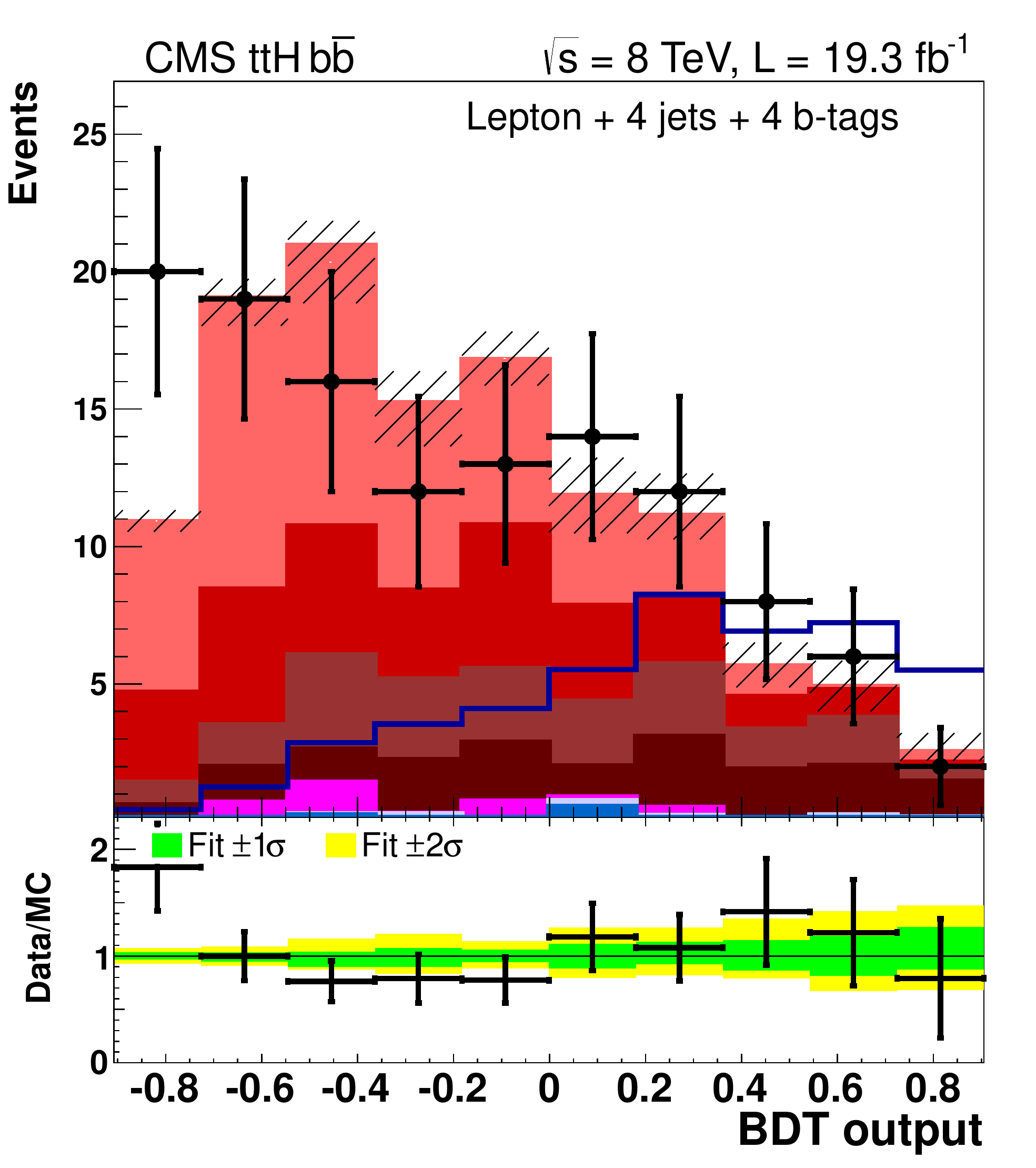}
   } \vspace{0.2cm}

   \subfigure{
   \label{bb_lj_finalMVA_8TeV_leg}
   \includegraphics[width=0.27\textwidth]{plots/tth-hadrons_finalMVA/ttH_legend_1columns.pdf}
   }
   \hspace{0.04\textwidth}
   \subfigure{
   \label{bb_lj_finalMVA_8TeV_3}
   \includegraphics[width=0.31\textwidth]{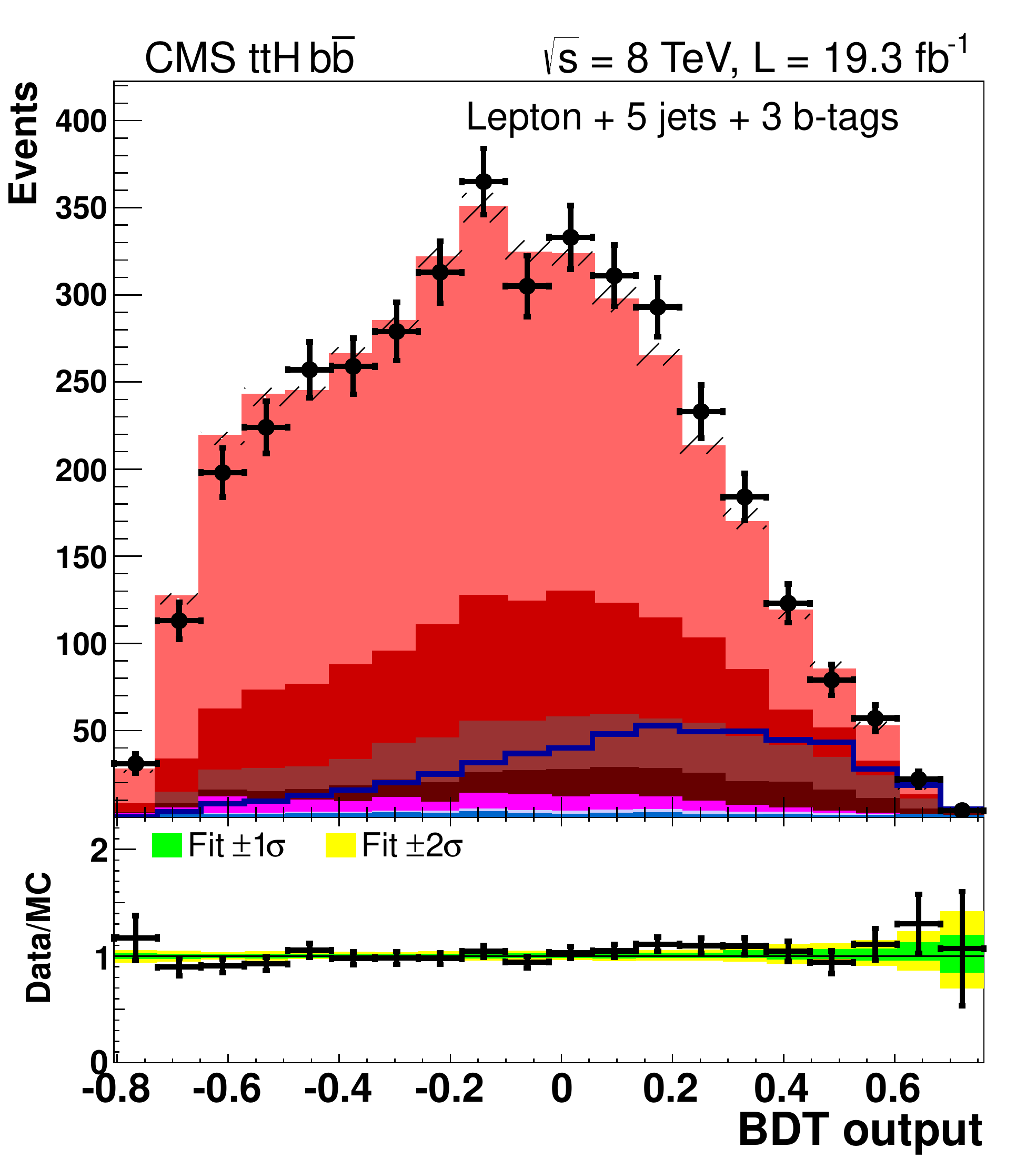}
   }
   \subfigure{
   \label{bb_lj_finalMVA_8TeV_4}
   \includegraphics[width=0.31\textwidth]{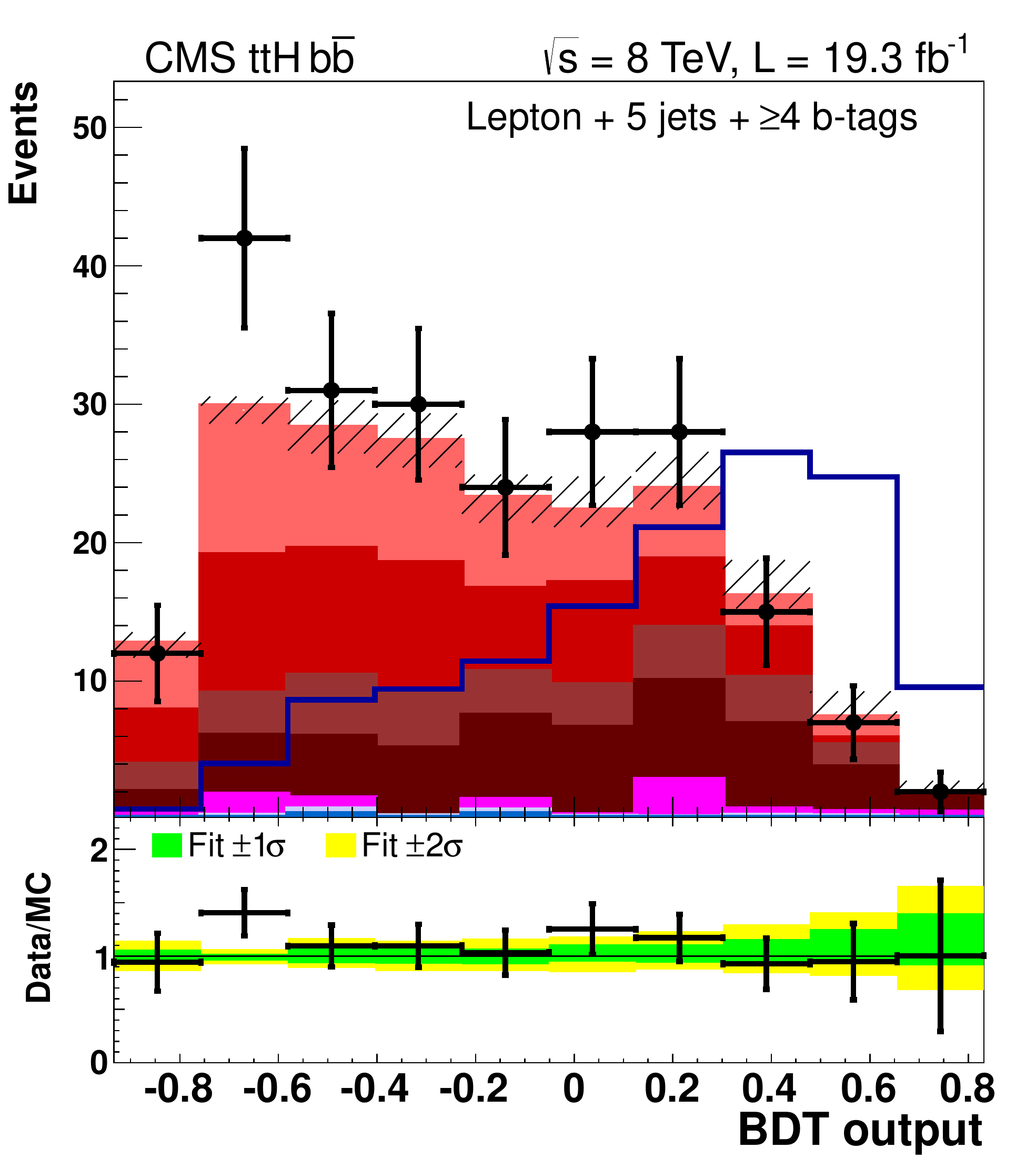}
   }
   \vspace{0.2cm}

   \subfigure{ \label{bb_lj_finalMVA_8TeV_5}
   \includegraphics[width=0.31\textwidth]{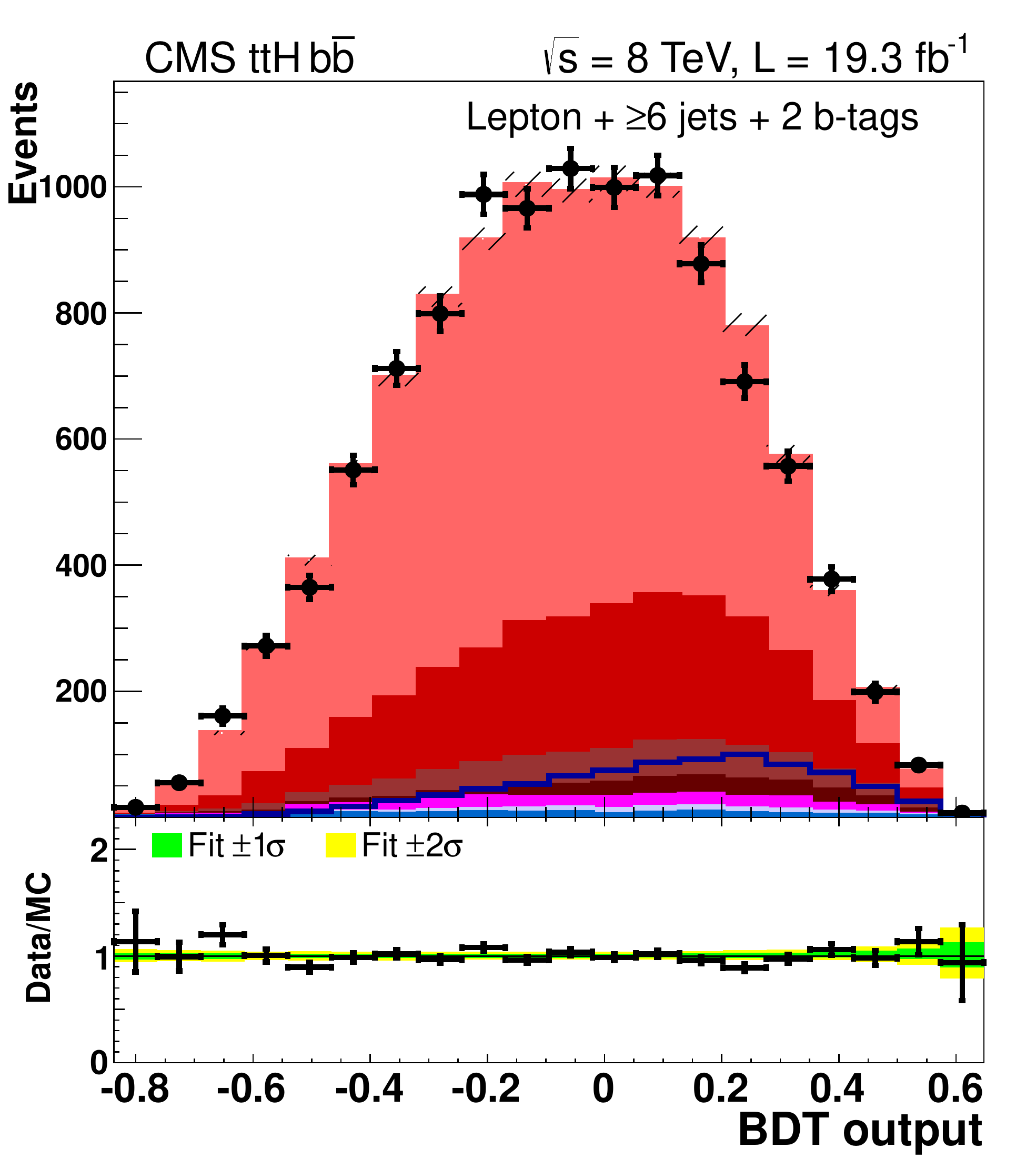}
   } \subfigure{ \label{bb_lj_finalMVA_8TeV_6}
   \includegraphics[width=0.31\textwidth]{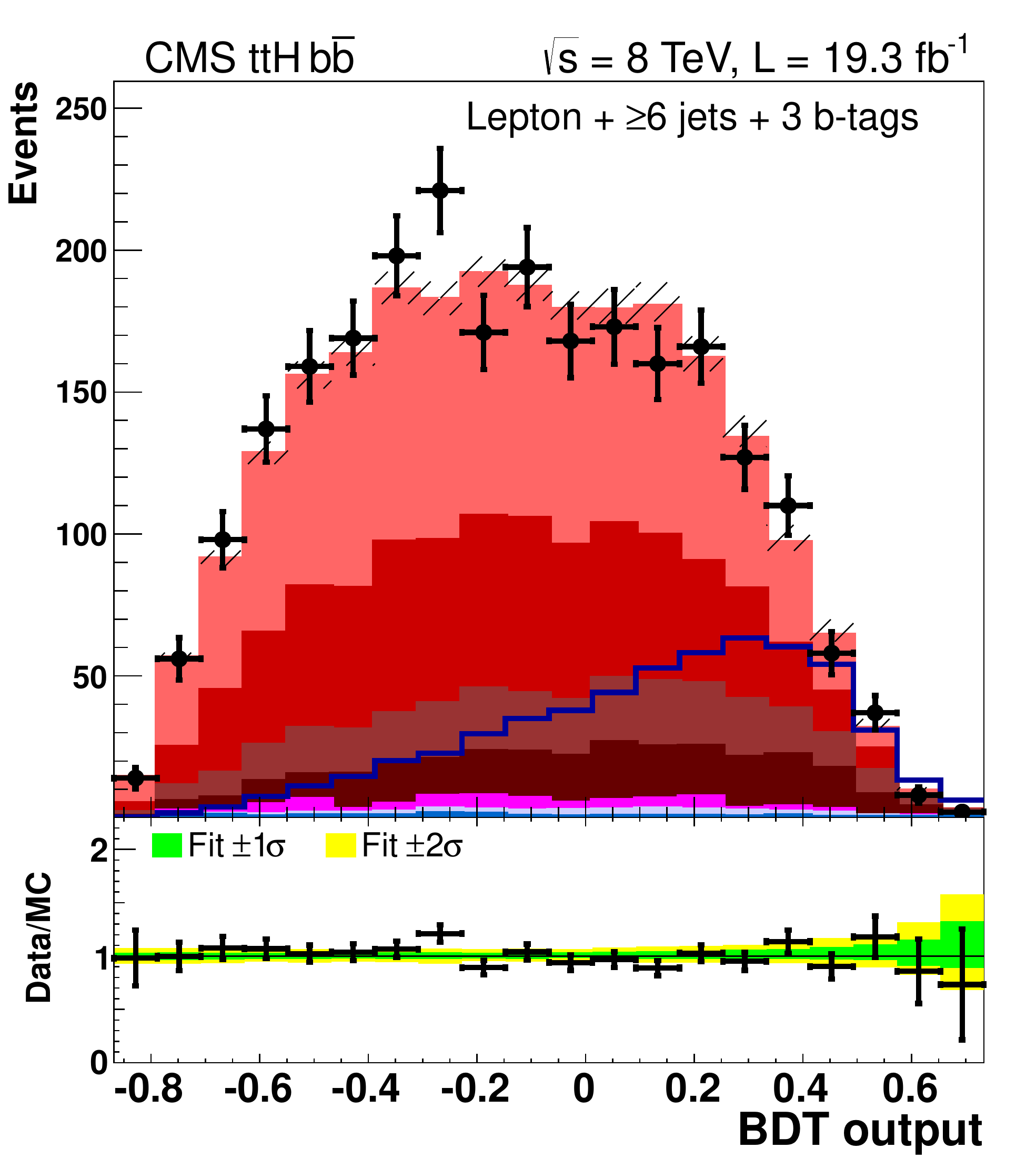}
   } \subfigure{ \label{bb_lj_finalMVA_8TeV_7}
   \includegraphics[width=0.31\textwidth]{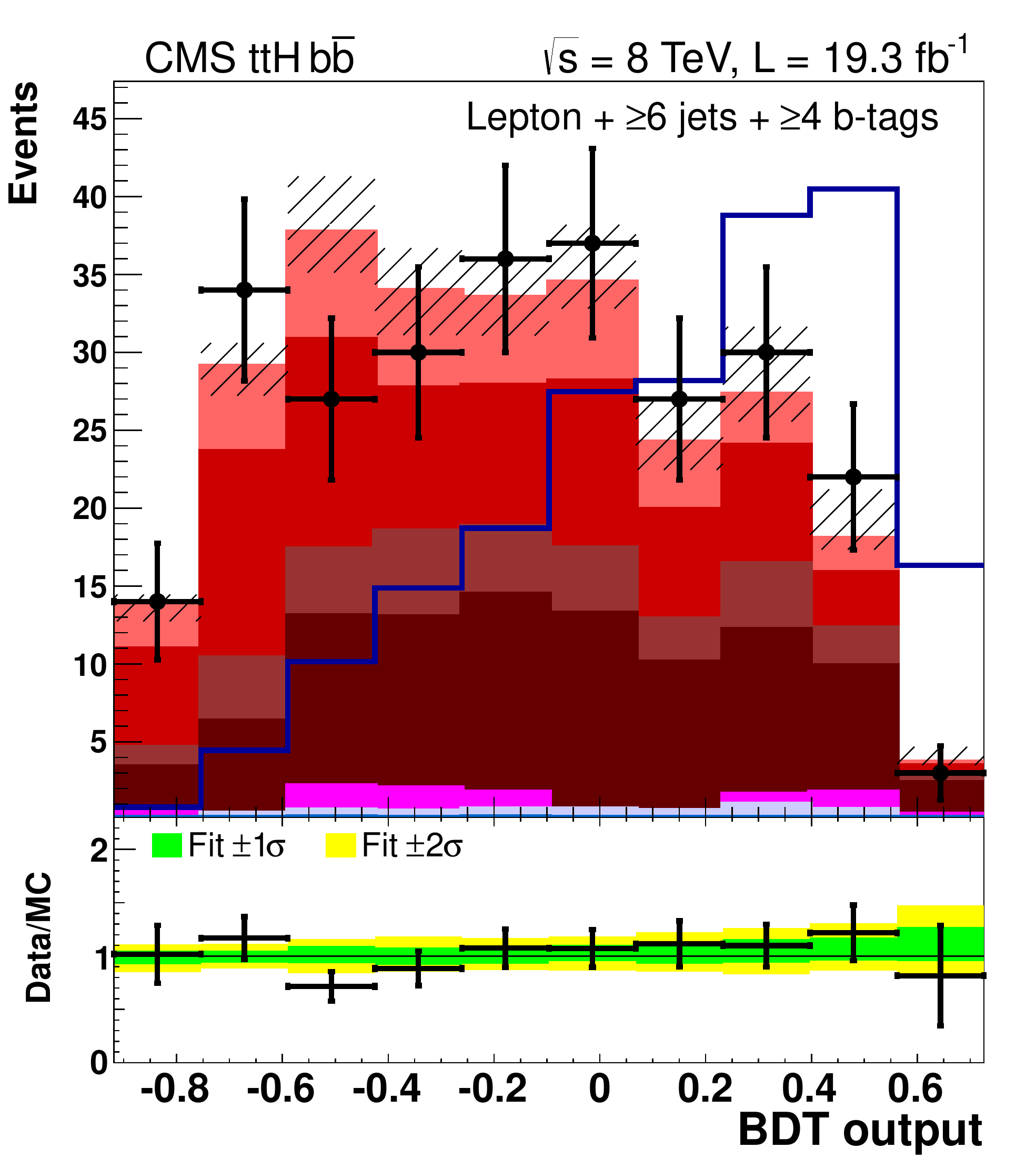}
 } \caption{Final BDT output for lepton+jets events.  The top, middle
 and, bottom rows are events with 4, 5, and $\ge$6 jets, respectively,
 while the left, middle, and right columns are events with 2, 3, and
 $\ge$4 b-tags, respectively.  Details regarding signal and background
 normalizations are described in the text.}
 \label{fig:bb_lj_finalMVA_8TeV}
   \end{center}
\end{figure}

\begin{figure}[!hbtp]
\begin{center}
  \includegraphics[width=0.40\textwidth]{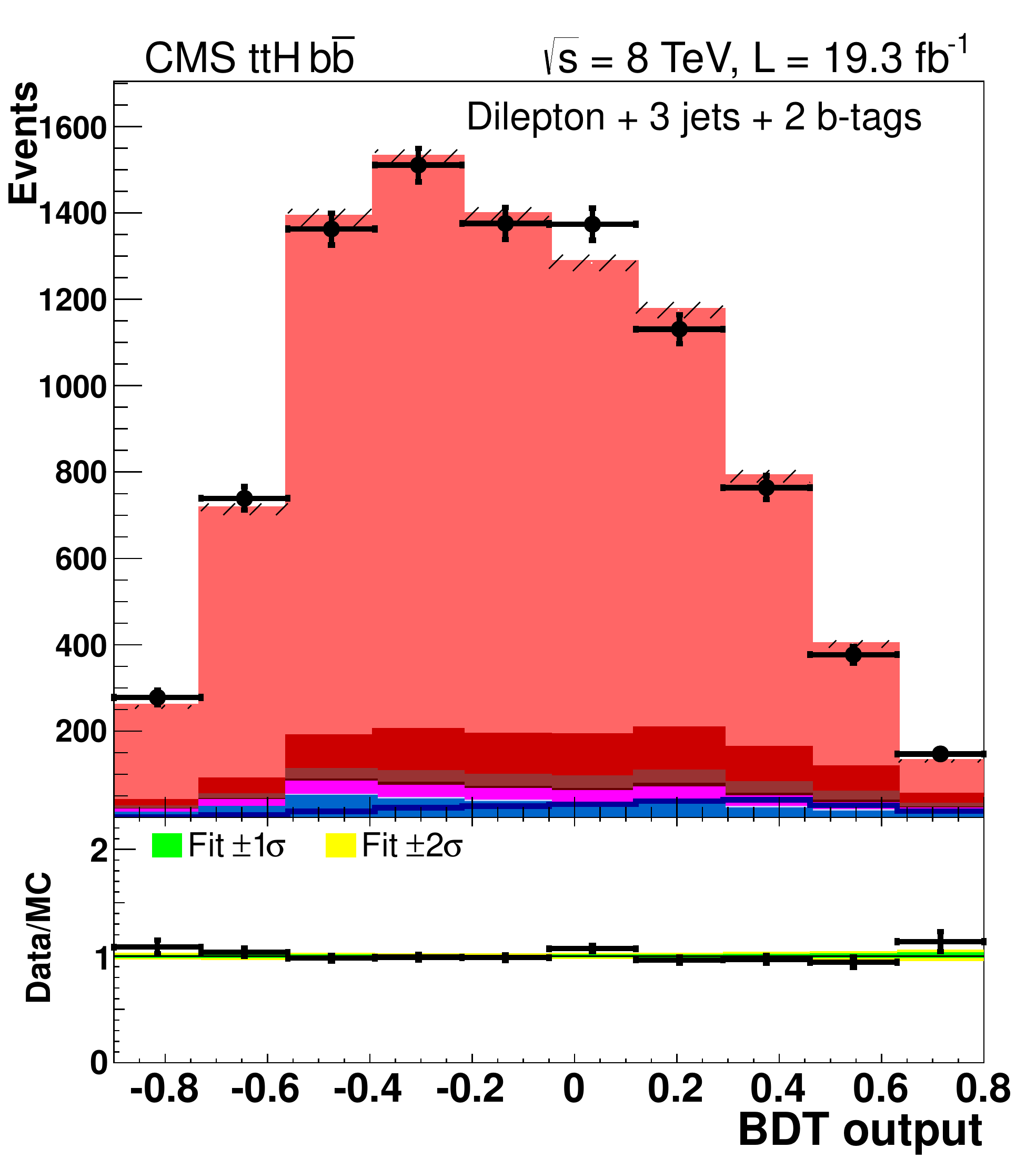}
  \includegraphics[width=0.40\textwidth]{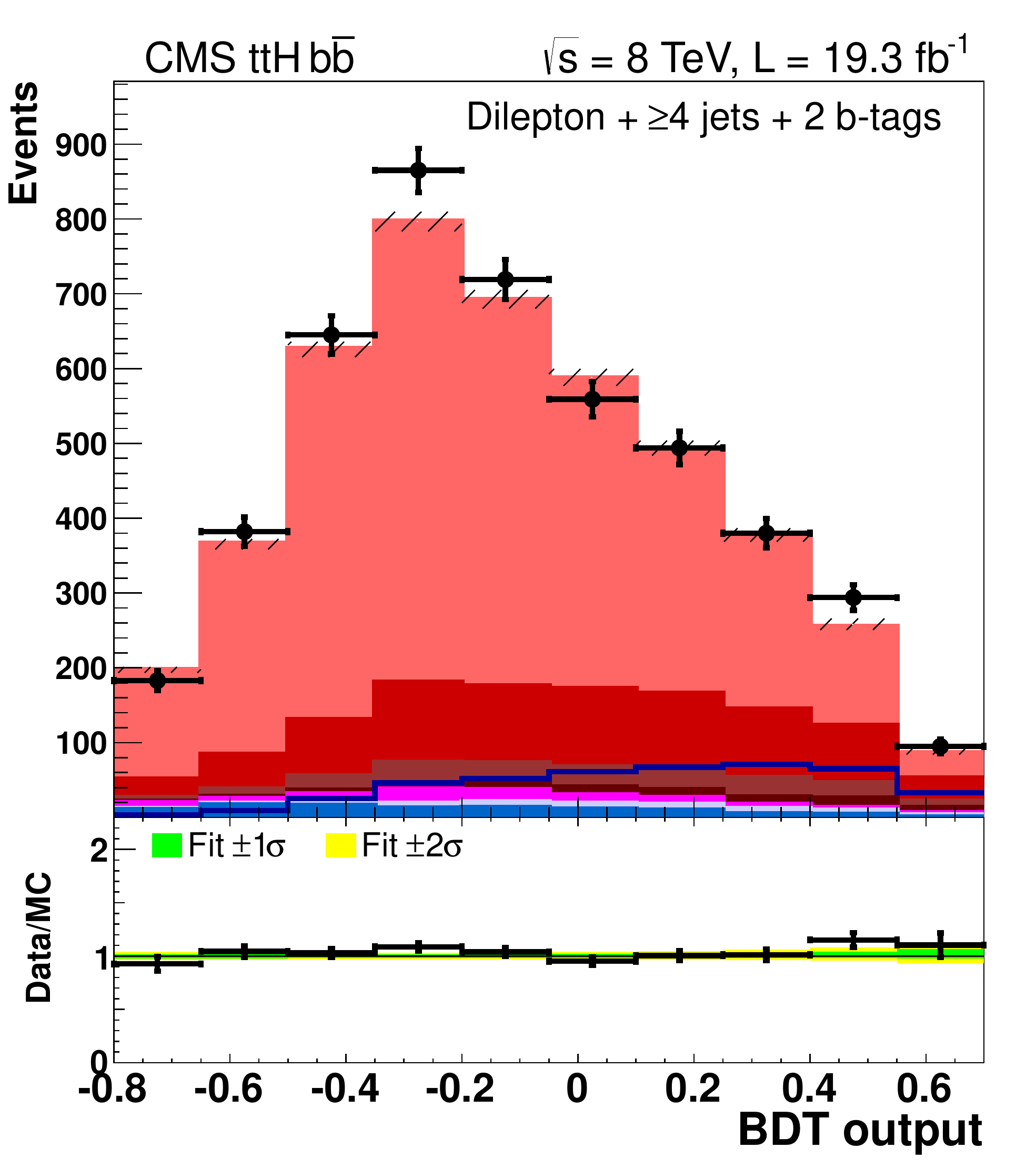}
  \includegraphics[width=0.40\textwidth]{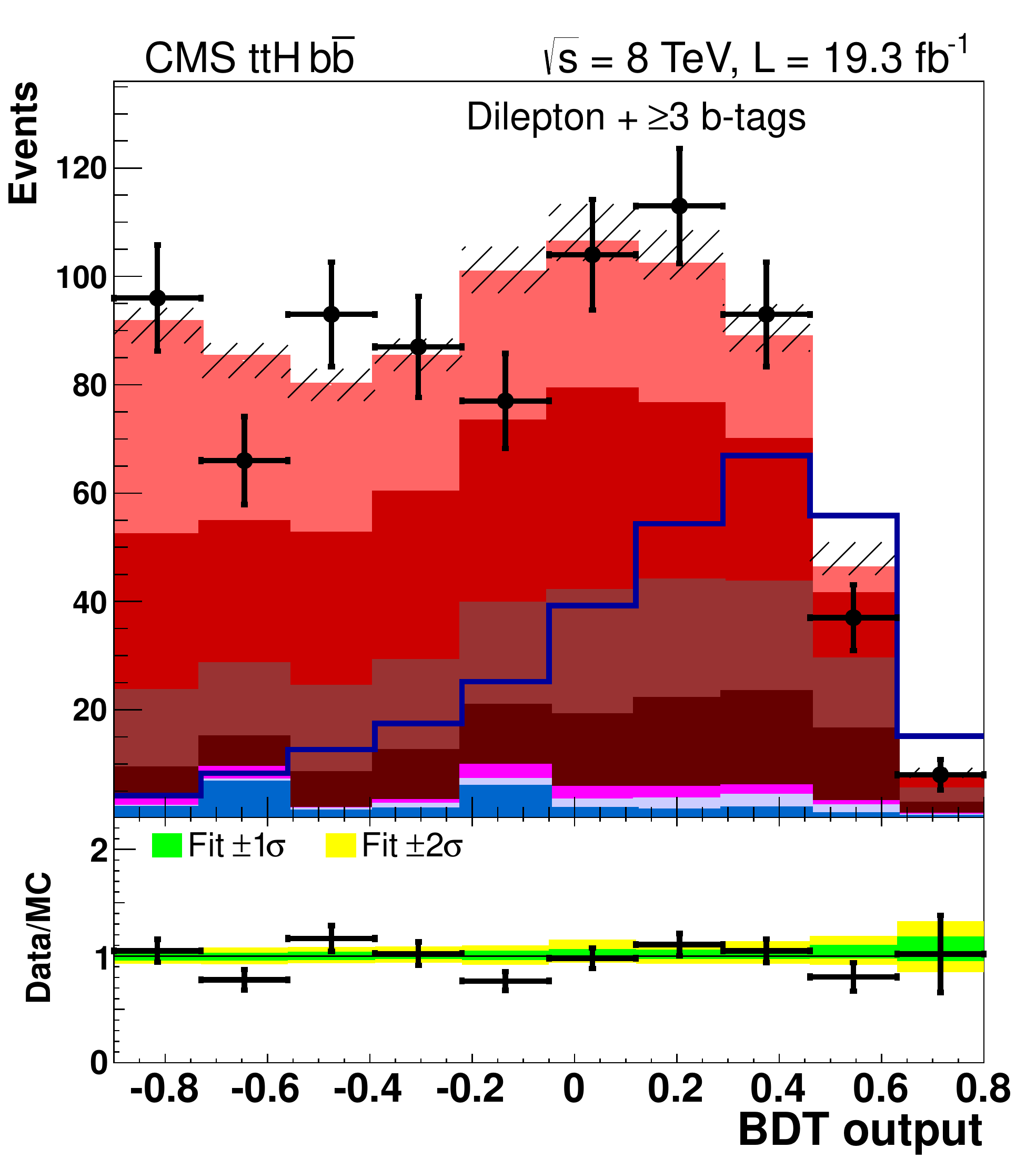}
  \hspace{0.05\textwidth}
  \includegraphics[width=0.33\textwidth]{plots/tth-hadrons_finalMVA/ttH_legend_1columns.pdf}

  \caption{Final BDT output for dilepton events.  The upper left, upper right, and lower left
    plots are events with 3 jets + 2 b-tags, $\ge$4 jets + 2 b-tags,
    and $\ge$3 b-tags, respectively.  Details regarding signal and
    background normalizations are described in the text.}
\label{fig:bb_dil_finalMVA_8TeV} \end{center}
\end{figure}

\begin{figure}[!hbtp]
  \begin{center}
    \hspace{0.32\textwidth}
    \subfigure{\includegraphics[width=0.31\textwidth]{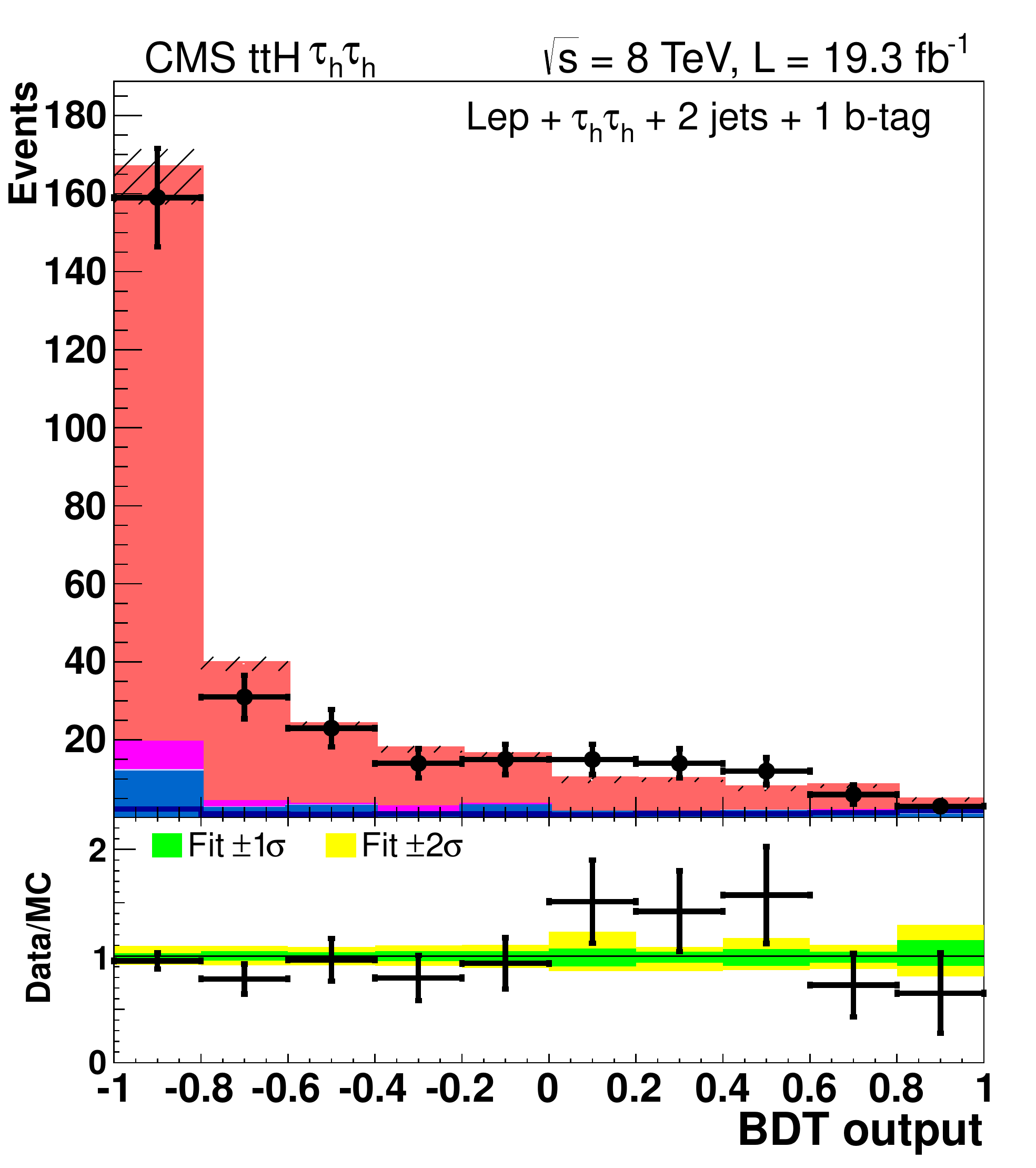}}
    \subfigure{\includegraphics[width=0.31\textwidth]{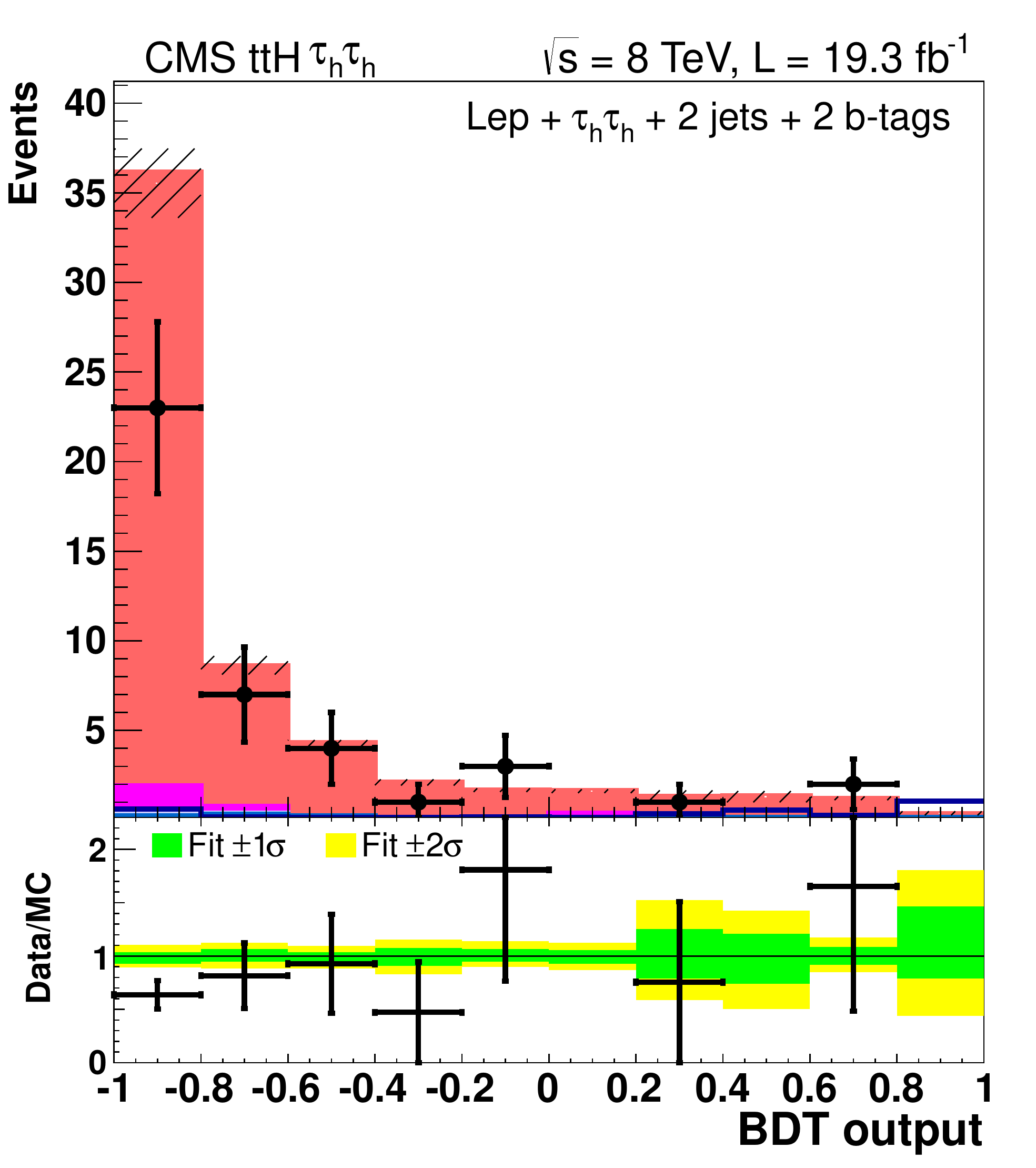}}
    \vspace{0.2cm}

    \hspace{0.32\textwidth}
    \subfigure{\includegraphics[width=0.31\textwidth]{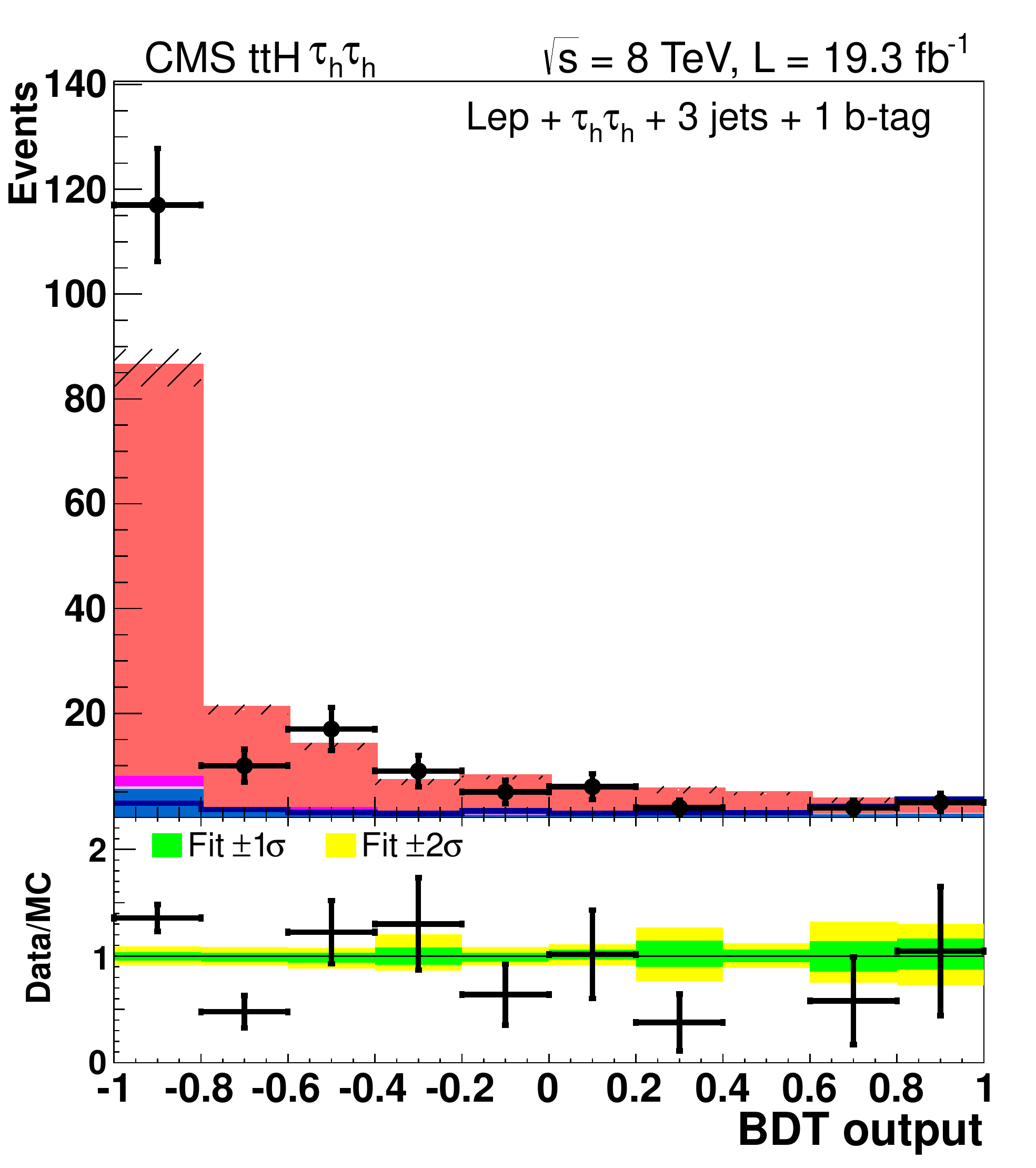}}
    \subfigure{\includegraphics[width=0.31\textwidth]{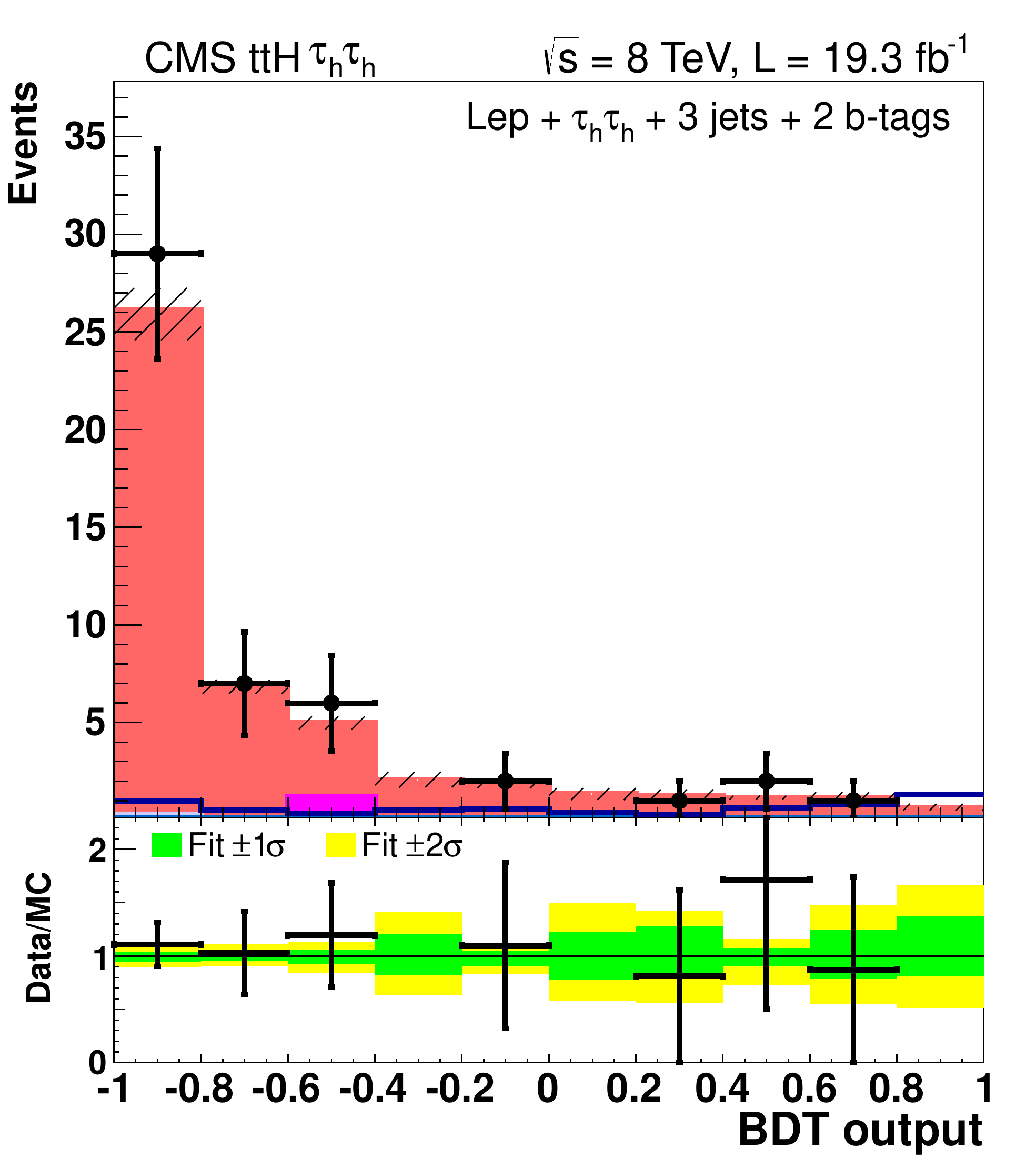}}
    \vspace{0.2cm}

    \subfigure{\includegraphics[width=0.21\textwidth]{plots/tth-hadrons_finalMVA/ttH_legend_1columns_forTaus.pdf}}
    \hspace{0.11\textwidth}
    \subfigure{\includegraphics[width=0.31\textwidth]{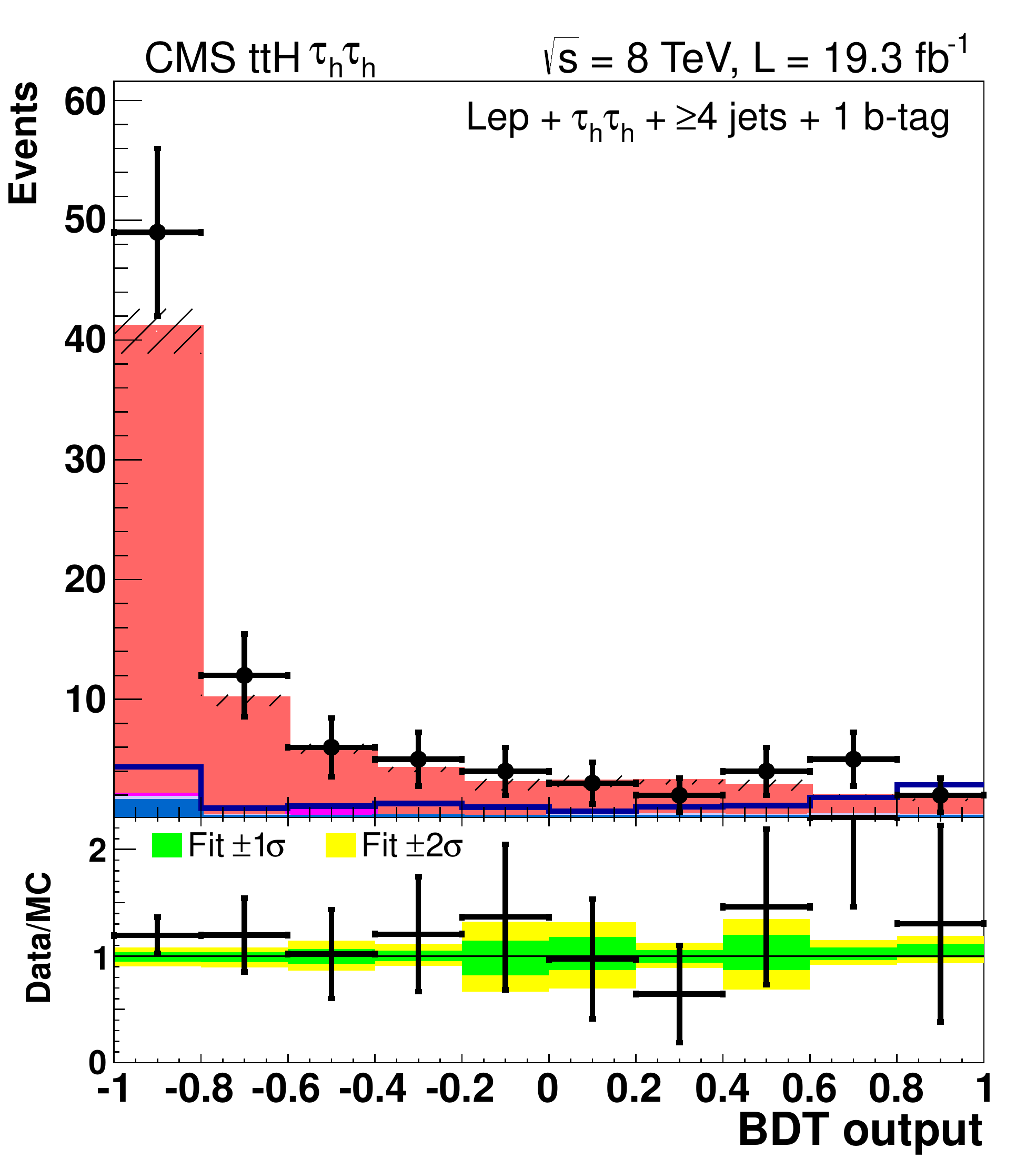}}
    \subfigure{\includegraphics[width=0.31\textwidth]{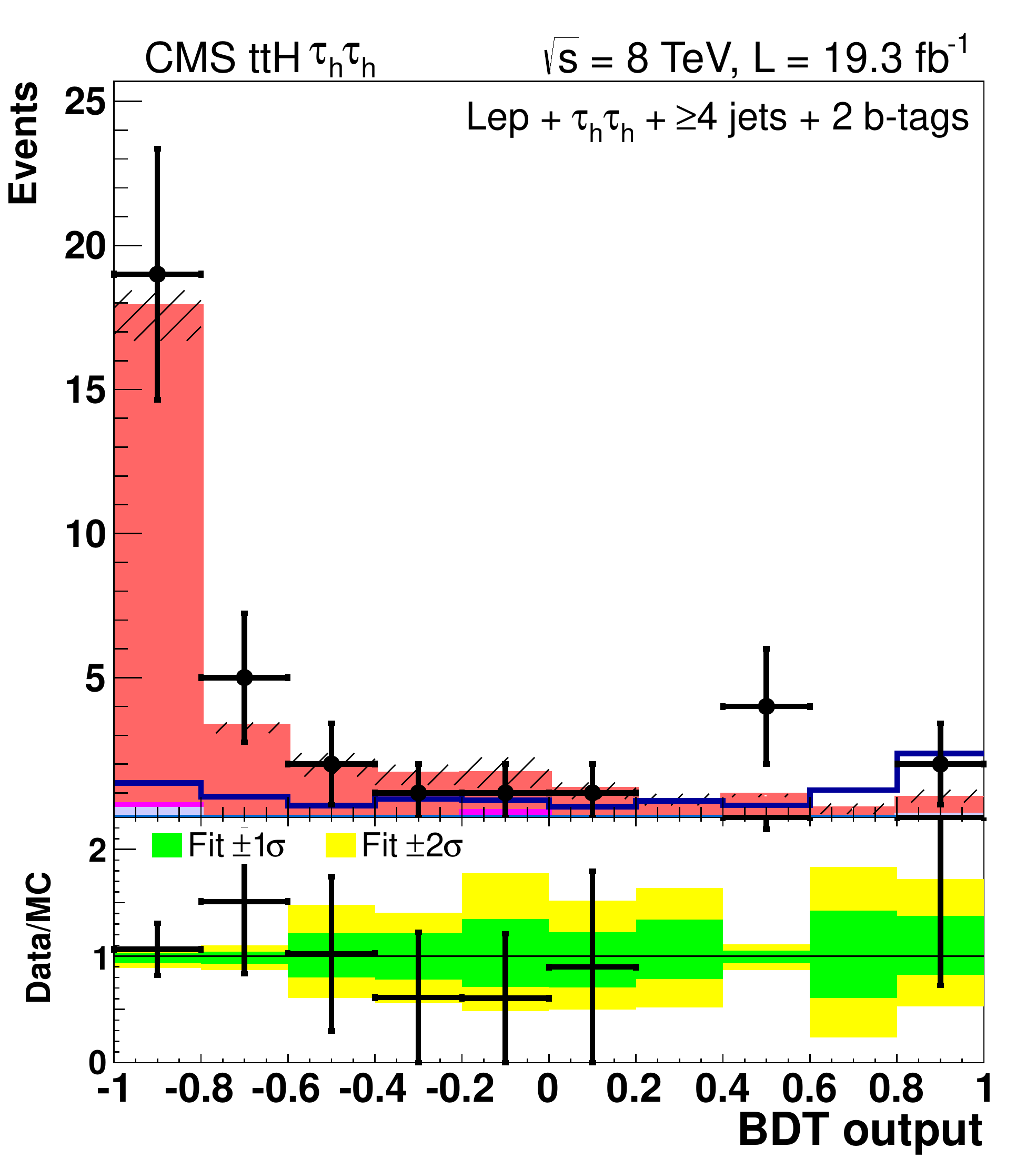}}
    \caption{Final BDT output for events in the $\tauh$
             channel.  The top row is the 2 jet categories, while the
             second and third rows are for the categories with 3 jets
             and $\geq$4 jets, respectively. In each row, the columns
             are for the categories with 1 b-tag (left) and 2 b-tags
             (right).  Details regarding signal and background
             normalizations are described in the text.}
    \label{fig:bb_tau_finalMVA_8TeV} \end{center} \end{figure}

Figures~\ref{fig:bb_lj_finalMVA_8TeV},~\ref{fig:bb_dil_finalMVA_8TeV},
and~\ref{fig:bb_tau_finalMVA_8TeV} show the final BDT output
distributions for the lepton+jets, dilepton, and $\tauh$ channels,
respectively.  Background-like events have a low BDT output value,
while signal-like events have a high BDT output value.  The background
distributions use the best-fit values of all nuisance parameters, with
$\mu$ fixed at 1, and the uncertainty bands are constructed using
the post-fit nuisance parameter uncertainties.  The fit is described
in section~\ref{sec:results}.  The $\ttbar\PH$ signal ($m_{\PH} =
125.6\GeV$) is not included in the stacked histogram, but is shown as
a separate open histogram normalized to 30 times the SM expectation
($\mu = 30$).  For the ratio plots shown below each BDT distribution,
only the background expectation (and not the signal) is included in
the denominator of the ratio.  The final BDT outputs provide better
discrimination between signal and background than any of the input
variables individually. The BDT output distributions are used to set
limits on the Higgs boson production cross section, as described in
section~\ref{sec:results}.

\section{\texorpdfstring{$\PH \to$~photons}{H to photons}}
\label{sec:tth-gamma}

The diphoton analysis selects events using the diphoton system to
identify the presence of a Higgs boson, and a loose selection on the
remaining objects to accept all possible $\ttbar$ decays, while
rejecting other Higgs boson production modes that are not directly
sensitive to the top-quark Yukawa coupling.  The background is
extracted directly from the diphoton invariant mass distribution
$m_{\Pgg\Pgg}$, exploiting the fact that a signal around 125\GeV
will be characterized by a narrow peak.

The event selection starts from the requirement of two photons, where
the leading photon is required to have a $\pt > m_{\Pgg\Pgg}/2$ and
the second photon to have a $\pt > 25\GeV$.  The variable threshold on
the leading photon $\pt$ increases the efficiency while minimizing
trigger turn-on effects.  The photon identification and energy
measurement is the same as that used in Ref.~\cite{hgg_inclusive} with
the only exception being that the primary vertex selection is done as
described in section~\ref{sec:objects} of this paper.  The presence of
at least one $\cPqb$-tagged jet according to the medium working point
of the CSV algorithm is required, consistent with the presence of
$\cPqb$ jets from top quark decays in the final state. Muons must lie
in the pseudorapidity range $\abs{\eta} < 2.4$, and electrons within
$\abs{\eta} < 2.5$. Both muons and electrons are required to have $\pt$
greater than $20\GeV$.

Events are categorized in two subsamples: the leptonic and hadronic
channels.  The hadronic channel requires, in addition to the two
photons in the event, at least four jets of which at least one is
b-tagged and no identified high-$\pt$ charged leptons, whereas the
leptonic channel requires at least two jets of which at least one is
b-tagged and at least one charged lepton, where $\ell= \Pe,\Pgm$, with
$\pt > 20\GeV$. The 7\TeV dataset is too small to perform an
optimization on each signal decay mode; thus events passing the
hadronic and leptonic selections are combined in a single category.

Unlike the $\PH \to$ hadrons and $\PH \to$ leptons channels, the
contribution from Higgs boson production modes other than $\ttbar \PH$
must be treated with care for this channel. This is because this
analysis is designed to have very loose requirements on the jet and
lepton activity, and the other Higgs boson production modes will peak
at the same location in the diphoton invariant mass distribution as
the $\ttbar \PH$ signal. This is in contrast with the situation for
the $\PH \to$ hadrons and $\PH \to$ leptons analyses, where the
non-$\ttbar \PH$ production modes tend to populate the most
background-rich region of the phase space investigated, thus a very
small contamination of non-$\ttbar \PH$ Higgs boson production has
almost no impact on those analyses.  The event selection for the
$\ttbar \PH$, $\PH \to$ photons channel is thus designed to minimize
the contribution from other Higgs boson production modes.  The expected
signal yields for the various production processes for the SM Higgs
boson of mass 125.6\GeV in this channel are shown in
table~\ref{tab:signal_yield}, after selection in the $100 \le m_{\Pgg
  \Pgg} \le 180\GeV$ range. As can be seen the contribution of
production modes other than $\ttbar \PH$ is minor.  The contribution
of single-top-quark-plus-Higgs-boson production has not been
explicitly estimated but its cross section is expected to be only
about 1/10 of the $\ttbar \PH$ cross section and the events have
different kinematics\,\cite{Biswas:2013xva}, so its contribution to
the sample is expected to be small.

The main backgrounds are the production of top quarks and either
genuine or misidentified photons in the final state, and the
production of high-$\pt$ photons in association with many jets,
including heavy-flavor jets. Because the background will be estimated
by fitting the data which is a mixture of these processes, it is
useful to test the background modeling in an independent control
sample defined using collision data.  The control sample is
constructed using events that have been recorded with the
single-photon trigger paths, and inverting the photon identification
requirements on one of the two photons used to reconstruct the Higgs
boson signal. To take into account the fact that the efficiency of the
photon isolation requirement is not constant as a function of the
photon $\pt$ and $\eta$, a two-dimensional reweighting procedure is
applied to the leading and subleading photon candidates in such
events. The reweighting is performed so as to match the photon $\pt$
and $\eta$ spectra to the ones of photons populating the signal
region. A control sample with similar kinematic properties as the
data, yet statistically independent, is thus obtained.

\begin{figure}[!hbtp]
  \begin{center}
    \includegraphics[width=0.45\textwidth]{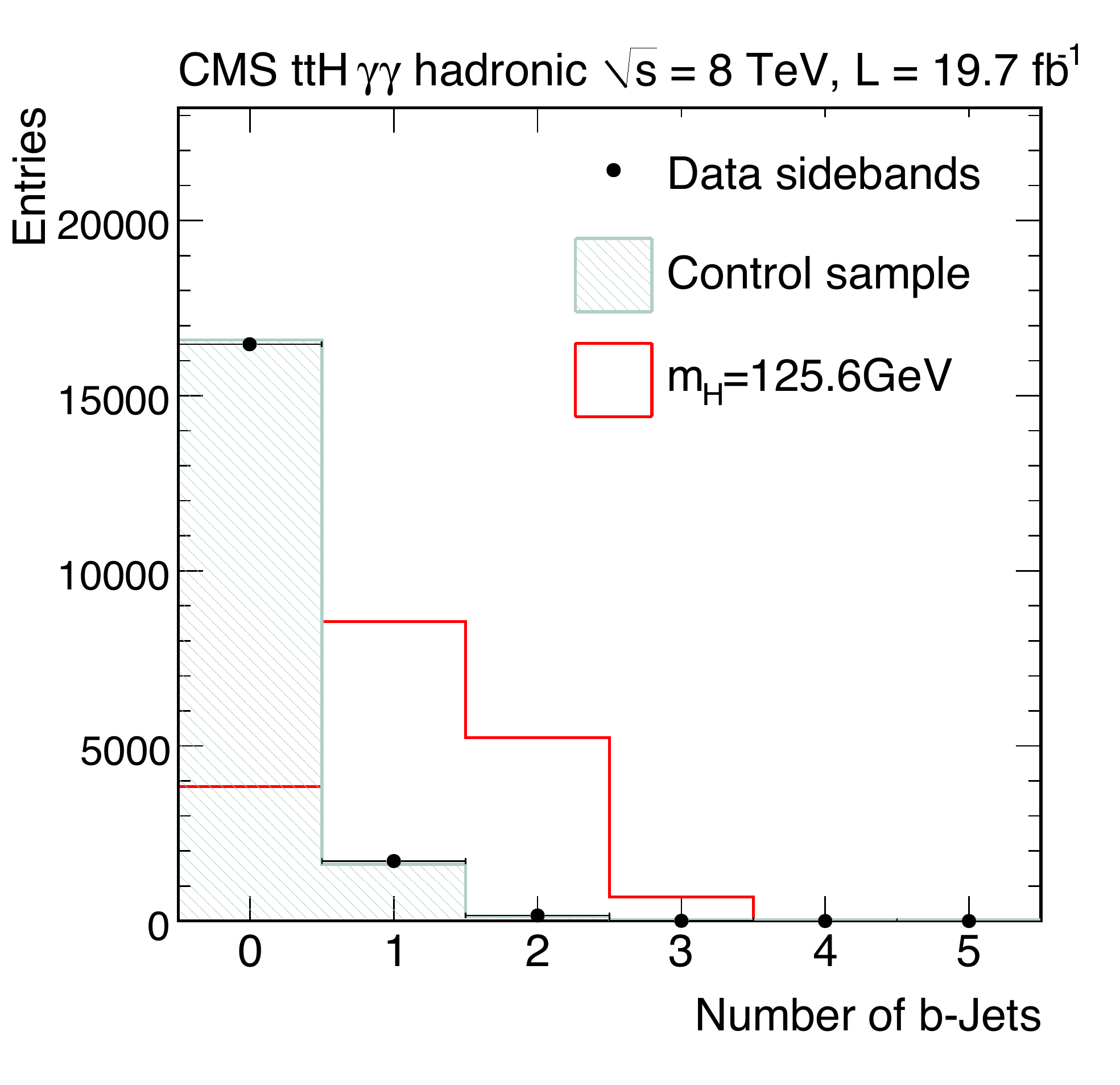}
    \includegraphics[width=0.45\textwidth]{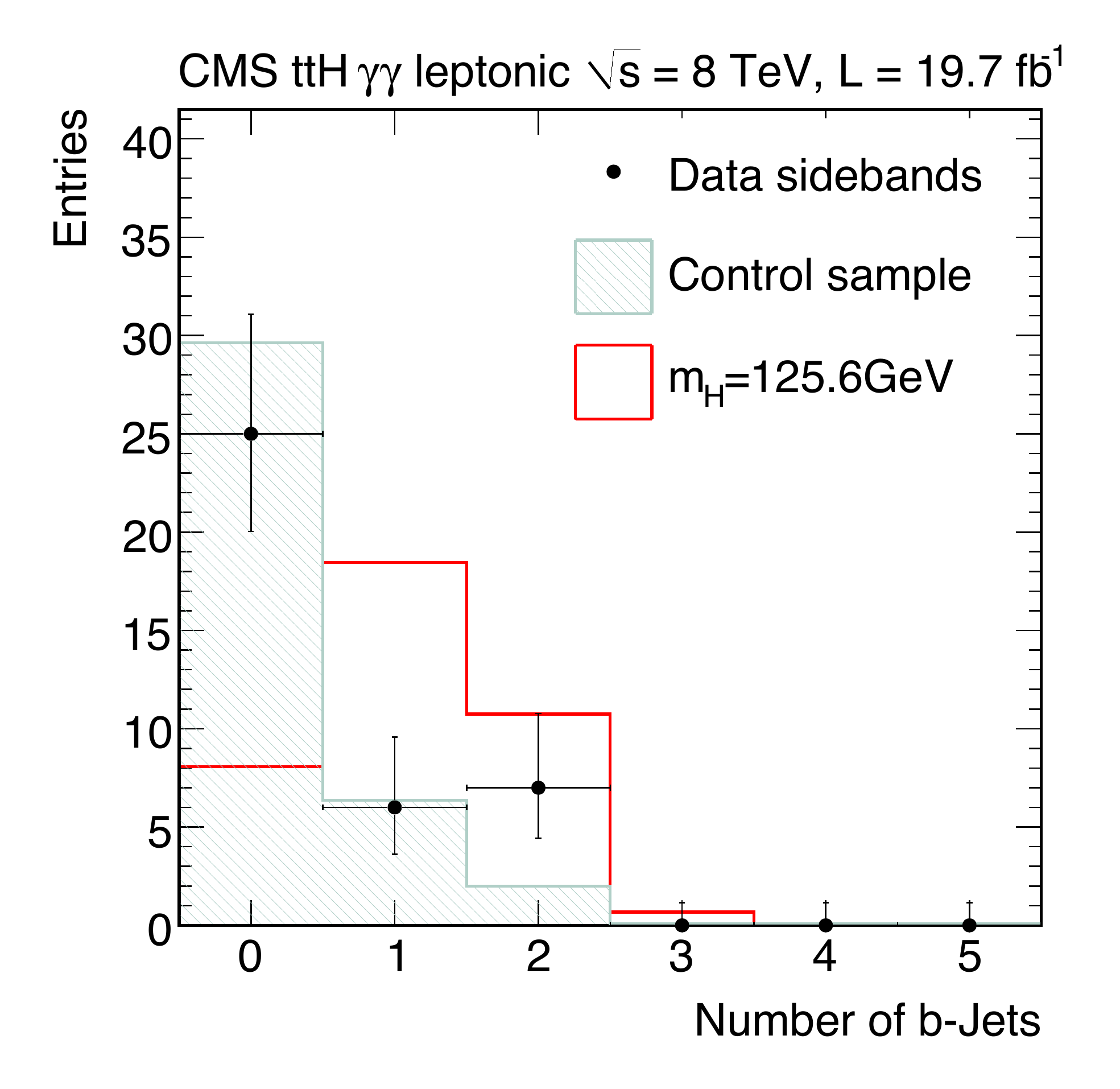}
    \includegraphics[width=0.45\textwidth]{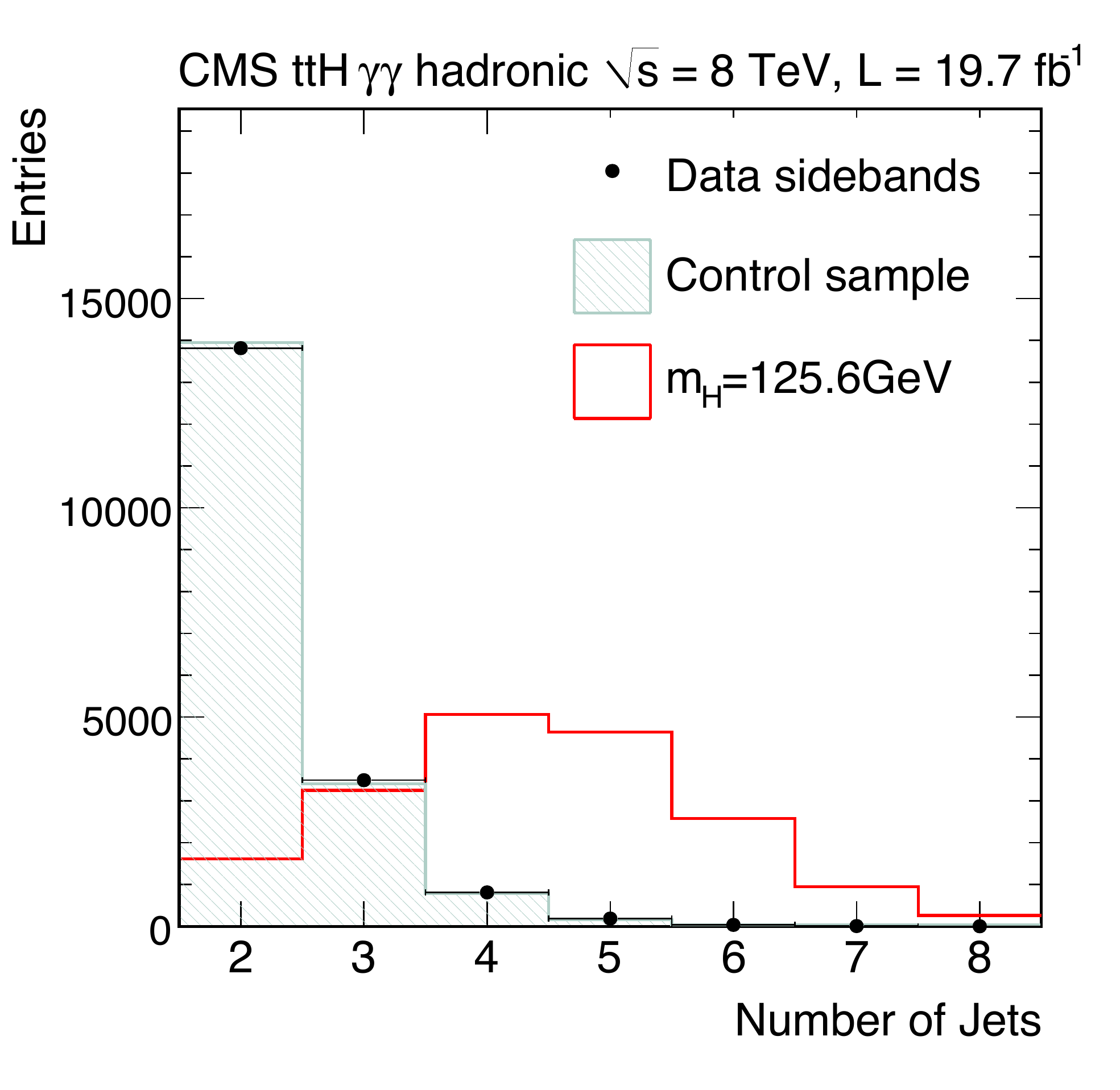}
    \includegraphics[width=0.45\textwidth]{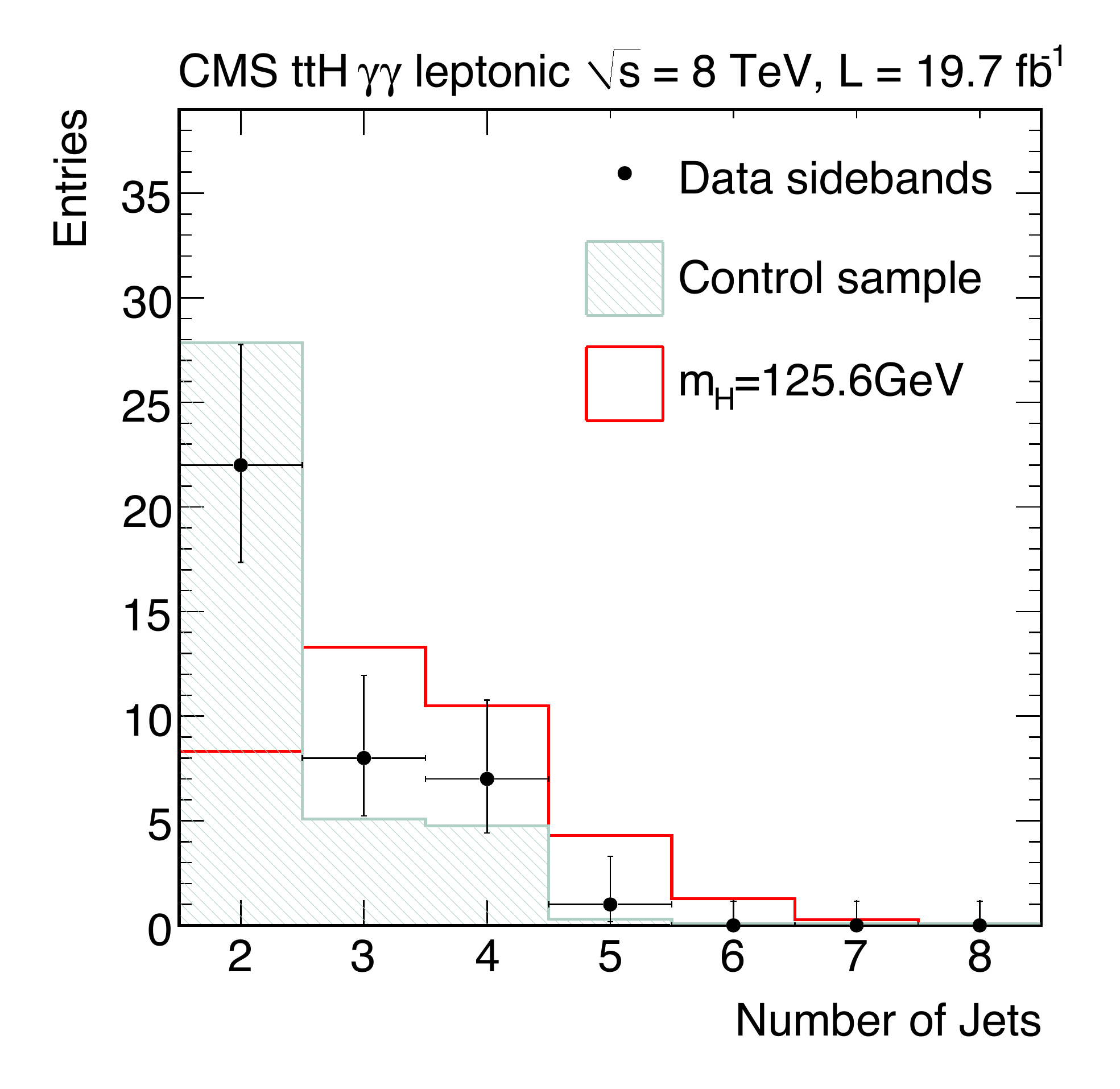}
    \caption{Distributions of the b-tagged jet multiplicity (top row)
      and jet multiplicity (bottom row) for events passing a relaxed
      selection in the hadronic (left) and leptonic (right) channels,
      but removing events where the diphoton invariant mass is
      consistent with the Higgs boson mass within a 10\GeV window.
      The relaxed selection applies the standard photon and lepton
      requirements but allows events with any number of jets.  The
      plots compare the data events with two photons and at least two
      jets (black markers) and the data from the control sample (green
      filled histogram) to simulated $\ttbar \PH$ events (red open
      histogram).  Both signal and background histograms are
      normalized to the total number of data events observed in this
      region to allow for a shape comparison.}
    \label{fig:CR}
  \end{center}
\end{figure}

The extent to which the control sample is well-modeled is tested
using events passing the photon selections, and the requirement of at least
two high-$\pt$ jets.  The sample is further split into events with and
without charged leptons, to test the kinematic properties of the
model against data. A few key kinematic distributions are shown in
figure~\ref{fig:CR}, where the black markers show the signal sample, the
green histogram is the control sample data, and the red line displays the
signal kinematics. All distributions are normalized to the number of
events observed in data.

Even after the dedicated event selection, the dataset is still largely
dominated by backgrounds. The strategy adopted in this analysis is to
fit for the amount of signal in the diphoton mass spectrum, as this
provides a powerful discriminating variable due to the excellent
photon energy resolution, in the region surrounding the Higgs boson
mass.  The background is obtained by fitting this distribution in each
channel (hadronic or leptonic) over the range $100\GeV < m_{\gamma
  \gamma} < 180\GeV$.  The actual functional form used to fit the
background, in any particular channel, is included as a discrete
nuisance parameter in the likelihood functions used to extract the
results; exponentials, power-law functions, polynomials (in the
Bernstein basis), and Laurent series are considered for this analysis.
When fitting the background by minimizing the value of twice the
negative logarithm of the likelihood (2NLL), all functions in these
families are tried, with a penalty term added to 2NLL to account for
the number of free parameters in the fitted function.
Pseudoexperiments have shown that this ``envelope'' method provides
good coverage of the uncertainty associated with the choice of the
function, for all the functions considered for the background, and
provides an estimate of the signal strength with negligible
bias~\cite{hgg_inclusive}.

\begin{table}[!htbp]
    \begin{center}
      \topcaption{Expected signal yields after event selections in
        the $100\GeV <m_{\Pgg \Pgg} < 180\GeV$ diphoton mass window.
        Different Higgs boson production processes are shown
        separately. The total number of data events present in each
        channel is displayed at the bottom of the table. A Higgs boson
        mass of 125.6\GeV is assumed.}
      \begin{tabular}{|l|c|c|c|}
      \hline
         & 7\TeV & \multicolumn{2}{c|}{8\TeV} \\
                     & All decays & Hadronic channel & Leptonic channel \\ \hline\hline
        $\ttbar \PH$ & 0.21 & 0.51  & 0.45  \\ \hline
        $\Pg\Pg \to \PH$ & 0.01 & 0.02 & 0 \\
        VBF $\PH$ & 0 & 0 & 0  \\
        $\PW\PH/\cPZ\PH$ & 0.01 & 0.01 & 0.01  \\
        \hline
        Total $\PH$ & 0.23 & 0.54 & 0.46 \\
       \hline
       Data & 9 & 32 & 11 \\
       \hline
      \end{tabular}
    \label{tab:signal_yield}
    \end{center}
  \end{table}

\begin{figure}[!hbtp]
  \begin{center}
    \includegraphics[width=0.31\textwidth]{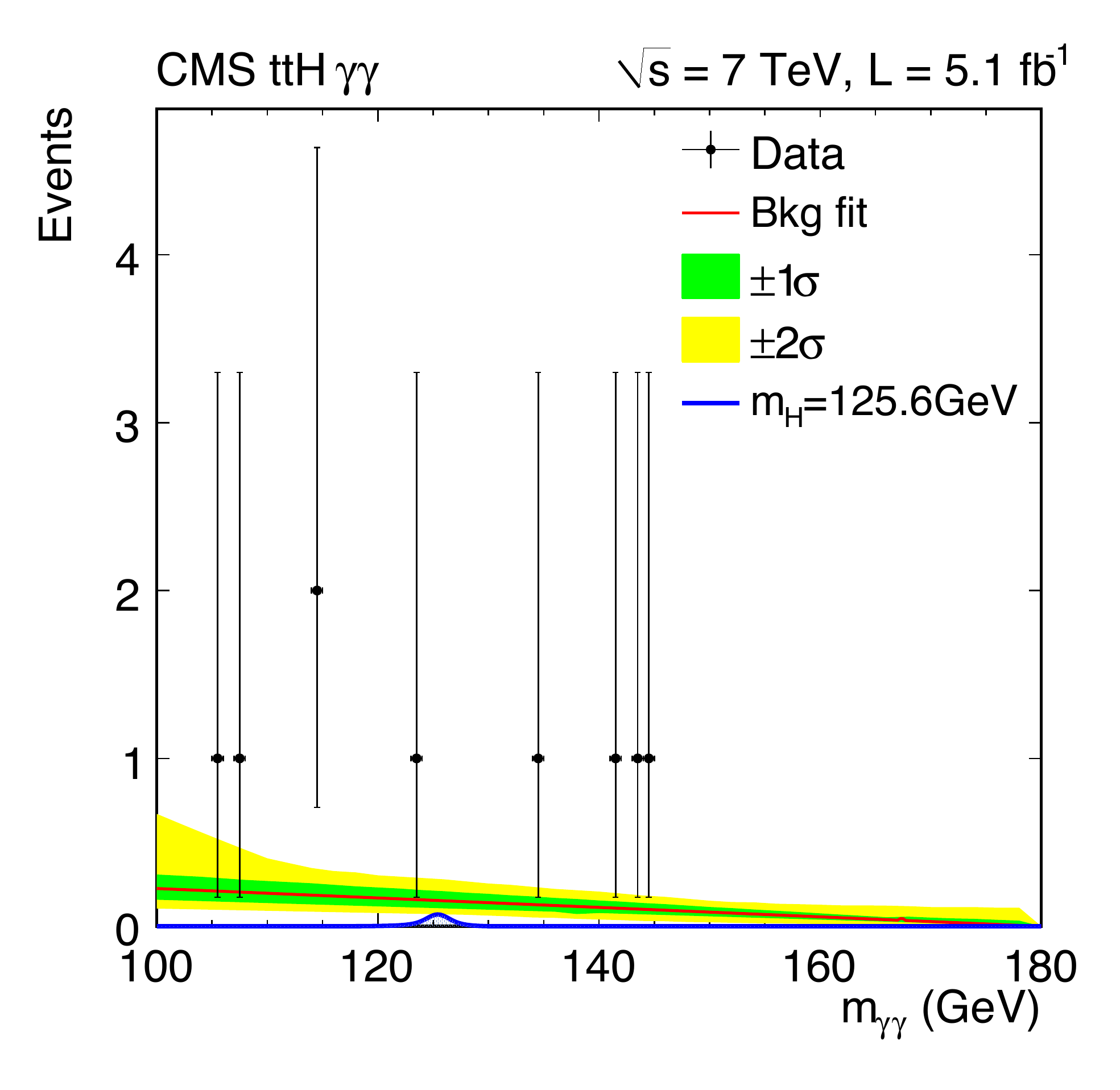}
    \includegraphics[width=0.31\textwidth]{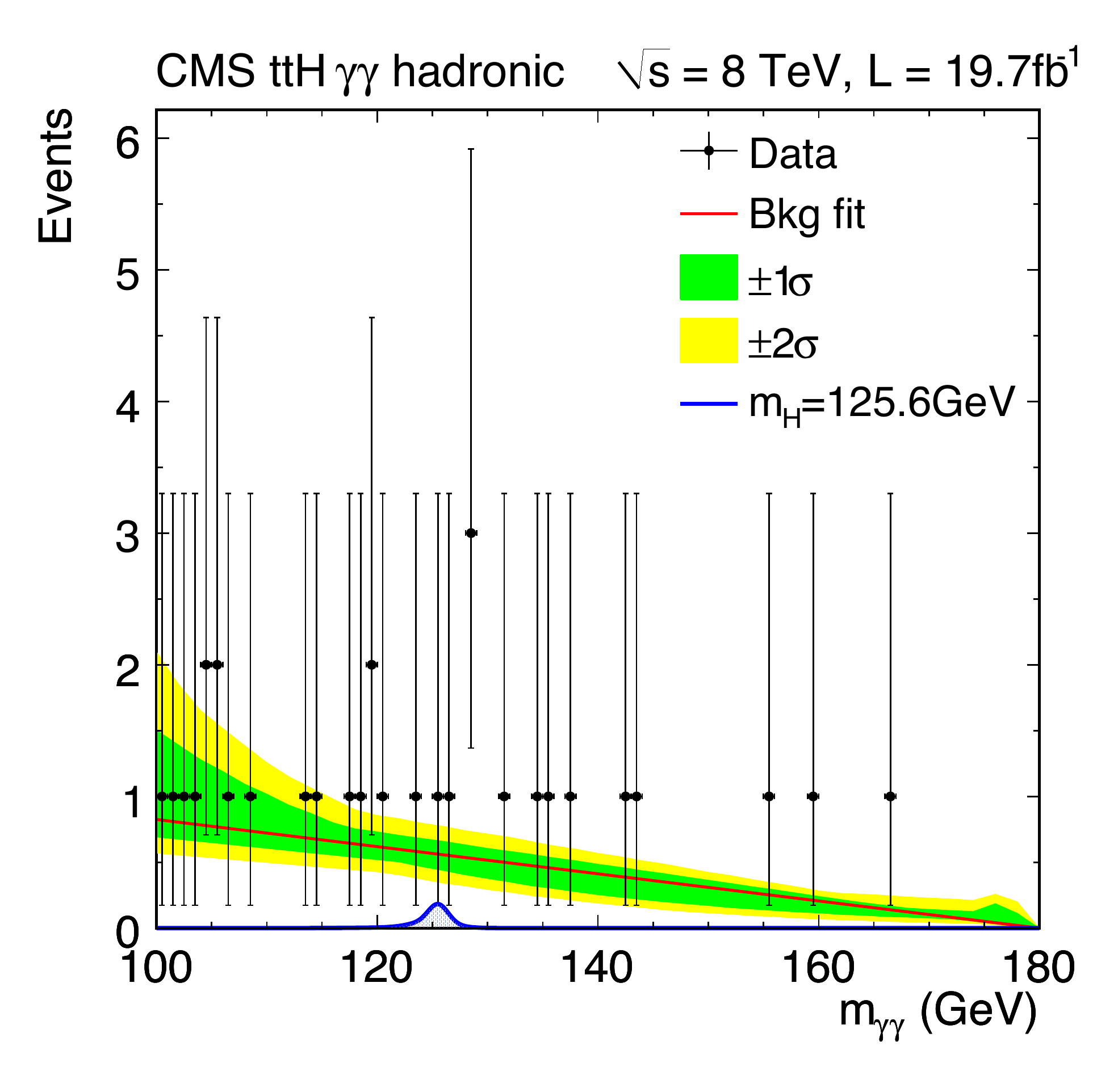}
    \includegraphics[width=0.31\textwidth]{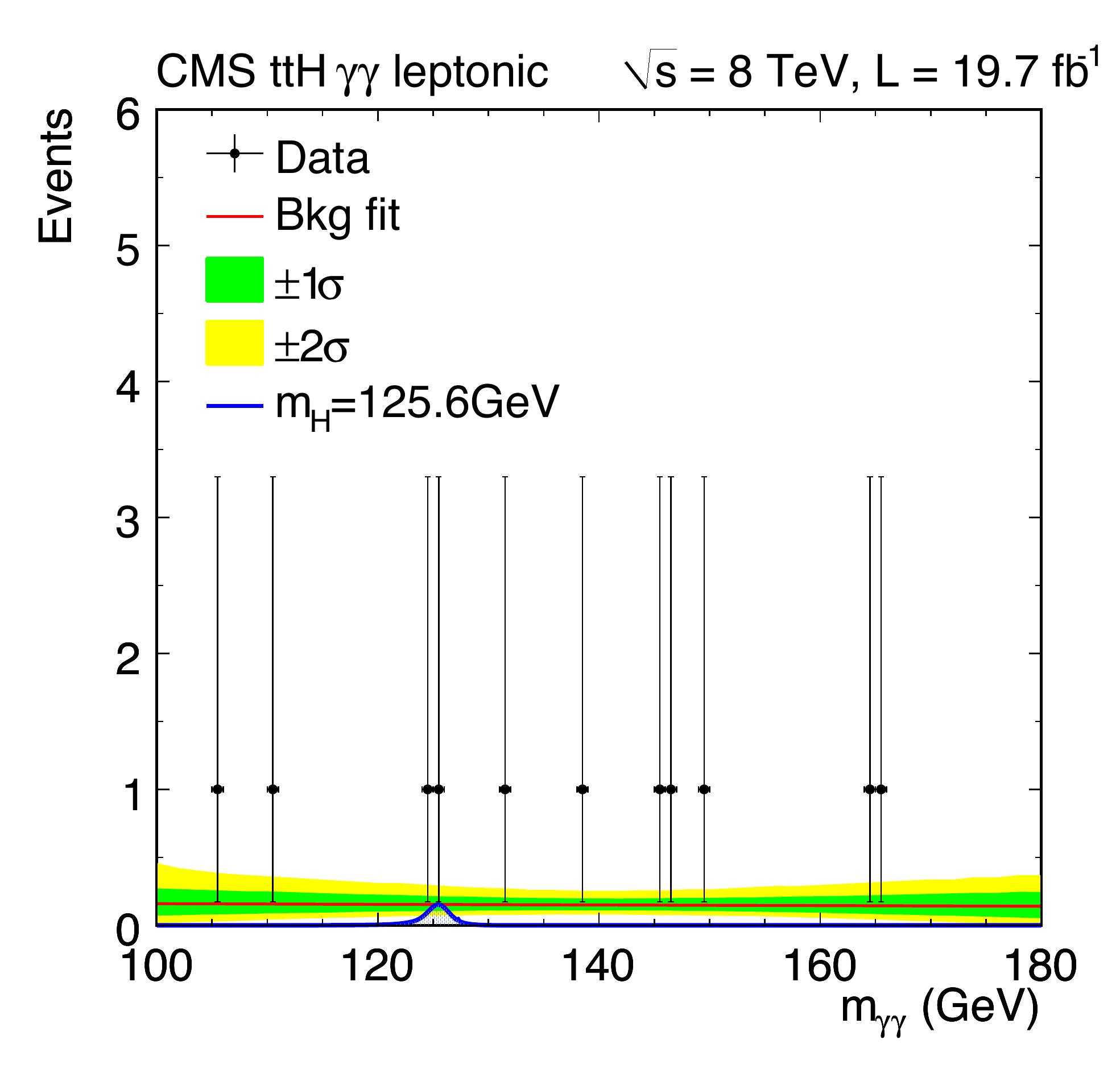}
    \caption{Diphoton invariant mass distribution for $\sqrt{s} = 7
     \TeV$ data events for the combined hadronic and leptonic
      selections on the left, and for $\sqrt{s} = 8\TeV$ data events
      passing the hadronic (middle), and leptonic (right) selections.
      The red line represents the fit to the data, while the green
      (yellow) band show the $1 \sigma$ ($2 \sigma$) uncertainty band.
      The theoretical prediction for the signal contribution (in blue)
      includes the main Higgs boson production modes. }
    \label{fig:mgg}
  \end{center}
\end{figure}

The diphoton invariant mass spectra for data, the expected signal
contribution, and the background estimate from data are shown in
figure~\ref{fig:mgg} for the combination of hadronic and leptonic
selections on the $\sqrt{s} = 7\TeV$ data (left), the hadronic
(middle) and leptonic (right) channels separately using $\sqrt{s} =
8\TeV$ data. The expected signal contribution of the dominant SM Higgs
boson production modes is shown as a blue histogram.  The result of
the fit is shown in the plots as a red line, together with the
uncertainty bands corresponding to $1 \sigma$ (green) and $2 \sigma$
(yellow) coverage.  The observed diphoton mass spectra agree well with
the background estimates.

\section{\texorpdfstring{$\PH \to$~leptons}{H to leptons}}
\label{sec:tth-leptons}
\subsection{Object identification}

In this channel the signal has multiple prompt leptons from W, Z, or
$\tau$ decays.  The largest backgrounds have at least one non-prompt
lepton, usually from the decay of a b hadron (in $\ttbar$+jets,
$\cPZ$+jets, and $\PW$+jets events). The analysis begins with a
preselection of electron and muon objects using loose criteria with
very high efficiency for prompt leptons and moderate non-prompt lepton
rejection.  In addition to the basic cuts from section~\ref{sec:objects},
the lepton is required to be associated with the event vertex.  The
distance between the lepton track and the event vertex along the
$z$-axis and perpendicular to it ($d_z$ and $d_{xy}$) must be less
than $1\cm$ and $0.5\cm$, respectively.  The $S_{\mathrm{IP}}$ (defined
as the ratio of the IP to its uncertainty) is required to be less
than 10, a fairly loose cut intended to retain efficiency for
leptons coming from $\Pgt$ decays.  Next, a multivariate discriminator based on
BDT techniques is used to distinguish prompt from non-prompt leptons.
This discriminator, referred to as the lepton MVA, is trained with simulated
prompt leptons from the $\ttbar \PH$ MC sample and non-prompt leptons from
the $\ttbar$+jets MC sample, separately for electrons and muons and
for several bins in $\pt$ and $\eta$.

The lepton MVA input variables relate to the lepton IP, isolation,
and the properties of the nearest jet,
within $\Delta R < 0.5$.  A tight working point on the lepton
MVA output is used for the search in the
dilepton and trilepton final states, and a loose working point is used
for the four-lepton final state.  For the tight working point, the
efficiency to select prompt electrons is of order 35\% for
$\PT^{\Pe} \sim 10\GeV$ and reaches a plateau of 85\%
at $\PT^{\Pe} \sim 45\GeV$; for prompt muons it is of order
55\% for $\PT^{\Pgm} \sim 10\GeV$, and reaches a plateau of
about 97\% at $\PT^{\Pgm} \sim 45\GeV$.  The efficiency to select electrons (muons)
from the decay of b hadrons is between 5--10\% (around 5\%).

To suppress electrons from photon conversions, tight
electrons with missing tracker hits before the
first reconstructed hit, or associated with a successfully reconstructed
conversion vertex, are rejected~\cite{Khachatryan:2010pw}.

Additional cuts are used to suppress incorrect charge reconstruction in the
dilepton final states.  For electrons, the tracker and ECAL charge
measurements must agree, where the ECAL charge is measured by
comparing the position of the energy deposits in the ECAL to a
straight-line trajectory formed from the electron hits in the pixel
detector~\cite{Chatrchyan:2011wba,CMS_DPS_2011-003}. For muons, the
relative uncertainty in the track $\PT$ must be less than 20\%.

The agreement between data and simulation for the input variables and
the final lepton MVA is validated in dedicated control regions.  For
prompt leptons, high-purity control samples are selected with same-flavor,
opposite-sign pairs of leptons with an invariant mass close to that of
the $\Z$ boson and little $\ETmiss$. In these events, tight isolation and $\pt$ selection are
applied to the leading lepton, and the trailing lepton is used to
check the agreement between simulation and data.  High-purity $\tau$
leptons are selected by requiring opposite-flavor, opposite-sign pairs
of electrons and muons with an invariant mass between $20\GeV$ and
$80\GeV$. In these events, tight isolation, $\pt$, and
$S_{\mathrm{IP}}$ cuts are applied to one of the two leptons, and
the other lepton is used to compare simulation and data.  For
non-prompt leptons, samples enriched in leptons from the decay of
b hadrons are selected with three-lepton $\Z\to\ell\ell + \ell$ and
$\ttbar\to\ell\ell + \ell$ control regions.  The agreement is good;
small corrections to better match the data distributions of the input
variables are applied to the simulation before training the MVA
discriminant.  Efficiency scale factors for the tight and loose
lepton MVA working points are computed for prompt leptons with a
tag-and-probe technique in the $\Z\to\ell\ell$ control region.  Backgrounds with
non-prompt leptons are estimated directly from data, as described
in section \ref{sec:modeling}.

\subsection{Event selection}
\label{sec:EvtSelection}

The multilepton selection is optimized to accept $\ttbar \PH$ events
where the Higgs boson decays into $\PW\PW$, $\Z\Z$, or
$\Pgt\Pgt$, and at least one $\PW$ boson, $\Z$ boson, or $\Pgt$ decays
leptonically.  With at least one additional lepton from the top
decays, the events have one of the following three signatures:
\begin{itemize}
\item two same-sign leptons (electrons or muons) plus two b-quark jets;
\item three leptons plus two b-quark jets;
\item four leptons plus two b-quark jets.
\end{itemize}
The first three rows in table~\ref{tab:yields-sel} show the expected distribution of the
$\ttbar \PH$ signal among these different signatures.  The other rows in the table will be discussed below.

\begin{table}[!thbp]
  \centering
  \topcaption{Expected and observed yields after the selection in all five final states. For the expected yields, the total systematic uncertainty is also indicated.  The rare SM backgrounds include triboson production, $\mathrm{tbZ}$, $\PW^\pm\PW^\pm\mathrm{qq}$, and $\PW\PW$ produced in double parton interactions.  A '-' indicates a negligible yield. Non-prompt and charge-misidentification backgrounds are described in section~\ref{sec:modeling}. }
\begin{tabular}{|l@{\qquad}|r|r|r@{\qquad}|r@{\qquad}|r|}\hline
\multicolumn{1}{|c@{\qquad}}{} &
\multicolumn{1}{|c}{$\Pe\Pe$} &
\multicolumn{1}{|c@{\qquad}}{$\Pe\Pgm$} &
\multicolumn{1}{|c}{$\Pgm\Pgm$} &
\multicolumn{1}{|c@{\qquad}}{$3\ell$} &
\multicolumn{1}{|c|}{$4\ell$} \\ \hline\hline
$\ttH$, $\PH\to\PW\PW$          & $ 1.0 \pm  0.1$     & $ 3.2 \pm  0.4$     & $ 2.4 \pm  0.3$     & $ 3.4 \pm  0.5$     & $0.29 \pm 0.04$      \\
$\ttH$, $\PH\to\Z\Z$            & ---                   & $ 0.1 \pm  0.0$     & $ 0.1 \pm  0.0$     & $ 0.2 \pm  0.0$     & $0.09 \pm 0.02$      \\
$\ttH$, $\PH\to\Pgt\Pgt$        & $ 0.3 \pm  0.0$     & $ 1.0 \pm  0.1$     & $ 0.7 \pm  0.1$     & $ 1.1 \pm  0.2$     & $0.15 \pm 0.02$      \\ \hline
$\ttbar\,\PW$                   & $ 4.3 \pm  0.6$     & $16.5 \pm  2.3$     & $10.4 \pm  1.5$     & $10.3 \pm  1.9$     & ---                    \\
$\ttbar\,\Z\!/\!\gamma^*$       & $ 1.8 \pm  0.4$     & $ 4.9 \pm  0.9$     & $ 2.9 \pm  0.5$     & $ 8.4 \pm  1.7$     & $1.12 \pm 0.62$      \\
$\ttbar\,\PW\PW$                & $ 0.1 \pm  0.0$     & $ 0.4 \pm  0.1$     & $ 0.3 \pm  0.0$     & $ 0.4 \pm  0.1$     & $0.04 \pm 0.02$      \\
$\ttbar\,\gamma$                & $ 1.3 \pm  0.3$     & $ 1.9 \pm  0.5$     & ---                   & $ 2.6 \pm  0.6$     & ---                    \\ \hline
$\PW\Z$                         & $ 0.6 \pm  0.6$     & $ 1.5 \pm  1.7$     & $ 1.0 \pm  1.1$     & $ 3.9 \pm  0.7$     & ---                    \\
$\Z\Z$                          & ---                   & $ 0.1 \pm  0.1$     & $ 0.1 \pm  0.0$     & $ 0.3 \pm  0.1$     & $0.47 \pm 0.10$      \\
Rare SM bkg.                    & $ 0.4 \pm  0.1$     & $ 1.6 \pm  0.4$     & $ 1.1 \pm  0.3$     & $ 0.8 \pm  0.3$     & $0.01 \pm 0.00$      \\ \hline
Non-prompt                      & $ 7.6 \pm  2.5$     & $20.0 \pm  4.4$     & $11.9 \pm  4.2$     & $33.3 \pm  7.5$     & $0.43 \pm 0.22$      \\
Charge misidentified            & $ 1.8 \pm  0.5$     & $ 2.3 \pm  0.7$     & ---                   & ---                   & ---                    \\ \hline
All signals                     & $ 1.4 \pm  0.2$     & $ 4.3 \pm  0.6$     & $ 3.1 \pm  0.4$     & $ 4.7 \pm  0.7$     & $0.54 \pm 0.08$      \\
All backgrounds                 & $18.0 \pm  2.7$     & $49.3 \pm  5.4$     & $27.7 \pm  4.7$     & $59.8 \pm  8.0$     & $2.07 \pm 0.67$      \\ \hline
Data            & 19   & 51    & 41    & 68   & 1 \\ \hline
\end{tabular}
\label{tab:yields-sel}
\end{table}

Candidate events that match one of these signal signatures are selected by
requiring combinations of reconstructed objects. Three
features are common to all three decay signatures:

\begin{itemize}
\item Each event is required to have one lepton with
$\pt>20\GeV$ and another with $\pt>10\GeV$
to satisfy the dilepton trigger requirements.
\item If an event has any pair of leptons, regardless of charge or
  flavor, that form an invariant mass less than 12\GeV, that event
  is rejected.  This requirement reduces contamination from $\PgU$
  and $\JPsi$, as well as very low-mass Drell--Yan events that are not
  included in the simulation.
\item Since signal events have two top quarks, each event is
required to have at least two jets, where at least
two jets satisfy the loose CSV working point or
one jet satisfies the medium CSV working point.

\end{itemize}

In addition, pairs of leptons with the same flavor whose invariant
mass is within $10\GeV$ of the $\Z$ boson mass are rejected to
suppress background events with a $\Z$ boson decay. Same-sign
dielectron events are rejected if they contain any such pair. Events
in the $3\ell$ and $4\ell$ categories are rejected only if the two
leptons in the pair have opposite charges.

Same-sign dilepton events are required to have exactly two leptons
with identical charges and at least four hadronic jets. Each lepton
must pass the lepton preselection, the tight working point of the
lepton MVA discriminant, and the charge quality requirements.  To
reject events from backgrounds with a $\Z$ boson, $L_\mathrm{D} >
30\GeV$ is required for dielectron events, where $L_\mathrm{D}$ is
defined in section~\ref{sec:objects}, equation~\ref{eq:reconstruction_isolation}. To further suppress reducible
backgrounds, especially non-$\ttbar$ backgrounds, the threshold on the
$\pt$ of the second lepton is raised to $20\GeV$, and the scalar sum
of the $\pt$ of the two leptons and of the $\ETmiss$ is required to be
above $100\GeV$.

The three-lepton candidate selection requires exactly three leptons
that pass the lepton preselection and the tight working point for the
lepton MVA discriminant.  To further reject events from backgrounds
with a $\Z$ boson, an $L_\mathrm{D}$ requirement is applied,
with a tighter threshold if the event has a pair of leptons with the same
flavor and opposite charge.  For events with large jet multiplicity
($\geq$ 4 jets), where contamination from the $\Z$-boson background is
smaller, the $L_\mathrm{D}$ requirement is not applied.

The four-lepton candidate selection requires exactly four leptons that
each pass the lepton preselection and the loose working point of the
lepton MVA discriminant.

The observed event yields in data for each final state and the
expectations from the different physical processes after event
selection are summarized in table~\ref{tab:yields-sel}. The details of
the calculations of the signal and background yields are discussed in
the next section.

\subsection{Signal and background modeling}
\label{sec:modeling}
Three categories of backgrounds are identified in this search: $\ttbar
\mathrm{V}$ backgrounds from the associated production of a $\ttbar$
pair and one or more $\PW$ or $\Z$ bosons; diboson or multiboson
production associated with multiple hadronic jets; and reducible
backgrounds from events with non-prompt leptons, or opposite-sign
dilepton events in which the charge of one of the leptons is
misidentified.  These three background classes are estimated
separately with different methods, described below.  The systematic
uncertainties associated with each background estimate are discussed
in section~\ref{sec:systematics}.

The $\ttbar \PH$ signal and backgrounds from $\ttbar\PW$ and
$\ttbar\Z$, as well as minor backgrounds like $\ttbar\PW\PW$ and
triboson processes, are estimated from simulation, normalized to the
NLO inclusive cross sections for each process~\cite{Raitio:1978pt,
  Ng:1983jm, Kunszt:1984ri, Beenakker:2001rj, Beenakker:2002nc,
  Dawson:2002tg,
  Dawson:2003zu,Garzelli:2011vp,YellowReport,YellowReport3,Campbell:2012dh,Garzelli:2012bn,Alwall:2014hca,Melnikov:2011ta,ArkaniHamed2001232}.
The combined cross section of $\ttbar\PW$ and $\ttbar\Z$ has been
measured by the CMS Collaboration in 7\TeV data \cite{CMS_TTV_XSEC}.
The results are consistent with theory but have larger uncertainties.
The prediction for the $\ttbar\Z$ process is also tested directly in a
trilepton control region requiring two of the leptons to have the same
flavor, opposite charge, and invariant mass within $10\GeV$
of the nominal $\Z$ boson mass~\cite{Beringer:1900zz}.  Agreement is
observed in this control region, though the precision of the test is
dominated by the statistical uncertainty of about 35\%.
Agreement was also observed in a
$\ttbar\to\Pe^\pm\Pgm^\mp\,\cPqb\cPaqb\,\Pgn\Pagn$ sample, indicating
good simulation of prompt leptons and real b-quark jets.

The $\PW\Z$ and $\Z\Z$ production processes with the gauge bosons
decaying to electrons, muons, or taus can yield the same leptonic
final states as the signal. These processes are predicted
theoretically at NLO accuracy, but the uncertainty in the production
cross section of diboson with additional partons can be large. To
reduce this uncertainty, a low-signal control sample of $\PW\Z$ or
$\Z\Z$ plus at least two jets is selected by vetoing any event with a
loose $\cPqb$ tag, as well as inverting the $\Z\to\ell\ell$ veto. The
diboson background in the signal region is normalized according to the
event yield observed in this control region times an extrapolation
factor, taken from MC simulation, associated with going from the
control region to the signal region.

The expected flavor composition in simulation for $\PW\Z$ events
after the full selection in the trilepton final state is approximately
$50\%$ from $\PW\Z$ production in association with mistagged jets from
light quarks or gluons, 35\% from events with one jet originating
from a c quark, and 15\% from events with b quarks.  For
$\Z\Z$ in the four-lepton final state, the expectation is about 40\%
events with jets from gluons or light quarks, 35\% from events with
$\cPqb$ quarks and 25\% from events with $\cPqc$ quarks.

\begin{figure}[!htbp]
\centering

\includegraphics[width=0.32\textwidth]{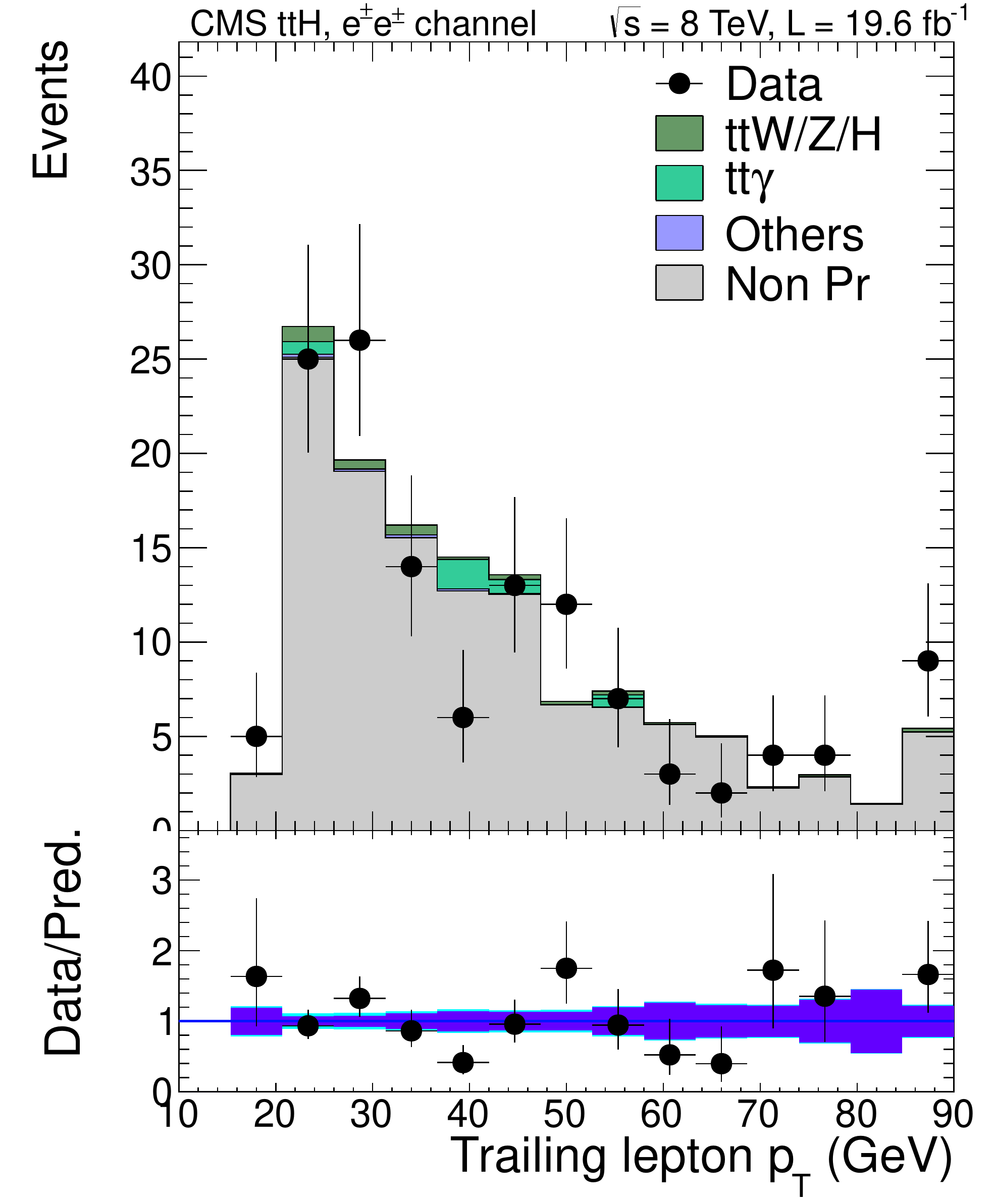}
\includegraphics[width=0.32\textwidth]{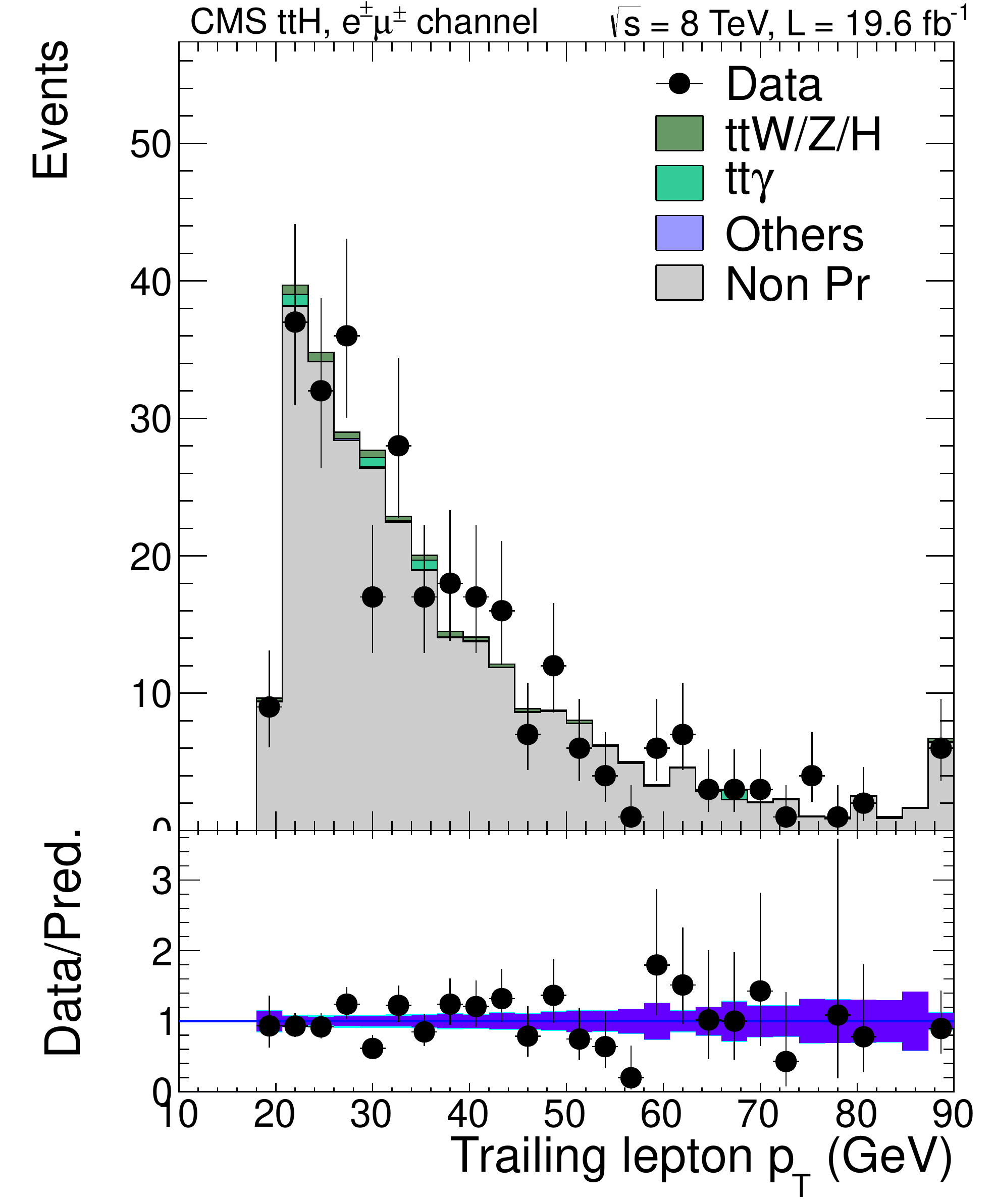}
\includegraphics[width=0.32\textwidth]{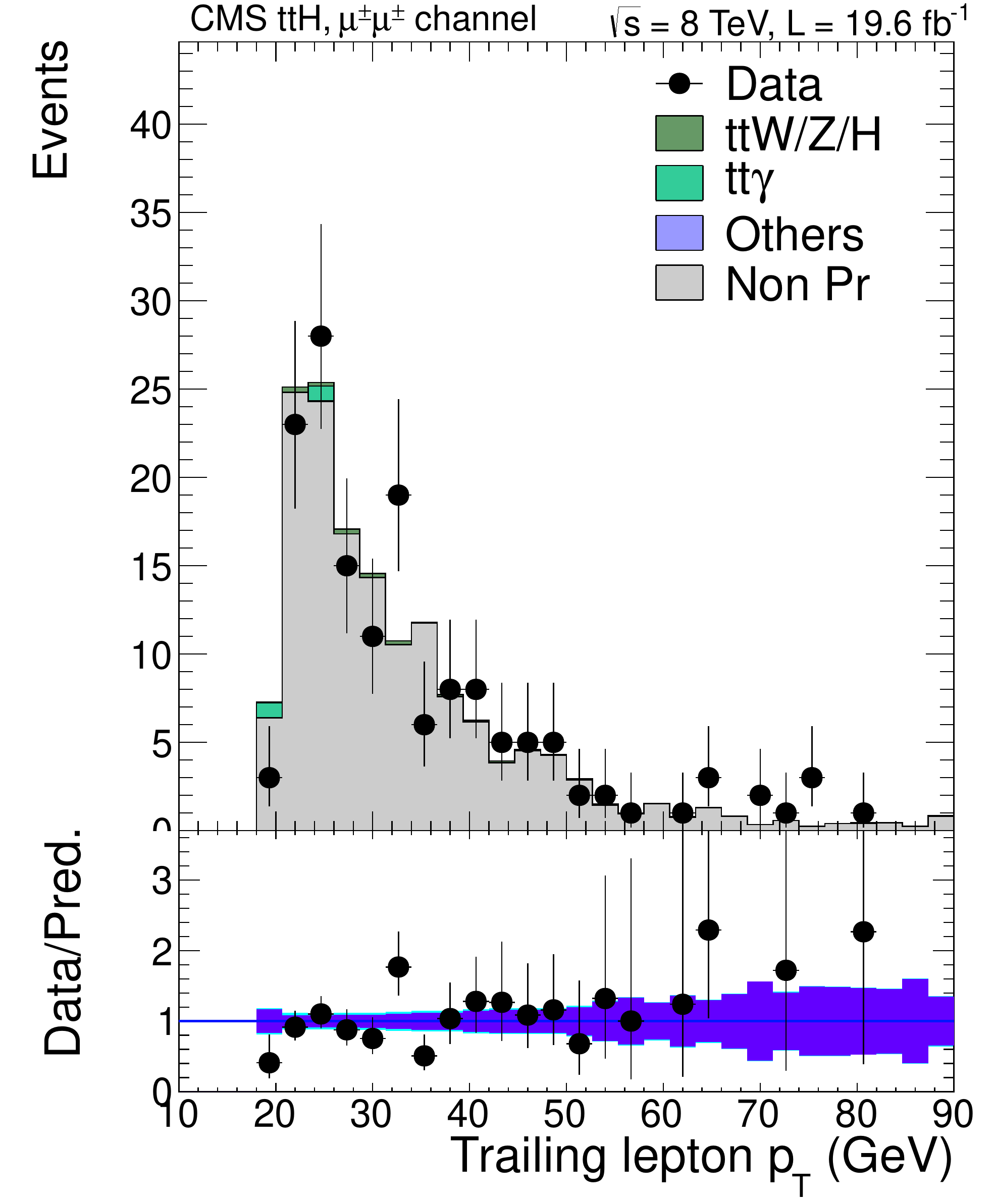} \\

\includegraphics[width=0.32\textwidth]{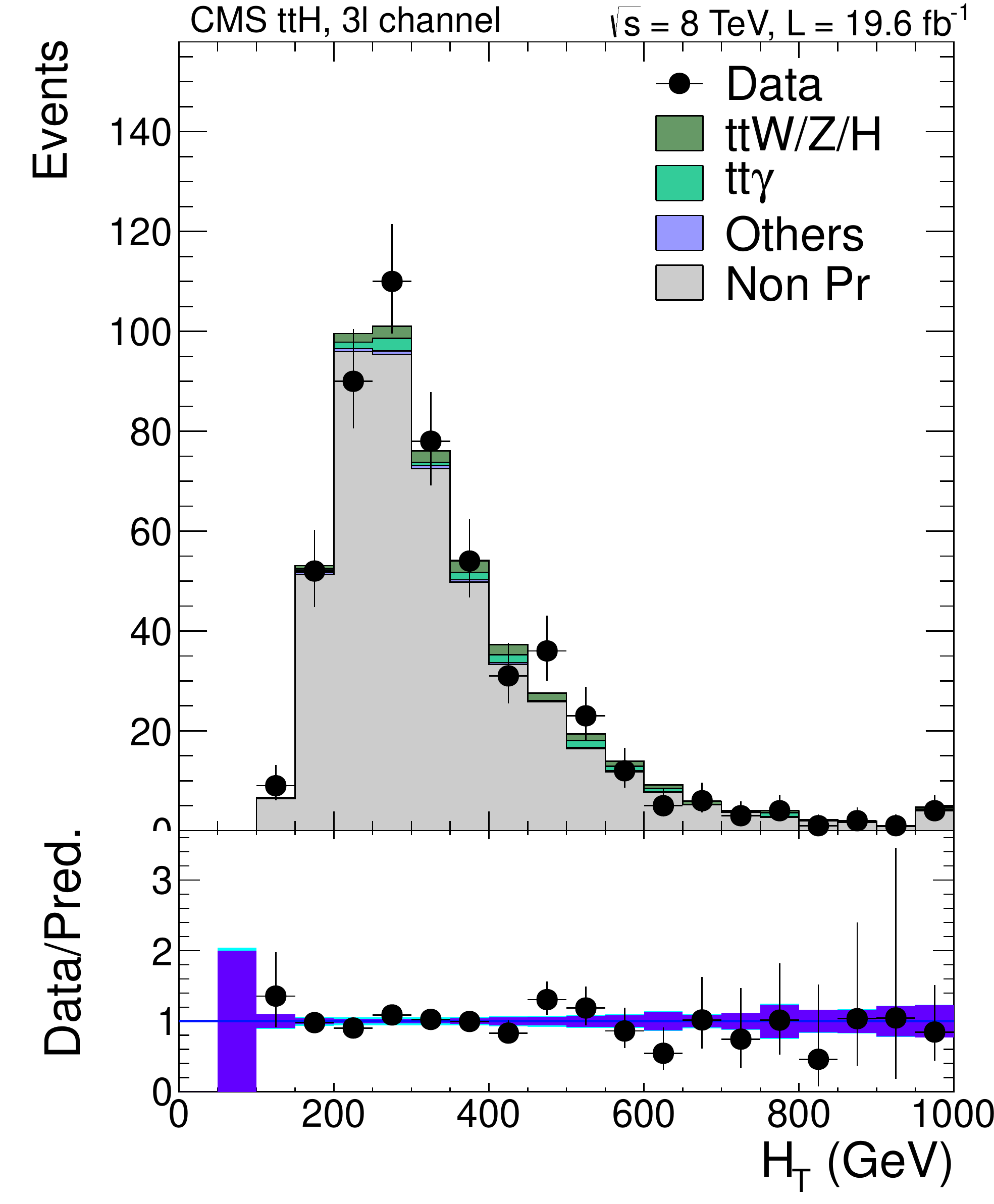}
\includegraphics[width=0.32\textwidth]{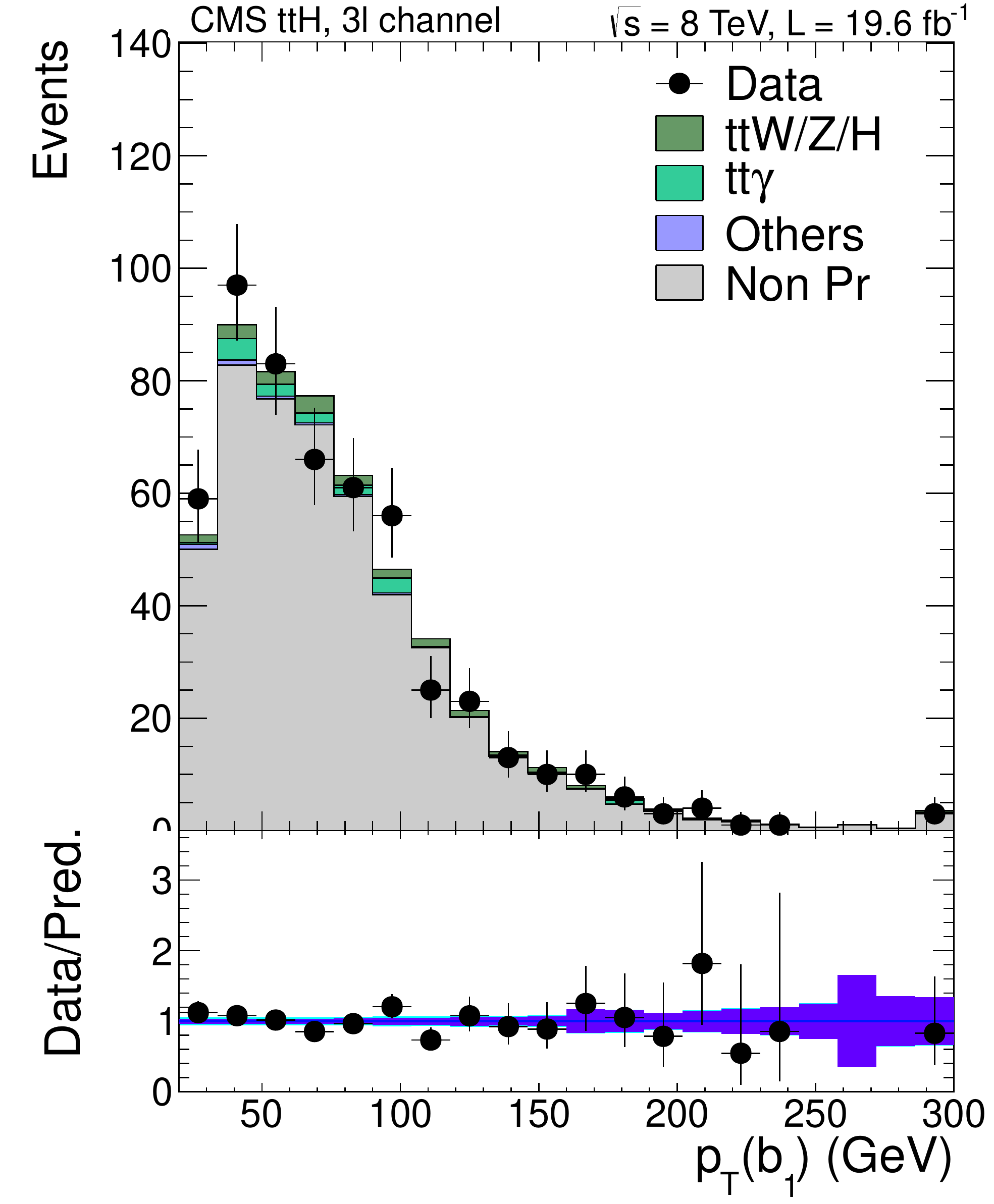}
\includegraphics[width=0.32\textwidth]{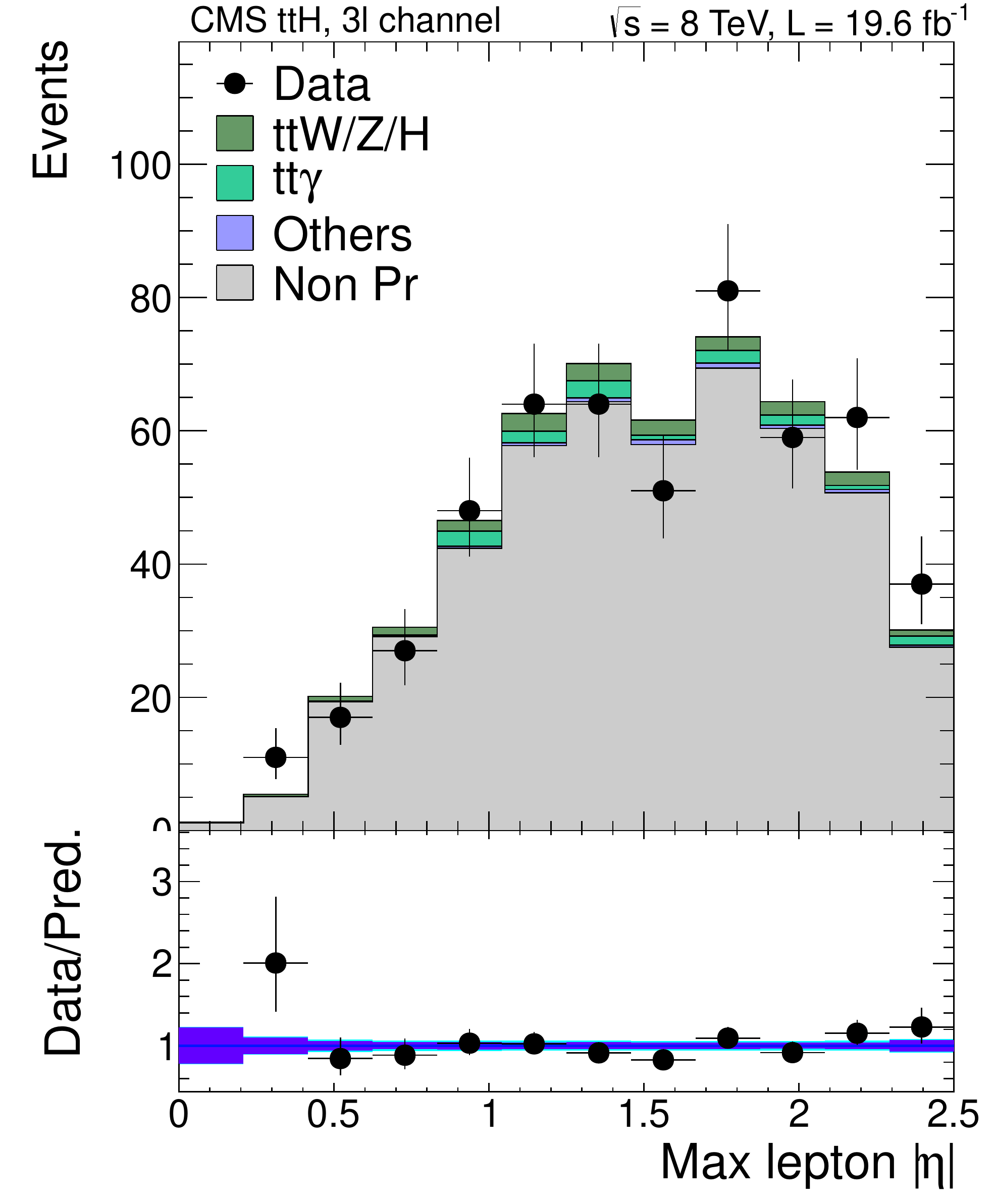}

\caption{These plots show the distribution of key discriminating
  variables for events where one lepton fails the lepton MVA
  requirement. The expected distribution for the non-prompt background
  is taken from simulation (mostly $\ttbar$+jets), and the yield is
  fitted from the data. The bottom panel of each plot shows the ratio
  between data and predictions as well as the overall uncertainties
  after the fit (blue). In the first row the distributions of the
  trailing lepton $\pt$ for the $\Pe^{\pm}\Pe^{\pm}$ (left),
  $\Pe^{\pm}\mu^{\pm}$ (center), and $\mu^{\pm}\mu^{\pm}$ (right)
  final states are shown. In the second row the distributions of the
  $H_\mathrm{T}$ (left), the $\pt$ of the jet with highest b-tagging
  discriminator (center), and the lepton maximum $\abs{\eta}$ (right) are
  shown for the trilepton channel.}

\label{fig:2lnoMVA-3lnoMVA}
\end{figure}

The reducible backgrounds with at least one non-prompt lepton
are estimated from data. A control region dominated by reducible
backgrounds is defined by selecting events with the same kinematics as
the signal region, but for which at least one of the leptons fails the
requirement on the lepton MVA. The kinematic distributions for data in this region
are consistent with MC, mostly $\ttbar$+jets with one non-prompt lepton,
as shown in figure~\ref{fig:2lnoMVA-3lnoMVA}.  Extrapolation to the signal region is
then performed by weighting events in the control region by the
probability for non-prompt leptons to pass the lepton MVA
selection, measured from same-sign dilepton and lepton+b-tagged jet data in control
regions with fewer jets than the signal region, as a function of the
lepton $\pt$ and $\eta$, separately for muons and electrons.

Events in which a single lepton fails the lepton MVA requirement enter
the signal region prediction with weight $\epsilon/(1-\epsilon)$,
where $\epsilon$ denotes the aforementioned probability computed for
the $\pt$, $\eta$, and flavor of the lepton failing the selection.
Events with two leptons failing the requirement are also used, but
with a negative weight
$-\epsilon_1\epsilon_2/[(1-\epsilon_1)(1-\epsilon_2)]$; this small
correction is necessary to account for events with two background-like
leptons contaminating the sample of events with a single lepton
failing the requirement.

The measurement of the probability for non-prompt leptons to pass the
lepton MVA cuts, and the weighting of events in the control region, are performed
separately for events with at most one jet satisfying the medium
CSV requirement and for events with at least
two, to account for the different flavor composition and kinematics
of the two samples.

Charge misidentification probabilities are determined as function of
the lepton $\pt$ and $\eta$ from the observed yields of same-sign and
opposite-sign dilepton pairs with mass within 10\GeV of the Z-boson
mass.  For electrons, this probability varies from 0.03\% in the
barrel to 0.3\% in the endcaps, while for muons the probability is
found to be negligible.

The prediction for background dilepton events with misidentified
electron charge in the signal region is computed from opposite-sign
dilepton events passing the full selection, except for the charge
requirement: events with a single electron enter the prediction with a
weight equal to the charge misidentification probability for that
electron, while dielectron events enter the prediction with a weight
equal to the sum of the charge misidentification probabilities for the
two electrons.

\subsection{Signal extraction}
\label{sec:Sigextraction}
After the event selection,
overall yields are still dominated by background.
The strategy adopted in this search is to fit for the amount of signal
in the distribution of a suitable discriminating variable.

In the dilepton analysis, a BDT output is used as discriminating variable.
The BDT is trained with simulated $\ttbar\PH$ signal and $\ttbar$+jets
background events, with six input variables: the $\pt$ and
$\abs{\eta}$ of the trailing lepton, the minimal angular separation
between the trailing lepton and the closest jet, the transverse mass
of the leading lepton and $\ETmiss$, $H_\mathrm{T}$, and
$H_\mathrm{T}^\mathrm{miss}$. The same training is used for the
$\Pe\Pe$, $\Pe\Pgm$, and $\Pgm\Pgm$ final states, as the gain in
performance from dedicated trainings in each final state is found to
be negligible.

In the trilepton analysis, a BDT output is also used as the final
discriminant. The BDT is trained with simulated $\ttbar\PH$ signal and
a mix of $\ttbar$+jets, $\ttbar\PW$, and $\ttbar\Z$ background events,
with seven discriminating variables: the number of hadronic jets, the
$\pt$ of the jet with the highest $\cPqb$-tagging discriminant value,
the scalar sum of lepton and jet $\pt$ ($H_\mathrm{T}$), the fraction
of $H_\mathrm{T}$ from jets and leptons with $\abs{\eta}<1.2$, the maximum
of the $\abs{\eta}$ values of the three leptons, the minimum $\Delta R$
separation between any pair of opposite-sign leptons, and the mass of
three jets, two close to the $\PW$-boson mass and a b-tagged jet,
closest to the nominal top quark mass~\cite{Beringer:1900zz}.

As a cross-check in both the dilepton and the trilepton final states,
the number of hadronic jets was used instead of the BDT as the discriminating
variable.  The gain in signal strength precision from the
multivariate analysis compared to this simpler cross-check is about
$10\%$.

In the four-lepton analysis, only the number of hadronic jets is
used: the sensitivity of this channel is limited by the very
small branching fraction, and the estimation of the kinematic
distributions of the reducible backgrounds from data is also
challenging due to the low event yields.

In the dilepton and trilepton final states, events are divided into
categories by the sum of the electrical charges of
the leptons, to exploit the charge asymmetry present in several
SM background cross sections in pp collisions
($\ttbar\PW$, $\PW\Z$, single top quark t-channel, $\PW$+jets).
The gain in signal strength precision
from this categorization is approximately 5\%.

\begin{figure}[!htbp]
\centering

\includegraphics[width=0.32\textwidth]{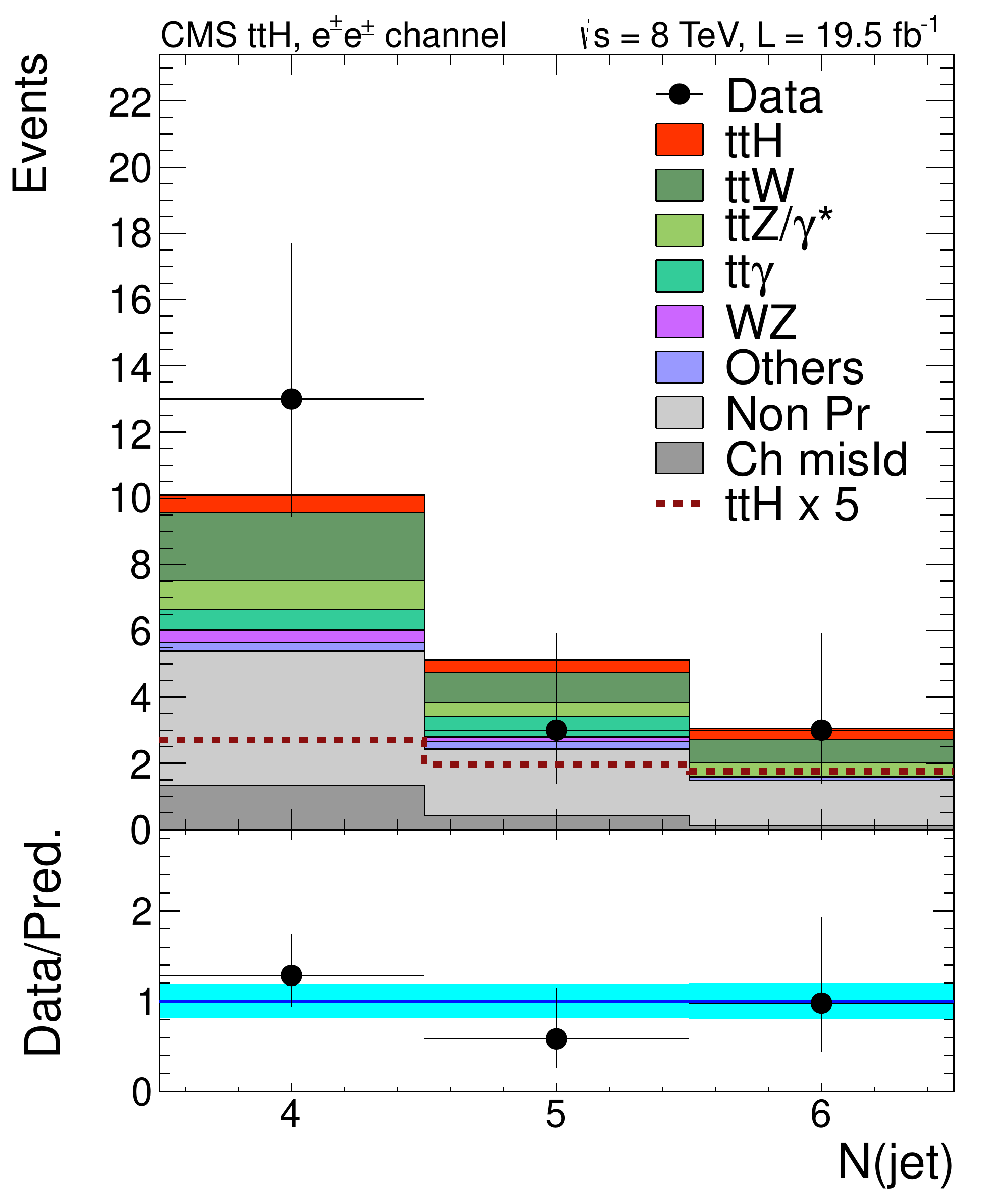}
\includegraphics[width=0.32\textwidth]{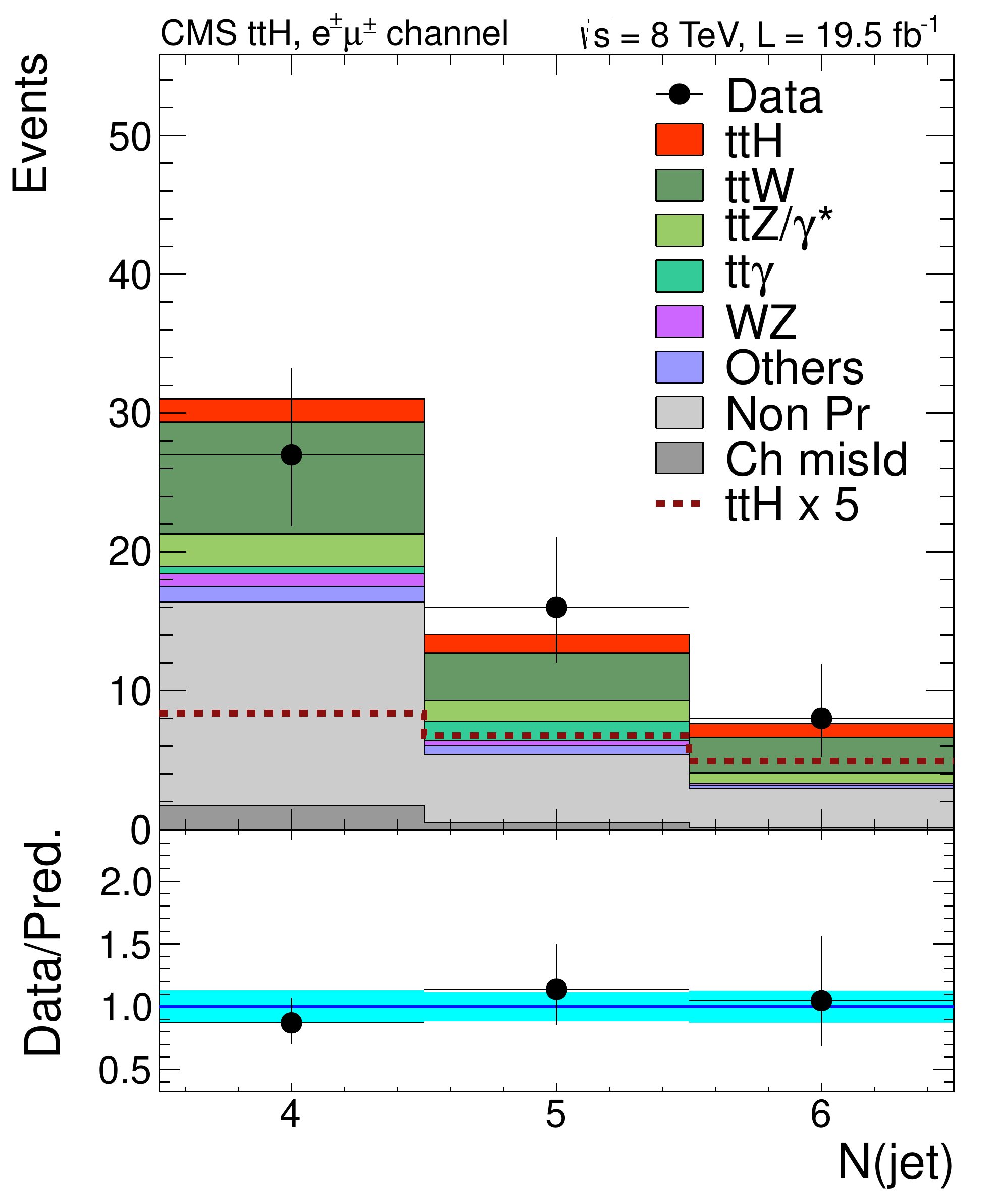}
\includegraphics[width=0.32\textwidth]{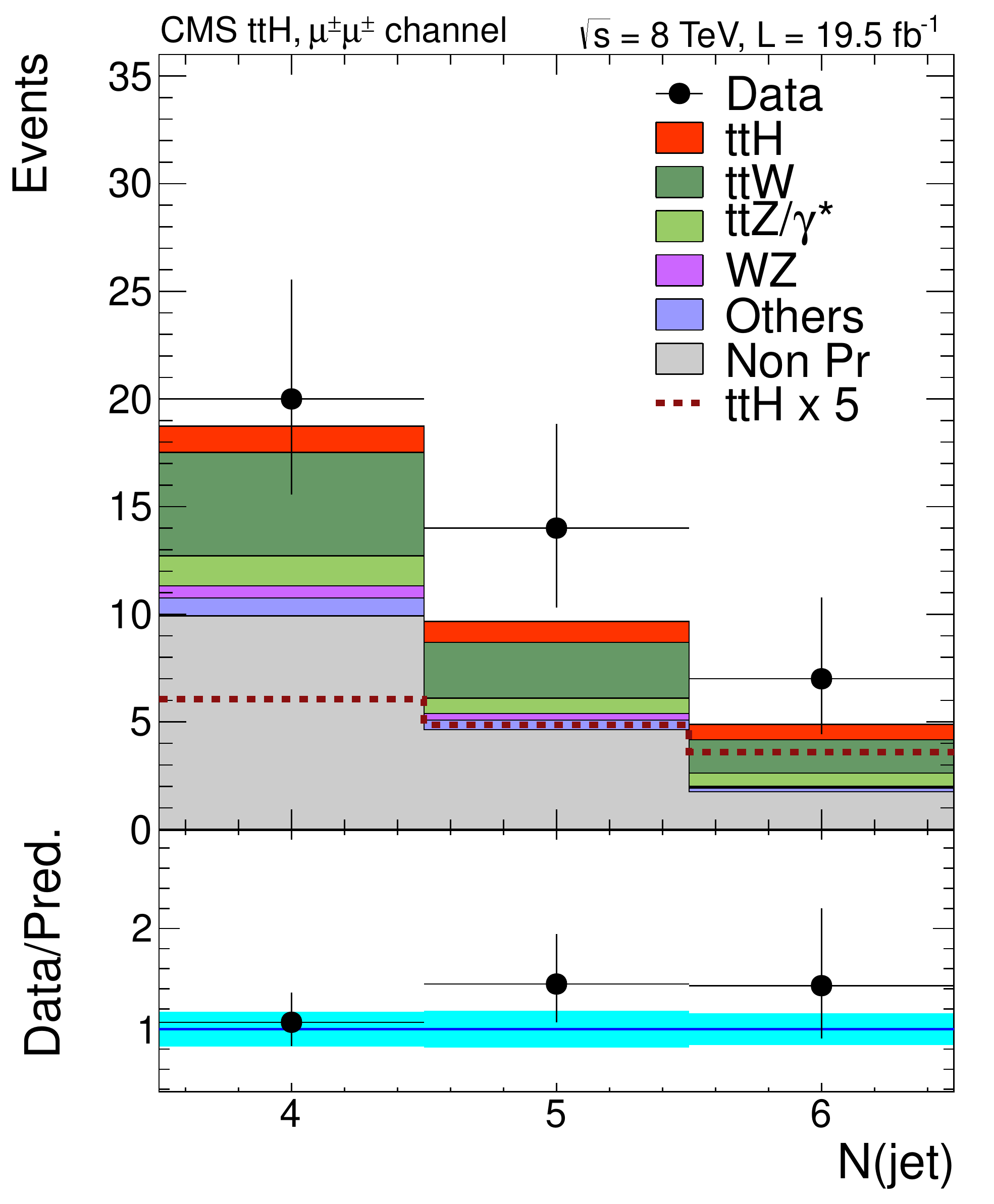}

\includegraphics[width=0.32\textwidth]{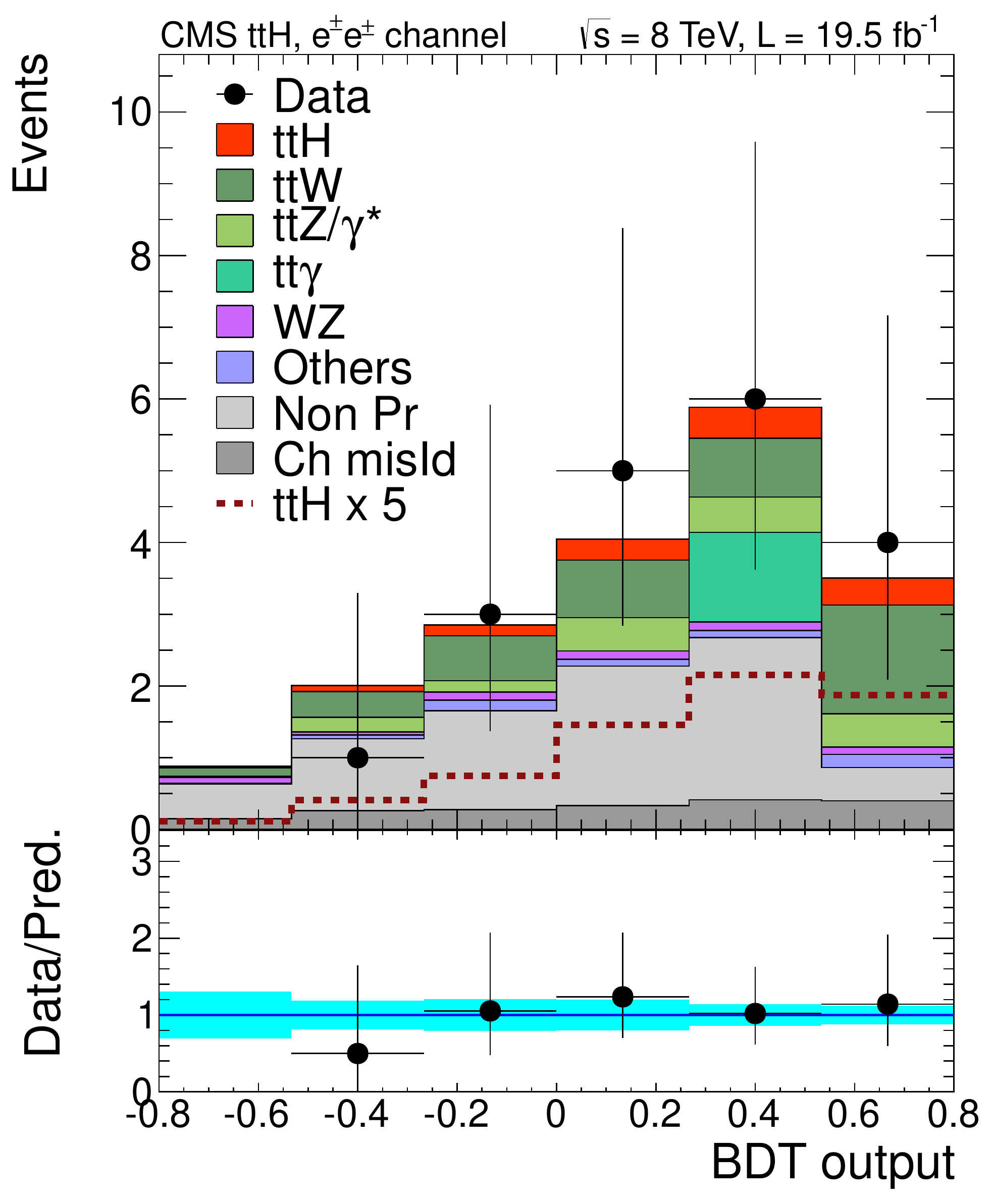}
\includegraphics[width=0.32\textwidth]{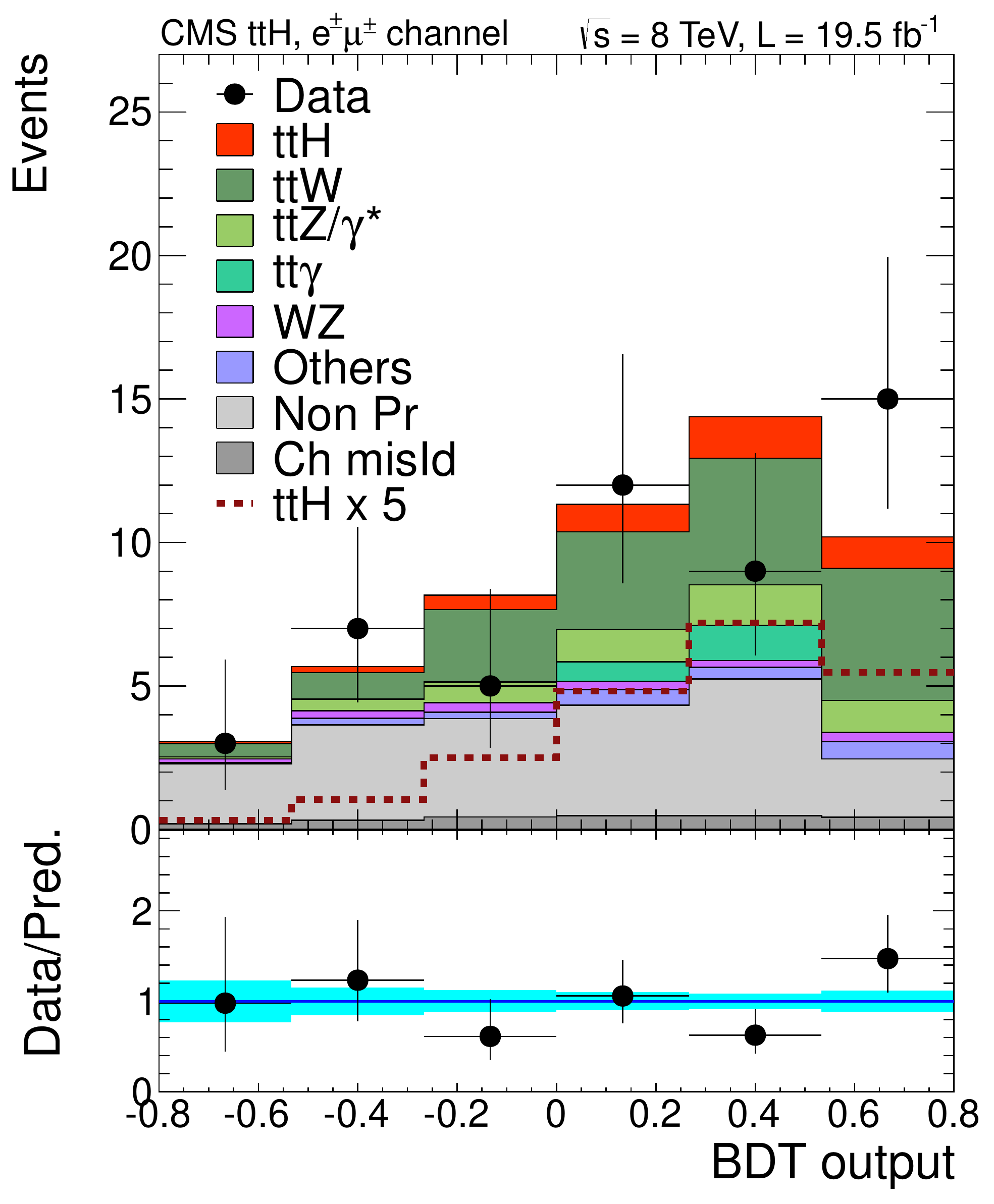}
\includegraphics[width=0.32\textwidth]{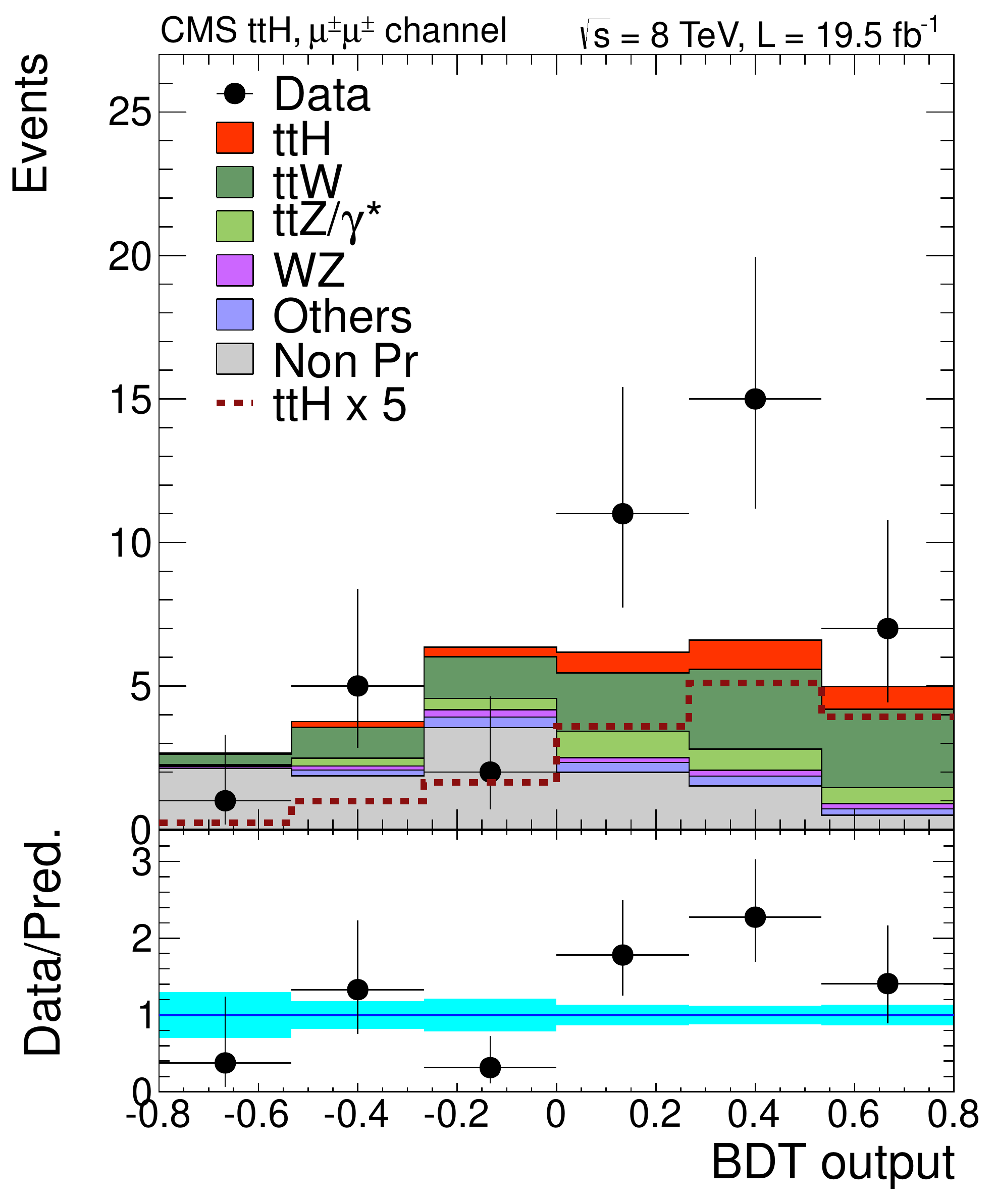} \\

\caption{Distribution of the jet multiplicity (top row) and the BDT
  discriminant (bottom row) for the same-sign dilepton search, for the
  final states $\Pe\Pe$ (left), $\Pe\Pgm$ (center), and $\Pgm\Pgm$
  (right).  Signal and background normalizations are explained in the
  text.  The b-tagged jets are included in the jet multiplicity.}

\label{fig:mva-2lss}
\end{figure}

\begin{figure}[!htbp]
\centering
\includegraphics[width=0.32\textwidth]{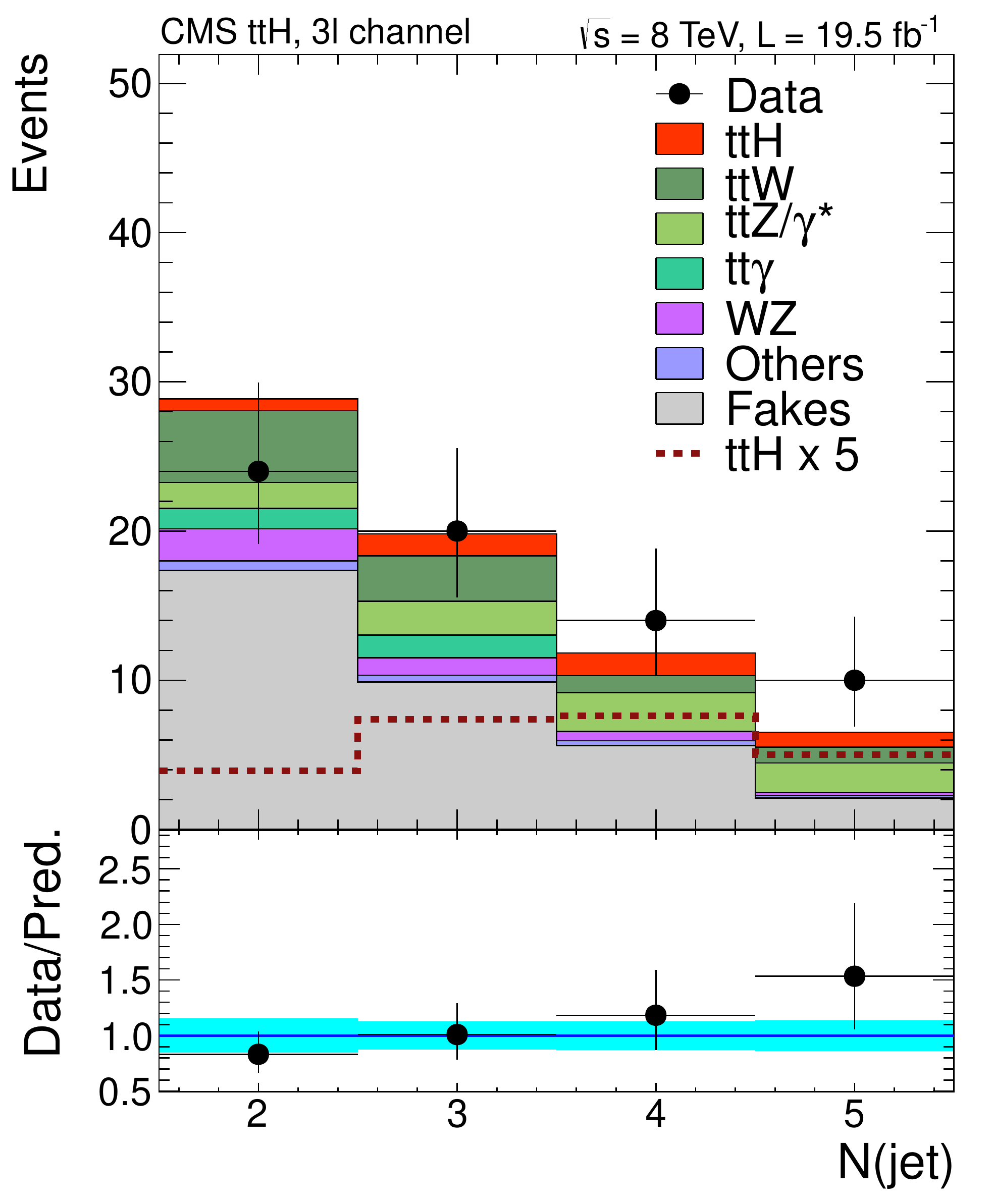}
\includegraphics[width=0.32\textwidth]{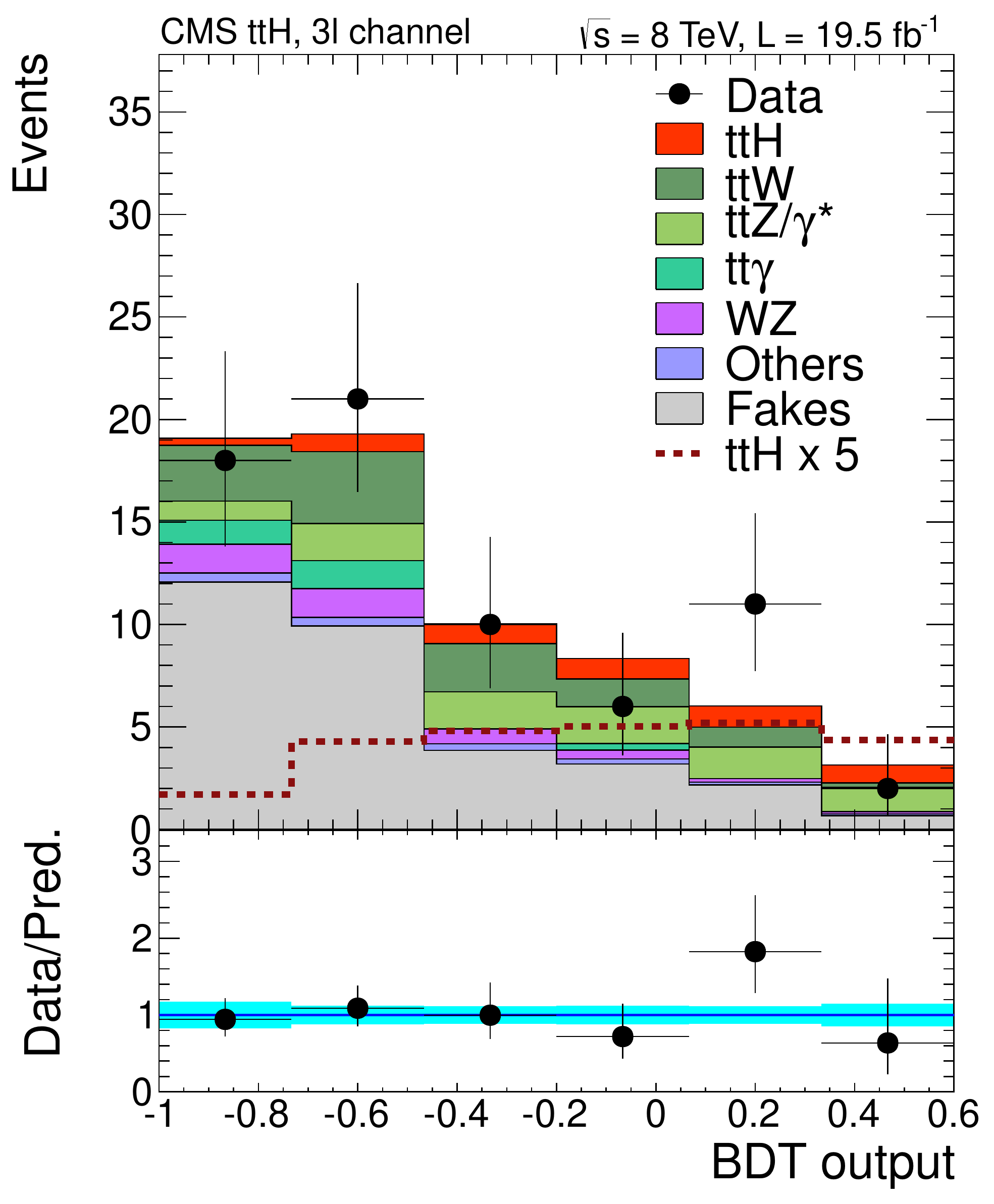}
\includegraphics[width=0.32\textwidth]{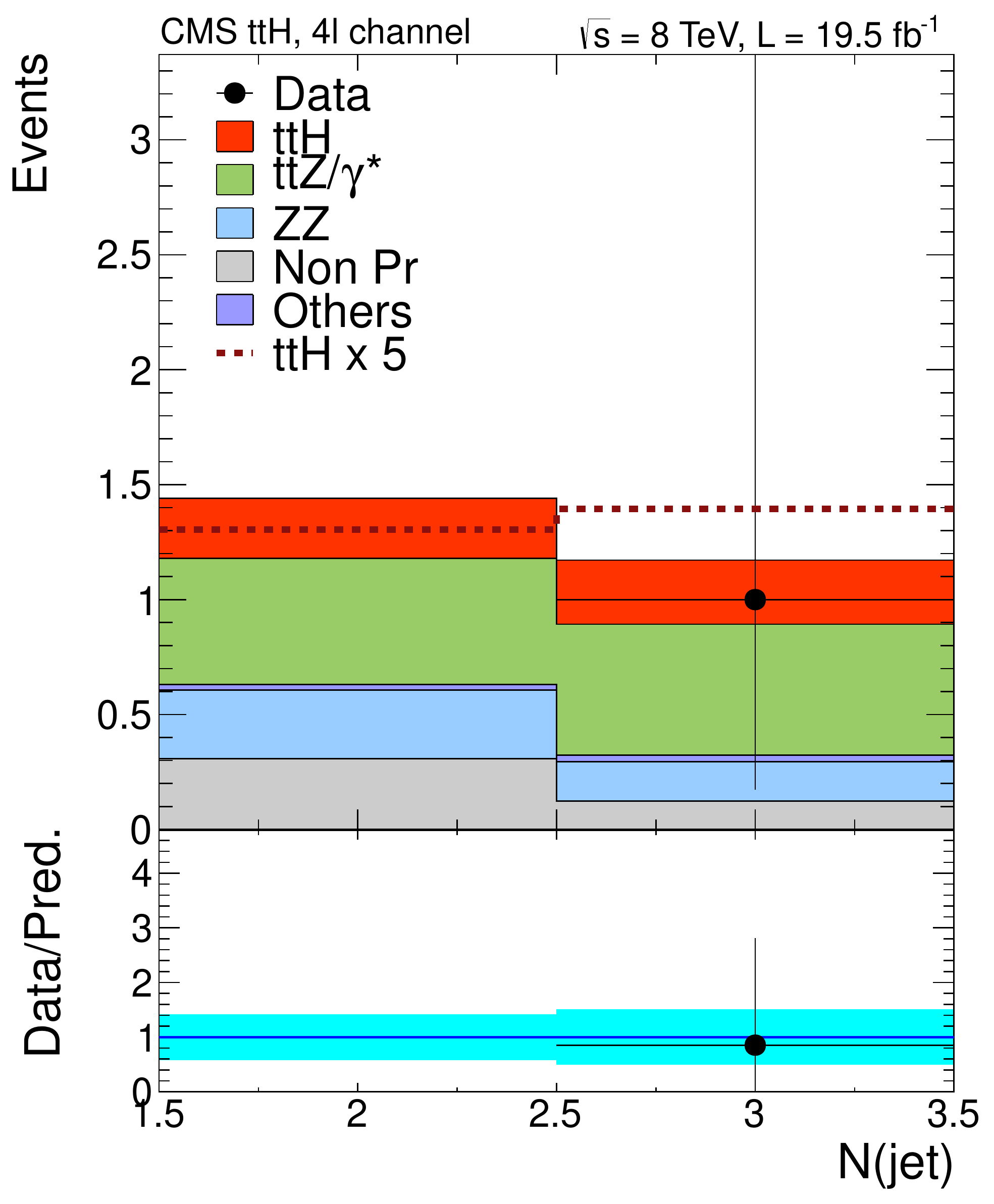}
\caption{Distribution of the jet multiplicity (left) and BDT
  discriminant (center) for the trilepton search.  Events with
  positive and negative charge are merged in these plots, but they are
  used separately in the signal extraction.  The plot on the right
  shows the jet multiplicity for the four-lepton search.  Signal and
  background normalizations are explained in the text.  The b-tagged
  jets are included in the jet multiplicity.}
\label{fig:mva-3l}
\end{figure}

The expected and observed distributions of the number of selected jets
and the BDT output, for the different final states of the dilepton
analysis, are shown in figure~\ref{fig:mva-2lss}.  The same
distributions are shown for the trilepton analysis in
figure~\ref{fig:mva-3l}. The distribution of the number of selected jets
is also shown for the four-lepton channel in figure~\ref{fig:mva-3l}.
The $\ttbar\PH$ signal yield in the stack is the SM prediction ($\mu
=1$); additionally, the signal yield for $\mu =5$ is shown as a dotted
line.  The background distributions use the best-fit values of all
nuisance parameters, with $\mu$ fixed at 1, and the uncertainty bands
are constructed using the nuisance parameter uncertainties.

The dilepton data are in good agreement with the predictions in the
$\Pe\Pe$ and $\Pe\Pgm$ channels, while an excess of signal-like events
is visible in the $\Pgm\Pgm$ final state. The details of this excess
are discussed below. In the trilepton channel the overall data yield
matches expectations.  The jet multiplicity in data is a bit higher,
but the distribution of the BDT discriminator matches the prediction.
In the four-lepton channel only one event is observed with respect to
an overall SM prediction (including expected $\ttbar \PH$
contribution) of about three events.

Because the excess of signal-like events is most pronounced in the
dimuon channel, additional cross-checks were performed.  The
agreement between expected and observed yields in the $\Pe \Pe$ and
$\Pe \Pgm$ channels suggests that the background estimates are
reasonable.  Detailed studies of various single-muon and dimuon
distributions did not reveal any potential additional source of
background.  Moreover, the analysis of the dimuon final state has been
repeated with different lepton selections, using looser working points
for the lepton MVA and also with traditional selections on individual
variables.  These approaches have sensitivities 10--50\% worse than the
nominal analysis and give compatible results.  The consistency of
these checks suggests this excess does not arise from a deficiency in
the estimation of the backgrounds.

\section{Systematic uncertainties}
\label{sec:systematics}
There are a number of systematic uncertainties that impact the
estimated signal or background rates, the shape of the final
discriminant, or both.  This section describes the various sources of
systematic uncertainty.  Section~\ref{sec:results} will explain how
the effects of these uncertainties are accounted for in the likelihood
function used to set limits and extract the best-fit Higgs boson
signal.

Different systematic uncertainties are relevant for different parts of
the overall $\ttbar \PH$ analysis.  Uncertainties related to MC
modeling affect all analysis channels, whereas systematic
uncertainties related to the background estimation or object
identification can be specific to particular channels.
Table~\ref{tab:systSummary} summarizes the impact of systematic
uncertainties on this analysis.  For each broad category,
table~\ref{tab:systSummary} shows the range of effects the systematic
uncertainties have on the signal and background rates, and notes
whether the uncertainty also has an effect on the shape of the final
discriminant.  Cases for which a systematic category only applies to
one analysis channel are noted in parentheses.  Further details are
given below.

 \begin{table}[!thbp]
 \begin{center}
 \topcaption{Summary of systematic uncertainties.  Each row in the table
   summarizes a category of systematic uncertainties from a common
   source or set of related sources.  In the statistical
   implementation, most of these uncertainties are treated via
   multiple nuisance parameters.  The table summarizes the impact of
   these uncertainties both in terms of the overall effect on signal
   and background rates, as well as on the shapes of the signal and
   background distributions.  The rate columns show a range of
   uncertainties, since the size of the rate effect varies both with
   the analysis channel as well as the specific event selection
   category within a channel.  The uncertainties quoted here are \textit{a
     priori} uncertainties; that is they are calculated prior to
   fitting the data, which leads to a reduction in the impact of the
   uncertainties as the data helps to constrain them.  }
     \begin{tabular}{|c|c|c|c|} \hline
                             & \multicolumn{2}{c|}{Rate uncertainty} &  \\
 Source                      & Signal         & Backgrounds         & Shape \\
 \hline\hline
 \multicolumn{4}{|c|}{Experimental} \\
 \hline
 Integrated luminosity       & 2.2--2.6\%    & 2.2--2.6\%          & No  \\
 Jet energy scale            & 0.0--8.4\%    & 0.1--11.5\%         & Yes \\
 CSV b-tagging               & 0.9--21.7\%   & 3.0--29.0\%         & Yes \\
 Lepton reco. and ID         & 0.3--14.0\%   & 1.4--14.0\%         & No  \\
 Lepton misidentification rate ($\PH \to$ leptons) & ---           & 35.1--45.7\%        & Yes \\
 Tau reco. and ID ($\PH \to$ hadrons) & 11.3--14.3\%  & 24.1--28.8\%         & Yes \\
 Photon reco. and ID ($\PH \to$ photons) & 1.6--3.2\%     & ---                 & Yes \\
 MC statistics               & ---           & 0.2--7.0\%          & Yes \\
 \hline
 \multicolumn{4}{|c|}{Theoretical} \\
 \hline
 NLO scales and PDF           & 9.7--14.8\%   & 3.4--14.7\%         & No  \\
 MC modeling                 & 2.3--5.1\%    & 0.9--16.8\%         & Yes \\
 Top quark $\pt$             & ---           & 1.4--6.9\%          & Yes \\
 Additional hf uncertainty  ($\PH \to$ hadrons)   & ---           & 50\%                & No  \\
 $\PH$ contamination ($\PH \to$ photons) & \multicolumn{2}{c|}{36.7--41.2\%} & No \\
 WZ (ZZ) uncertainty  ($\PH \to$ leptons) & ---     & 22\% (19\%)     & No \\
 \hline
 \end{tabular}
     \label{tab:systSummary}
 \end{center}
\end{table}

Global event uncertainties affect all the analysis channels.  The
integrated luminosity is varied by $\pm$2.2\% for the 7\TeV
dataset~\cite{CMS:lumi} and by $\pm$2.6\% for the 8\TeV
dataset~\cite{CMS-PAS-LUM-13-001} from its nominal value. The effect
of finite background MC statistics in the analysis is accounted for
using the approach described in
Refs.~\cite{BarlowBeeston,BarlowBeeston2}.  To avoid including
thousands of nuisance parameters that have no effect on the result,
this uncertainty is not evaluated for any bin in the BDT shapes for
which the MC statistical uncertainty is negligible compared to the
data statistics or where there is no appreciable contribution from
signal.  Tests show that the effect on the final result of neglecting
the MC statistical uncertainty for these bins is smaller than 2\%.  In
total, there are 190 nuisance parameters used to describe the
fluctuations in the bins of the BDT outputs.

The reconstructed objects in each event come with their own
uncertainties.  The uncertainty from the jet energy
scale~\cite{cmsJEC} is evaluated by varying the energy scale for all
jets in the signal and background simulation simultaneously either up
or down by one standard deviation as a function of jet $\pt$ and
$\eta$, and reevaluating the yields and discriminant shapes of all
processes.  These variations have a negligible effect on the $m_{\Pgg
  \Pgg}$ distribution and shape effects for the $\PH \to$ photons
channel are ignored.  The jet energy resolution uncertainty is found
to have a negligible impact for all channels.  The corrections for the
$\cPqb$-tagging efficiencies for light-flavored, c-, and b-quark jets
have associated uncertainties~\cite{CMS:2012hd}.  These uncertainties
are parameterized as a function of the $\pt$, $\eta$, and flavor of
the jets.  Their effect on the analysis is evaluated by shifting the
correction factor of each jet up and down by one standard deviation of
the appropriate uncertainty. Because the CSV distribution for jets in
the $\PH \to$ hadrons channel receives shape corrections, it requires
a different set of shape uncertainties.  In deriving the CSV shape
corrections, there are uncertainties from background contamination,
jet energy scales, and limited size of the data samples.  The
statistical uncertainty in the CSV shape corrections has the potential
to modify the shape of the CSV distribution in complicated ways.  To
parameterize this, the shape uncertainties are broken down into two
orthogonal components: one component can vary the overall slope of the
CSV distribution, while the other component changes the center of the
distribution relative to the ends.  These uncertainties are evaluated
separately for light-flavor and b-quark jets.  Twice the b-quark jet
uncertainties are also applied to c-quark jets, whose nominal scale
factor is 1.

Electron and muon identification and trigger efficiency uncertainties
are estimated by comparing variations in the difference in performance
between data and MC simulation using a high-purity sample of
$\cPZ$-boson decays. These uncertainties vary between 1\% and 6\%.
The systematic uncertainty associated with the MVA selection of prompt
leptons in the $\PH \to$ leptons channel uses tag-and-probe
measurements comparing data and simulation in dilepton $\Z$-boson
events in the dilepton channel. The overall uncertainty amounts to
about 5\% per lepton. The uncertainty in the misidentification
probabilities for non-prompt leptons is estimated from simulation.
The misidentification rate is estimated following the same approach
and parameterization used in the QCD dominated control region, but
instead using only MC samples with a similar composition.  This
simulation-based misidentification rate is then applied to MC samples
with the expected background composition in the signal region, and the
amount of disagreement between the number of non-prompt leptons predicted
by the parameterized misidentification rate and those actually
observed in this collection of MC samples is used to estimate the
systematic uncertainty.  The uncertainty is assessed separately for
different $\pt$, $\eta$ and b-tagged jet multiplicity bins for each
flavor.  The overall uncertainty amounts to about 40\%, which is
applied using linear and quadratic deformations of the $\pt$- and
$\eta$-dependent misidentification rate.

The uncertainties in the $\tauh$ identification consist of
electron and jet misidentification rates, as well as the uncertainty
in the $\tauh$ identification itself.  The last is applied
to the generator-level matched $\tauh$, and estimated to be
6\% per object, using a tag-and-probe technique with a
$\cPZ\to\Pgt\Pgt\to\mu\tauh$ process.  The
jet misidentification rate uncertainty is determined to be 20\%
comparing $\tauh$ misidentification rates in data and
simulated $\PW$+jets events, where the $\PW$ boson decays to $\mu\nu$.
Likewise, the electron misidentification rate uncertainty is found to
be 5\% from $\cPZ\to \Pe\Pe$ events using a tag-and-probe technique.
The $\tauh$ energy scale systematics are obtained from
studies involving $\cPZ\to\Pgt\Pgt$~\cite{CMS:utj}.

For photon identification, the uncertainty in the data--MC
efficiency scale factor from the fiducial region determines the
overall uncertainty, as measured using a tag-and-probe technique
applied to $Z\to \Pe\Pe$ events (3.0\% in the ECAL barrel, 4.0\% in
ECAL endcap). For the uncertainties related to the photon scale and
resolution, the photon energy is shifted and smeared respectively
within the known uncertainty for both photons.

Theoretical uncertainties may affect the yield of signal and background
contributions as well as the shape of distributions. Signal and
background rates are estimated using cross sections of at least NLO
accuracy, which have uncertainties arising primarily from the PDFs and
the choice of the factorization and renormalization scales. The cross
section uncertainties are each separated into their PDF and scale
components and correlated, where appropriate, between processes. For
example, the PDF uncertainty for processes originating primarily from
gluon-gluon initial states, \eg, $\ttbar$ and $\ttbar \PH$
production, are treated as completely correlated.

In addition to the rate uncertainties coming from the NLO or better cross
section calculations, the modeling of the $\ttbar+$jets (including
$\ttbar+\bbbar$ and $\ttbar+\ccbar$), $\ttbar$V, diboson+jets and
the $\PW/\Z+$jets processes are subject to MC modeling uncertainties
arising from the extrapolation from the inclusive rates to exclusive
rates for particular jet or tag categories using the \MADGRAPH
tree-level matrix element generator matched to the \PYTHIA parton
shower MC program.  Although \MADGRAPH incorporates contributions from
higher-order diagrams, it does so only at tree-level, and is subject
to fairly large uncertainties arising from the choice of scales.  These
uncertainties are evaluated using samples for which the factorization
and renormalization scales have been varied up and down by a factor of
two.  Scale variations are propagated to both the rate and (where
significant) the final discriminant shape.  Scale variations are
treated as uncorrelated for the $\ttbar+$light flavor,
$\ttbar+\bbbar$, and $\ttbar+\ccbar$ components. The scale variations
for $\PW+$jets and $\cPZ+$jets are treated as correlated; all other
scale variations are treated as uncorrelated.

A systematic uncertainty on the top quark $\pt$ reweighting for the
$\ttbar+$jets simulation is assessed using the uncorrected MC shapes
as a $-1$ standard deviation systematic uncertainty, and overcorrected MC shapes
as a $+1$ standard deviation uncertainty. The overcorrected shapes are calculated
by doubling the deviation of the top-quark $\pt$ scale factors from 1.
The $\ttbar+\bbbar$ and $\ttbar + \ccbar$ processes represent an
important source of irreducible background for the $\PH \to$ hadrons
analysis.  Neither control region studies nor higher-order theoretical
calculations~\cite{Bredenstein:2010rs} can currently constrain the
normalization of these contributions to better than 50\% accuracy.
Therefore, an extra 50\% uncorrelated rate uncertainty is
conservatively assigned to the $\ttbar + \bbbar$, $\ttbar$ + b and
$\ttbar + \ccbar$ processes.

In the $\PH \to$ photons analysis, to assess the contamination from
Higgs boson production from mechanisms other than $\ttbar \PH$, it is
necessary to extrapolate MC predictions to final states with several
jets beyond those included in the matrix elements used for the
calculation.  As these jets are modeled primarily with parton shower
techniques, the uncertainty in these predictions should be carefully
assessed.  As \POWHEG is used to model $\Pg\Pg \to \PH$ production,
the uncertainty on the rate of additional jets is estimated by taking
the observed difference between the \POWHEG predictions and data in
$\ttbar$ events which are dominated by gluon fusion production,
$\Pg\Pg \to \ttbar$~\cite{CMSttjets}. This uncertainty amounts to at
most 30\%, which includes the uncertainty in the fraction of $\Pg\Pg
\to \PH$ plus heavy-flavor jets. Furthermore, the fraction of $\Pg\Pg
\to \PH$ plus heavy-flavor jets is scaled by the difference observed
between data and the \POWHEG predictions~\cite{CMSbbtoqq} in $\ttbar
\bbbar$ and $\ttbar \Pq\Paq/\Pg\Pg$.  These large uncertainties apply
to a very small subset of the events falling into the signal region,
thus resulting in a very small uncertainty on the final sensitivity to
the signal itself.

In the $\PH \to$ leptons analysis, the normalization uncertainty in
the $\PW\Z$ ($\Z\Z$) process comes from a variety of sources.  Several
uncertainties are related to the control region used to estimate the
normalization, as described in section~\ref{sec:modeling}.  The
statistical uncertainty in the control region estimate results in
{10}\%\,(12\%) uncertainty in the normalization, while residual
backgrounds in the control region account for another 10\%\,(4\%).  Uncertainties in the $\cPqb$-tagging efficiencies result
in a 15\%\,(7.5\%) normalization uncertainty.  While
uncertainties in the PDFs~\cite{Alekhin:2011sk,Botje:2011sn} and on the
extrapolation from the control region to the signal region cause
normalization uncertainties of 4\%\,(3\%) and 5\%\,(12\%) respectively.  Taken together, the uncertainties described
above result in an overall $\PW\Z$ ($\Z\Z$) normalization uncertainty
of 22\%\,(19\%).

\section{Results}
\label{sec:results}
The statistical methodology employed for these results is identical to
that used for other CMS Higgs boson analyses.  More details can be
found in Ref.~\cite{Chatrchyan:2013lba}.  In brief, a binned
likelihood spanning all analysis channels included in a given result
is constructed.  The amount of signal is characterized by the signal
strength parameter $\mu$, which is the ratio of the observed cross
section for $\ttbar \PH$ production to the SM expectation.  In
extracting $\mu$ some assumption must be made about the branching
fractions of the Higgs boson.  Unless stated otherwise, $\mu$ is
extracted assuming SM branching fractions.  Under some circumstances
the branching fractions are parameterized in a more sophisticated
fashion, for example allowing separate scaling for the Higgs boson's
couplings to different particles in the SM.  Uncertainties in the
signal and background predictions are incorporated by means of
nuisance parameters.  Each distinct source of uncertainty is accounted
with its own nuisance parameter, and in the case where a given source
of uncertainty impacts more than one analysis channel, a single
nuisance parameter is used to capture the correlation in this
uncertainty between channels.  Nuisance parameters are profiled,
allowing high-statistics but signal-poor regions in the data to
constrain certain key nuisance parameters.

To assess the consistency of the data with different hypotheses, a
profile likelihood ratio test statistic is used: $q(\mu) = -2 \ln \left[
\mathcal{L}(\mu,\hat{\theta}_{\mu})/\mathcal{L}(\hat{\mu},\hat{\theta})\right]$,
where $\theta$ represents the full suit of nuisance parameters.  The
parameters $\hat{\mu}$ and $\hat{\theta}$ represent the values that
maximize the likelihood function globally, while the parameters
$\hat{\theta}_{\mu}$ are the nuisance parameter values that maximize
the likelihood function for a given $\mu$.  Results are reported both
in terms of the best-fit value for $\mu$ and its associated
uncertainty and in terms of upper limits on $\mu$ at 95\% confidence
level (CL).  Limits are computed using the modified frequentist
CL$_{\text{S}}$ method~\cite{Junk:1999kv,0954-3899-28-10-313}.
Results are obtained both independently for each of the distinct
$\ttbar \PH$ signatures ($\bbbar$, $\tauh \tauh$,
$\Pgg \Pgg$, same-sign $2l$, $3l$, and $4l$) as well as combined over
all channels.

\begin{figure}[!htbp]
  {\centering
    \includegraphics[width=0.49\textwidth]{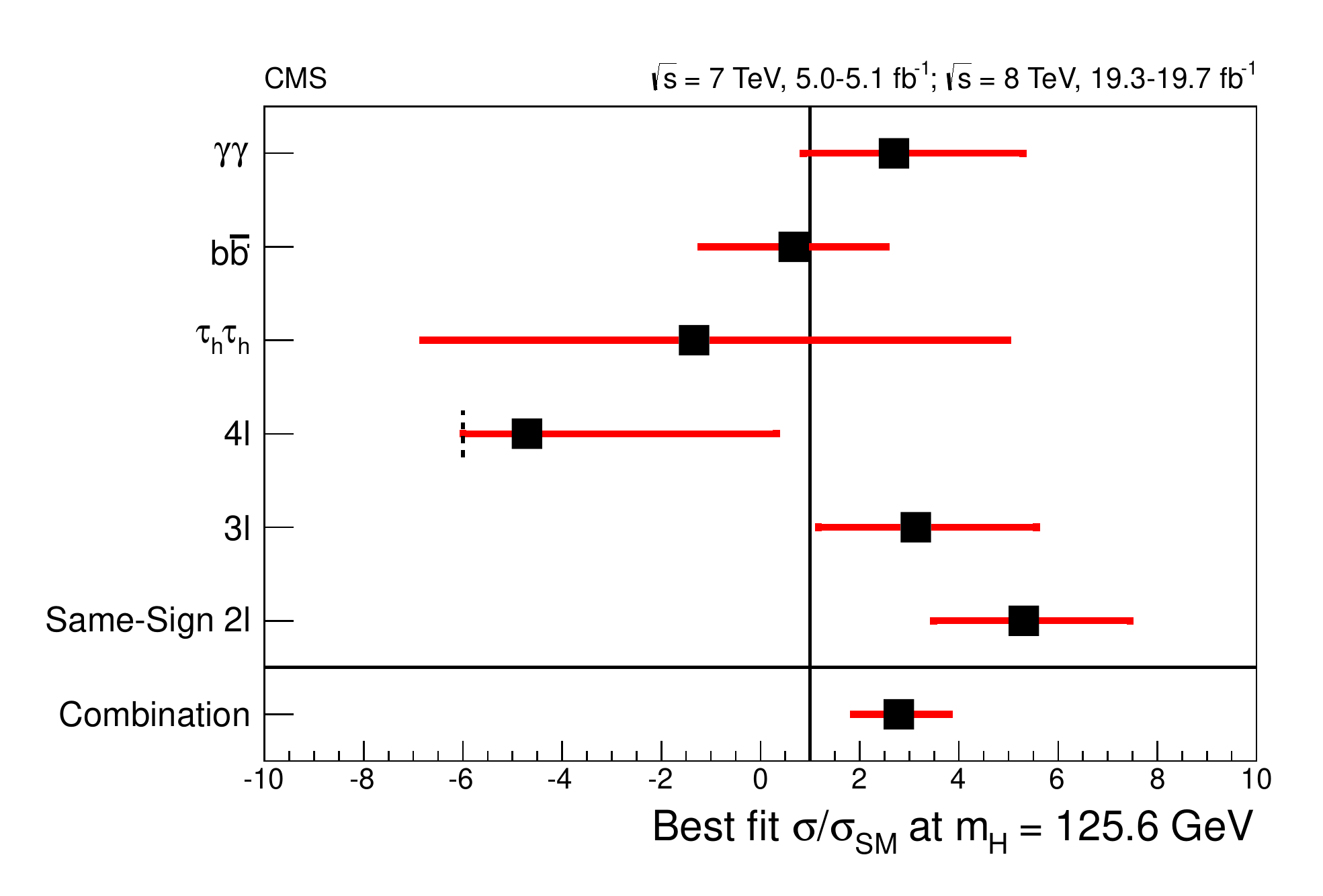}
    \includegraphics[width=0.49\textwidth]{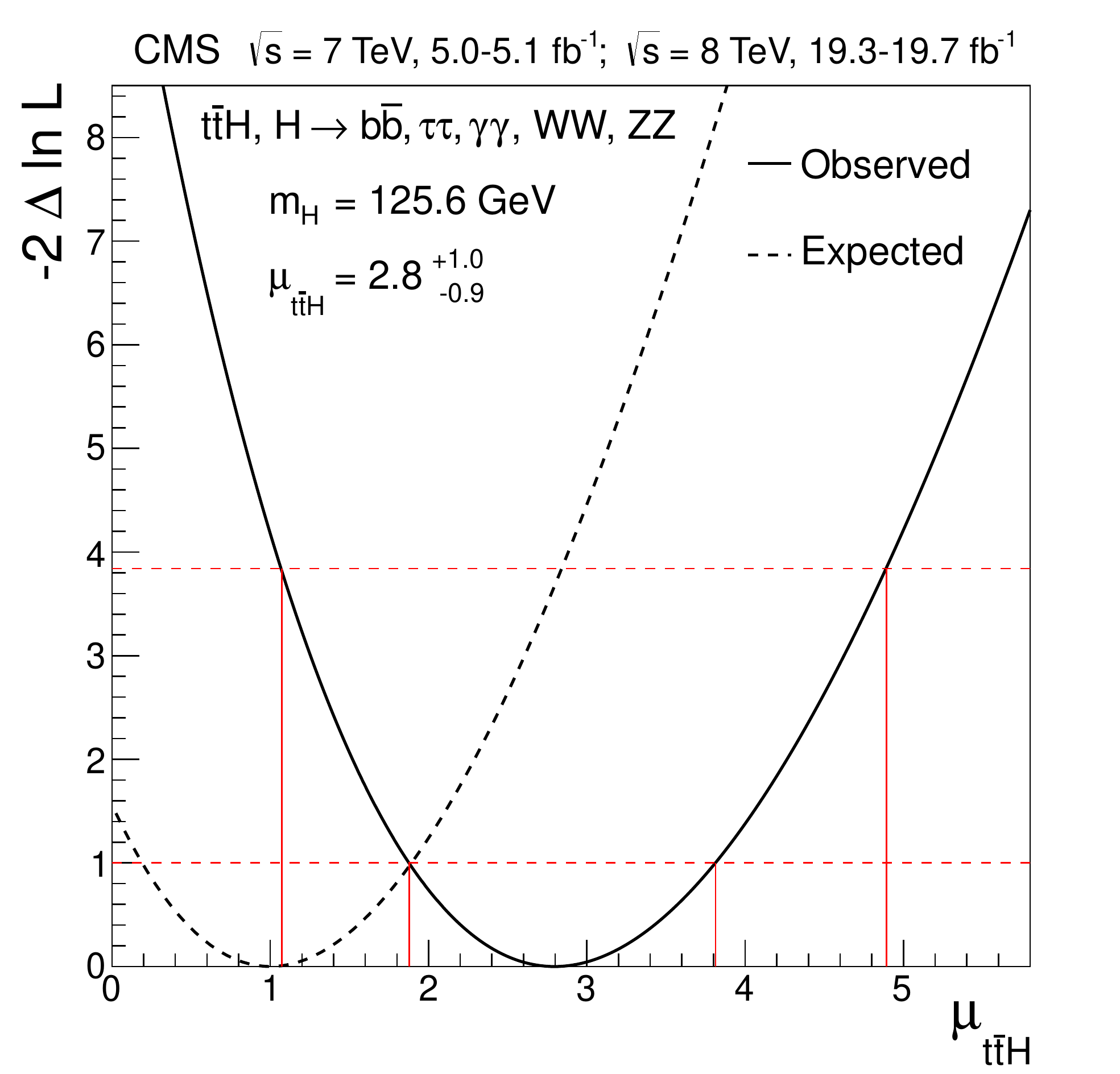}
    \caption{Left: The best-fit values of the signal strength parameter $\mu = \sigma/\sigma_{\mathrm{SM}}$ for each \ttH channel at $m_{\PH}$ = 125.6\GeV. The signal strength in the four-lepton final state is not allowed to be below approximately $-6$ by the requirement that the expected signal-plus-background event yield must not be negative in either of the two jet multiplicity bins.  Right: The 1D test statistic $q$(\muTTH) scan vs. the signal strength parameter for \ttH processes \muTTH, profiling all other nuisance parameters. The lower and upper horizontal lines correspond to the 68\% and 95\% CL, respectively. The \muTTH values where these lines intersect with the $q$(\muTTH) curve are shown by the vertical lines.}
    \label{fig:comb_limit_mu}}
\end{figure}

The best-fit signal strengths from the individual channels and from
the combined fit are given in table~\ref{tab:comb_limit_mu} and
figure~\ref{fig:comb_limit_mu}.  The internal consistency of the six
results with a common signal strength has been evaluated to be 29\%,
estimated from the asymptotic behavior of the profile likelihood
function~\cite{Chatrchyan:2013lba}.  Combining all channels, the best
fit value of the common signal strength is $\mu = 2.8^{+1.0}_{-0.9}$
(68\% CL).  For this fit, the rates of Higgs boson production from
mechanisms other than $\ttbar \PH$ production are fixed to their SM
expectations; however, allowing all Higgs boson contributions to float
with a common signal strength produces a negligible change in the fit
result.  Although the fit result shows an excess, within
uncertainties, the result is consistent with SM expectations.  The
$p$-value under the SM hypothesis ($\mu = 1$) is 2.0\%.  The $p$-value
for the background-only hypothesis ($\mu = 0$) is 0.04\%,
corresponding to a combined local significance of 3.4 standard
deviations.  Assuming SM Higgs boson production with $m_{\PH} = 125.6\GeV$~\cite{Chatrchyan:2013mxa}, the expected local significance is
1.2 standard deviations.

Throughout this paper, whenever a specific choice for Higgs boson mass
has been required, a mass of 125.6\GeV has been used, corresponding
to the most precise Higgs boson mass measurement by CMS at the time
these results were obtained~\cite{Chatrchyan:2013mxa}.  However, the
recent CMS measurement of inclusive Higgs boson production with the
Higgs boson decaying to a pair of photons~\cite{hgg_inclusive},
obtains a lower Higgs boson mass value.  The combination of CMS
Higgs boson mass measurements is expected to be very close to 125\GeV.  The combined $\ttbar \PH$ measurement is not very sensitive to
the Higgs boson mass value.  The combined best-fit signal strength
obtained assuming a Higgs boson mass of 125\GeV is $\mu =
2.9^{+1.1}_{-0.9}$.  This result corresponds to a 3.5 standard
deviation excess over the background-only ($\mu = 0$) hypothesis, and
represents a 2.1 standard deviation upward fluctuation on the SM
$\ttbar \PH$ ($\mu = 1$) expectation.  These values are very close to
the values quoted above for $m_{\PH} = 125.6\GeV$.

Although the observed signal strength is consistent with SM
expectations, it does represent a roughly 2 standard deviation upward
fluctuation.  Therefore, it is interesting to look more closely at how
the different channels contribute to the observed excess.  From
figure~\ref{fig:comb_limit_mu}, it can be seen that the same-sign
dilepton channel yields the largest signal strength.  Within that
channel, the same-sign dimuon subsample has the largest signal
strength, with $\mu = 8.5^{+3.3}_{-2.7}$ compared with $\mu =
2.7^{+4.6}_{-4.1}$ for the same-sign dielectron channel and $\mu =
1.8^{+2.5}_{-2.3}$ for the same-sign electron-muon channel.  The
internal consistency of these three channels, along with the three and
four lepton channels, is 16\%.  To characterize the impact of the
same-sign dimuon channel on the combined fit, the fit was repeated
with that channel omitted, resulting in a signal strength of $\mu =
1.9^{+1.0}_{-0.9}$.  This fit result corresponds to a $p$-value under
the SM hypothesis ($\mu = 1$) of 17\%.  The $p$-value under the
background-only hypothesis for this fit is 1.6\% corresponding to a
local significance of $2.2$ standard deviations.  Although removing
the same-sign dimuon channel does result in a lower fitted signal
strength, the overall conclusion is unchanged.

\begin{figure}[!htbp]
  {\centering
    \includegraphics[width=0.6\textwidth]{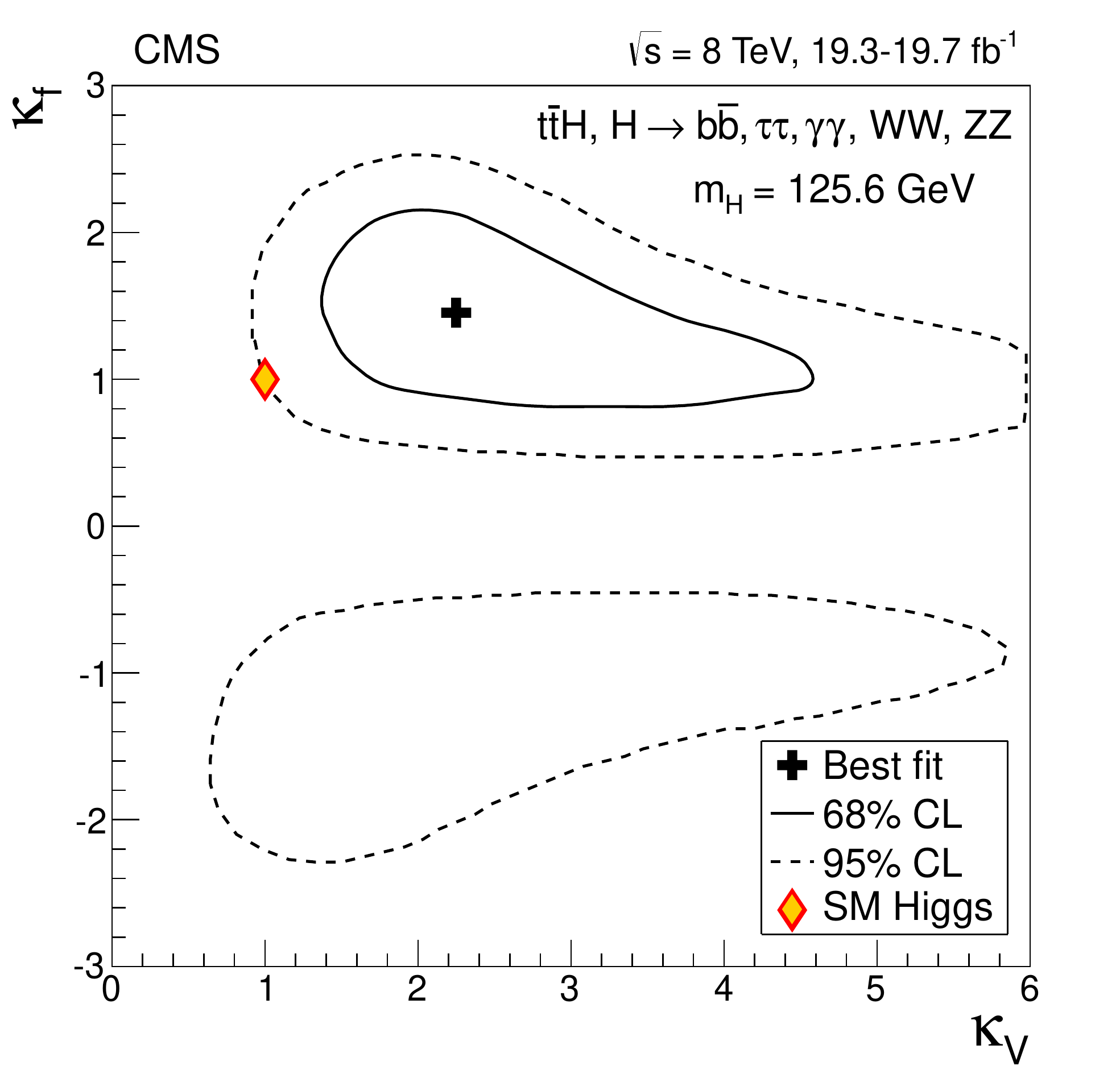}
    \caption{The 2D test statistic $q$(\kV, \kf) scan vs. the modifiers to the coupling of the Higgs boson to vector bosons (\kV) and fermions (\kf), profiling all other nuisances, extracted using only the $\ttbar \PH$ analysis channels. The contour lines at 68\% CL (solid line) and 95\% CL (dashed line) are shown. The best-fit and SM predicted values of the coupling modifiers (\kV, \kf) are given by the black cross and the open diamond, respectively.}
    \label{fig:comb_mu_cvcf}}
\end{figure}

In the above, consistency with SM expectations is assessed by varying
the $\ttbar \PH$ signal strength.  An alternative approach would be to
vary individual couplings between the Higgs boson and other particles.
The collected statistics are currently insufficient to allow
individual couplings to each SM particle to be probed.  However, it is
feasible to scale the couplings to vector bosons and fermions
separately.  This is a useful approach for testing whether the excess
observed is consistent with expectations from SM $\ttbar \PH$
production.  Following the methodology used to study the properties of
the new boson in the global CMS Higgs boson
analysis~\cite{Chatrchyan:2013lba}, the scale factors \kV and \kf are
introduced to modify the coupling of the Higgs boson to vector bosons
and fermions, respectively.  Figure~\ref{fig:comb_mu_cvcf} shows the 2D
likelihood scan over the (\kV,\kf) phase space using only the $\ttbar
\PH$ analysis channels.  The best-fit values of the coupling modifiers
are at (\kV,\kf) = (2.2,1.5), which is compatible at the 95\% CL with
the expectation from the SM Higgs boson (1,1).

\begin{figure}[!htbp]
  {\centering
    \includegraphics[width=0.49\textwidth]{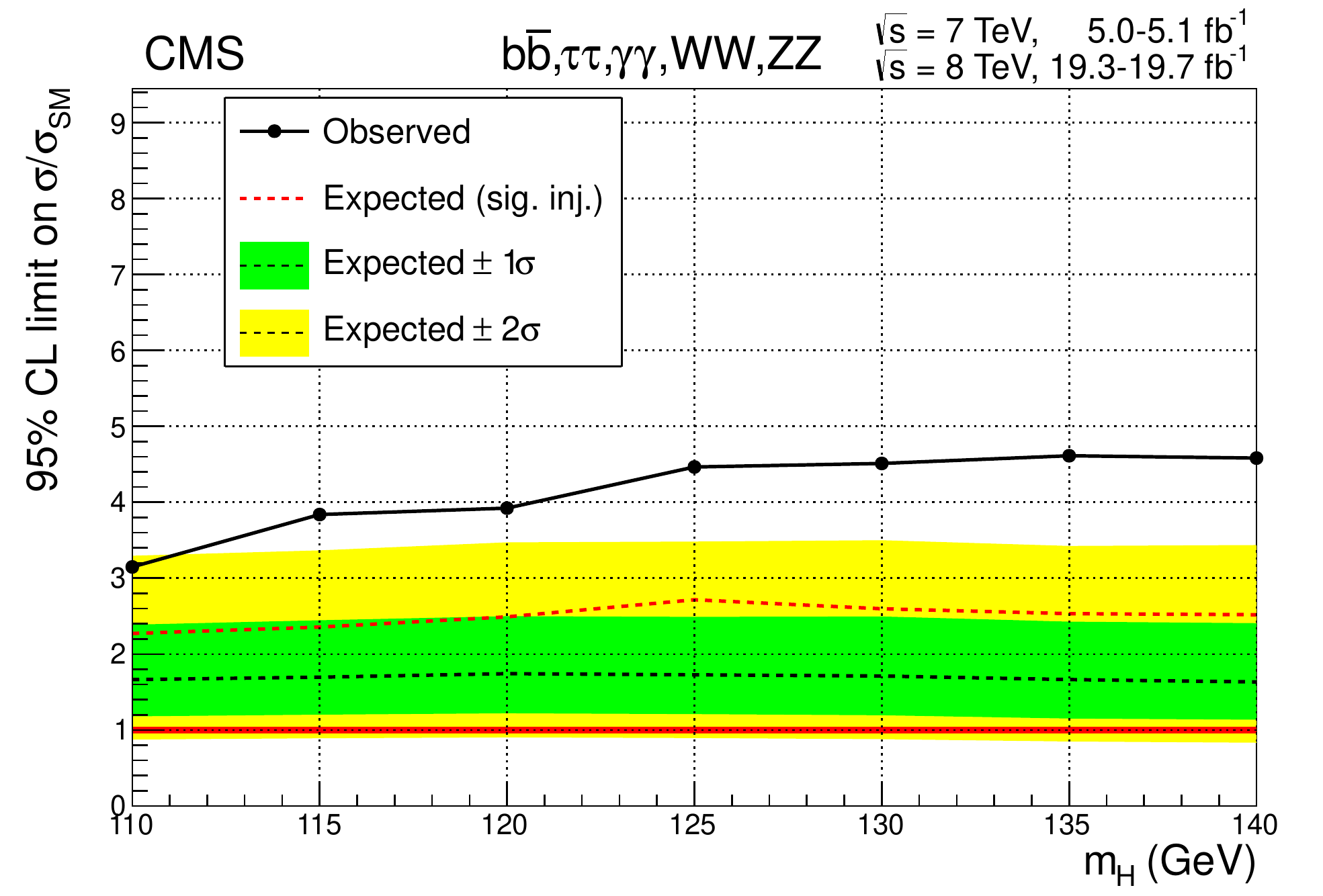}
    \includegraphics[width=0.49\textwidth]{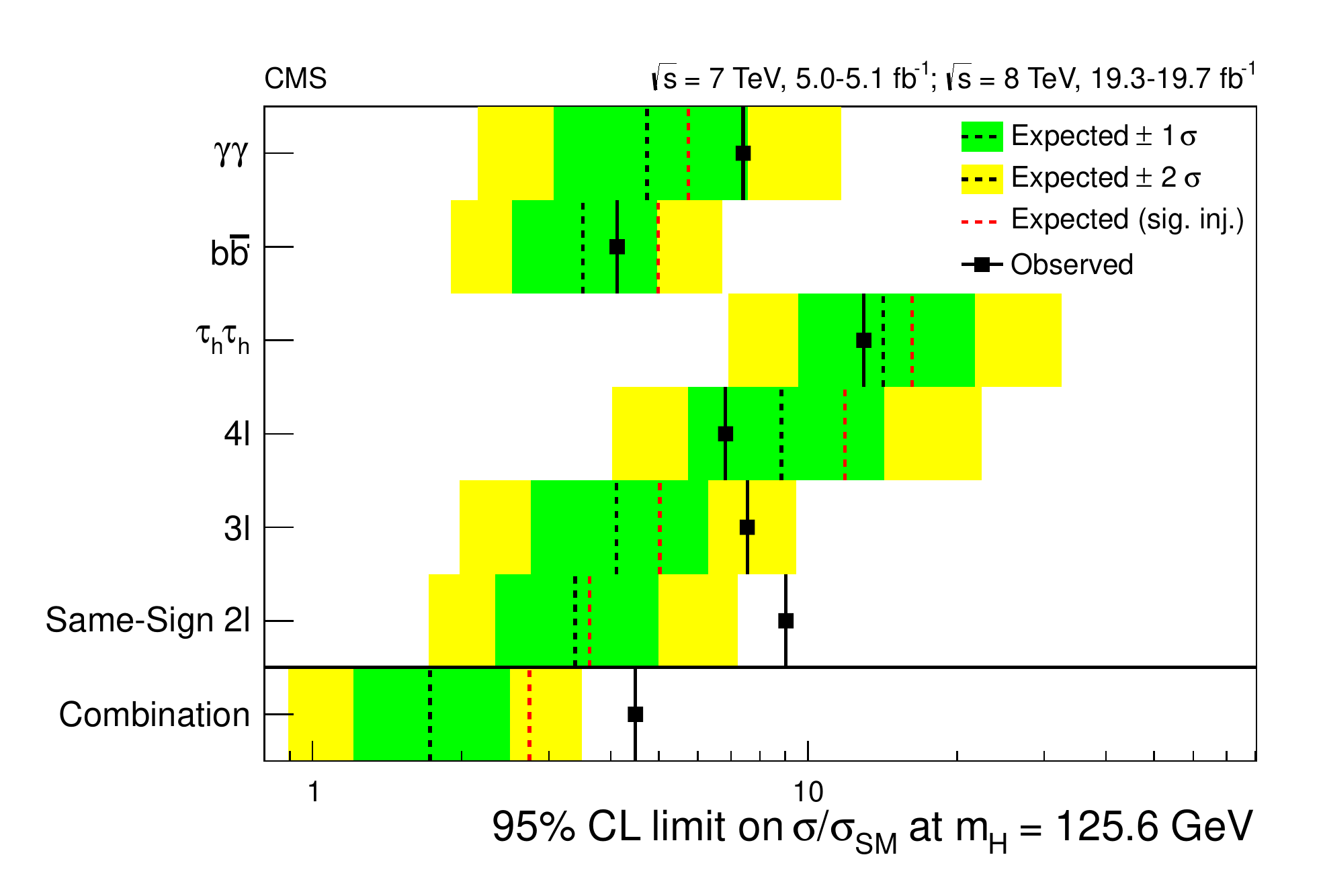}
    \caption{The 95\% CL upper limits on the signal strength parameter
      $\mu = \sigma/\sigma_{\mathrm{SM}}$. The black solid and dotted
      lines show the observed and background-only expected limits,
      respectively. The red dotted line shows the median expected
      limit for the SM Higgs boson with $m_{\PH}$ = 125.6\GeV. The green
      and yellow areas show the $1\sigma$ and $2\sigma$ bands, respectively. Left: limits as a function of $m_{\PH}$ for all channels combined. Right: limits for each channel at $m_{\PH}$ = 125.6\GeV.}
    \label{fig:comb_limit}}
\end{figure}

As BSM physics can enhance the production rate for the $\ttbar \PH$
and $\ttbar \PH + X$ final states, it is also useful to characterize
the upper limit on $\ttbar \PH$ production.  Furthermore, the expected
limit serves as a convenient gauge of the sensitivity of the analysis.
The 95\% CL expected and observed upper limits on $\mu$ are shown in
table~\ref{tab:comb_limit_mu} for $m_{\PH} = 125.6\GeV$ and as a
function of $m_{\PH}$ in figure~\ref{fig:comb_limit}, when combining
all channels.  Both the expected limit in the background-only
hypothesis and the hypothesis including the SM Higgs boson signal,
assuming the SM cross section, are quoted.  In addition to the median
expected limit under the background-only hypothesis, the bands that
contain the one and two standard deviation ranges around the median are
also quoted.  In the absence of a \ttH signal, the median expected
upper limit on $\mu$ from the combination of all channels is 1.7; the
corresponding median expectation under the hypothesis of SM \ttH
production with $m_{\PH} = 125.6\GeV$ is 2.7.  The observed upper
limit on $\mu$ is 4.5, larger than both expectations, compatible with
the observation that the best fit value of the signal strength
modifier $\mu$ is greater than one.  The limits for the individual
channels at $m_{\PH} = 125.6\GeV$ are given in the right panel of
figure~\ref{fig:comb_limit}.

\begin{table}[!htbp]
  \centering
\topcaption{The best-fit values of the signal strength parameter $\mu = \sigma/\sigma_{\mathrm{SM}}$ for each \ttH channel at $m_{\PH}$ = 125.6\GeV.  The signal strength in the four-lepton final state is not allowed to be below approximately $-6$ by the requirement that the expected signal-plus-background event yield must not be negative in either of the two jet multiplicity bins.  The observed and expected 95\% CL upper limits on the signal strength parameter $\mu = \sigma/\sigma_{\mathrm{SM}}$ for each \ttH channel at $m_{\PH}$ = 125.6\GeV are also shown. }
  {\footnotesize
    \begin{tabular}{|c|c|c|c|c|c|c|} \hline
\ttH channel & Best-fit $\mu$ &  \multicolumn{5}{c|}{95\% CL upper limits on $\mu = \sigma/\sigma_{\mathrm{SM}}$ ($m_{\PH} = 125.6$\GeV)} \\ \hline\hline
  &  &  &  \multicolumn{4}{c|}{Expected} \\
  & \multirow{2}{*}{Observed} &  \multirow{2}{*}{Observed}  &  Median  &  \multirow{2}{*}{Median}  &  \multirow{2}{*}{68\% CL range}  &  \multirow{2}{*}{95\% CL range}\\
  &  &   &  signal-injected  &   &   &  \\ \hline
\rule[-1.4ex]{0pt}{4ex} $\gamma\gamma$ & $+2.7^{+2.6}_{-1.8}$ & 7.4 & 5.7 & 4.7 & [3.1, 7.6] & [2.2, 11.7] \\
\rule[-1.4ex]{0pt}{4ex} $b\overline{b}$ & $+0.7^{+1.9}_{-1.9}$ & 4.1 & 5.0 & 3.5 & [2.5, 5.0] & [1.9, 6.7] \\
\rule[-1.4ex]{0pt}{4ex} $\tau_\mathrm{h}\tau_\mathrm{h}$ & $-1.3^{+6.3}_{-5.5}$ & 13.0 & 16.2 & 14.2 & [9.5, 21.7] & [6.9, 32.5] \\
\rule[-1.4ex]{0pt}{4ex} 4l & $-4.7^{+5.0}_{-1.3}$ & 6.8 & 11.9 & 8.8 & [5.7, 14.3] & [4.0, 22.5]  \\
\rule[-1.4ex]{0pt}{4ex} 3l & $+3.1^{+2.4}_{-2.0}$ & 7.5 & 5.0 & 4.1 & [2.8, 6.3] & [2.0, 9.5] \\
\rule[-1.4ex]{0pt}{4ex} Same-sign 2l & $+5.3^{+2.1}_{-1.8}$ & 9.0 & 3.6 & 3.4 & [2.3, 5.0] & [1.7, 7.2] \\ \hline
\rule[-1.4ex]{0pt}{4ex} Combined & $+2.8^{+1.0}_{-0.9}$  & 4.5 & 2.7 & 1.7 & [1.2, 2.5] & [0.9, 3.5] \\
\hline
\end{tabular}
}
  \label{tab:comb_limit_mu}
\end{table}

\section{Summary}
\label{sec:summary}
The production of the standard model Higgs boson in association with a
top-quark pair has been investigated using data recorded by the CMS
experiment in 2011 and 2012, corresponding to integrated luminosities
of up to 5.1\fbinv and 19.7\fbinv at $\sqrt{s} = 7\TeV$ and 8\TeV respectively. Signatures resulting from different combinations
of decay modes for the top-quark pair and the Higgs boson have been
analyzed. In particular, the searches have been optimized for the $\PH
\to \bbbar$, $\tauh\tauh$, $\Pgg \Pgg$, $\PW\PW$,
and $\Z\Z$ decay modes. The best-fit value for the signal strength
$\mu$ is $2.8 \pm 1.0$ at 68\% confidence level.  This result
represents an excess above the background-only expectation of 3.4
standard deviations.  Compared to the SM expectation including the
contribution from $\ttbar \PH$, the observed excess is equivalent to a
2-standard-deviation upward fluctuation.  These results are obtained
assuming a Higgs boson mass of 125.6\GeV but they do not vary
significantly for other choices of the mass in the vicinity of 125\GeV.  These results are more consistent with the SM $\ttbar \PH$
expectation than with the background-only hypothesis.

\section*{Acknowledgments}
\hyphenation{Bundes-ministerium Forschungs-gemeinschaft Forschungs-zentren} We congratulate our colleagues in the CERN accelerator departments for the excellent performance of the LHC and thank the technical and administrative staffs at CERN and at other CMS institutes for their contributions to the success of the CMS effort. In addition, we gratefully acknowledge the computing centres and personnel of the Worldwide LHC Computing Grid for delivering so effectively the computing infrastructure essential to our analyses. Finally, we acknowledge the enduring support for the construction and operation of the LHC and the CMS detector provided by the following funding agencies: the Austrian Federal Ministry of Science, Research and Economy and the Austrian Science Fund; the Belgian Fonds de la Recherche Scientifique, and Fonds voor Wetenschappelijk Onderzoek; the Brazilian Funding Agencies (CNPq, CAPES, FAPERJ, and FAPESP); the Bulgarian Ministry of Education and Science; CERN; the Chinese Academy of Sciences, Ministry of Science and Technology, and National Natural Science Foundation of China; the Colombian Funding Agency (COLCIENCIAS); the Croatian Ministry of Science, Education and Sport, and the Croatian Science Foundation; the Research Promotion Foundation, Cyprus; the Ministry of Education and Research, Estonian Research Council via IUT23-4 and IUT23-6 and European Regional Development Fund, Estonia; the Academy of Finland, Finnish Ministry of Education and Culture, and Helsinki Institute of Physics; the Institut National de Physique Nucl\'eaire et de Physique des Particules~/~CNRS, and Commissariat \`a l'\'Energie Atomique et aux \'Energies Alternatives~/~CEA, France; the Bundesministerium f\"ur Bildung und Forschung, Deutsche Forschungsgemeinschaft, and Helmholtz-Gemeinschaft Deutscher Forschungszentren, Germany; the General Secretariat for Research and Technology, Greece; the National Scientific Research Foundation, and National Innovation Office, Hungary; the Department of Atomic Energy and the Department of Science and Technology, India; the Institute for Studies in Theoretical Physics and Mathematics, Iran; the Science Foundation, Ireland; the Istituto Nazionale di Fisica Nucleare, Italy; the Korean Ministry of Education, Science and Technology and the World Class University program of NRF, Republic of Korea; the Lithuanian Academy of Sciences; the Ministry of Education, and University of Malaya (Malaysia); the Mexican Funding Agencies (CINVESTAV, CONACYT, SEP, and UASLP-FAI); the Ministry of Business, Innovation and Employment, New Zealand; the Pakistan Atomic Energy Commission; the Ministry of Science and Higher Education and the National Science Centre, Poland; the Funda\c{c}\~ao para a Ci\^encia e a Tecnologia, Portugal; JINR, Dubna; the Ministry of Education and Science of the Russian Federation, the Federal Agency of Atomic Energy of the Russian Federation, Russian Academy of Sciences, and the Russian Foundation for Basic Research; the Ministry of Education, Science and Technological Development of Serbia; the Secretar\'{\i}a de Estado de Investigaci\'on, Desarrollo e Innovaci\'on and Programa Consolider-Ingenio 2010, Spain; the Swiss Funding Agencies (ETH Board, ETH Zurich, PSI, SNF, UniZH, Canton Zurich, and SER); the Ministry of Science and Technology, Taipei; the Thailand Center of Excellence in Physics, the Institute for the Promotion of Teaching Science and Technology of Thailand, Special Task Force for Activating Research and the National Science and Technology Development Agency of Thailand; the Scientific and Technical Research Council of Turkey, and Turkish Atomic Energy Authority; the National Academy of Sciences of Ukraine, and State Fund for Fundamental Researches, Ukraine; the Science and Technology Facilities Council, UK; the US Department of Energy, and the US National Science Foundation.

Individuals have received support from the Marie-Curie programme and the European Research Council and EPLANET (European Union); the Leventis Foundation; the A. P. Sloan Foundation; the Alexander von Humboldt Foundation; the Belgian Federal Science Policy Office; the Fonds pour la Formation \`a la Recherche dans l'Industrie et dans l'Agriculture (FRIA-Belgium); the Agentschap voor Innovatie door Wetenschap en Technologie (IWT-Belgium); the Ministry of Education, Youth and Sports (MEYS) of the Czech Republic; the Council of Science and Industrial Research, India; the HOMING PLUS programme of Foundation for Polish Science, cofinanced from European Union, Regional Development Fund; the Compagnia di San Paolo (Torino); the Consorzio per la Fisica (Trieste); MIUR project 20108T4XTM (Italy); the Thalis and Aristeia programmes cofinanced by EU-ESF and the Greek NSRF; and the National Priorities Research Program by Qatar National Research Fund.

\bibliography{auto_generated}

\cleardoublepage \appendix\section{The CMS Collaboration \label{app:collab}}\begin{sloppypar}\hyphenpenalty=5000\widowpenalty=500\clubpenalty=5000\textbf{Yerevan Physics Institute,  Yerevan,  Armenia}\\*[0pt]
V.~Khachatryan, A.M.~Sirunyan, A.~Tumasyan
\vskip\cmsinstskip
\textbf{Institut f\"{u}r Hochenergiephysik der OeAW,  Wien,  Austria}\\*[0pt]
W.~Adam, T.~Bergauer, M.~Dragicevic, J.~Er\"{o}, C.~Fabjan\cmsAuthorMark{1}, M.~Friedl, R.~Fr\"{u}hwirth\cmsAuthorMark{1}, V.M.~Ghete, C.~Hartl, N.~H\"{o}rmann, J.~Hrubec, M.~Jeitler\cmsAuthorMark{1}, W.~Kiesenhofer, V.~Kn\"{u}nz, M.~Krammer\cmsAuthorMark{1}, I.~Kr\"{a}tschmer, D.~Liko, I.~Mikulec, D.~Rabady\cmsAuthorMark{2}, B.~Rahbaran, H.~Rohringer, R.~Sch\"{o}fbeck, J.~Strauss, A.~Taurok, W.~Treberer-Treberspurg, W.~Waltenberger, C.-E.~Wulz\cmsAuthorMark{1}
% ...
\vskip\cmsinstskip
% ...
1:~~Also at Vienna University of Technology, Vienna, Austria\\
2:~~Also at CERN, European Organization for Nuclear Research, Geneva, Switzerland\\
% ...

\end{sloppypar}
\end{document}